\documentclass[useAMS,usenatbib]{mnras}
\usepackage[fleqn]{amsmath}

\usepackage{newtxtext,newtxmath}

\usepackage[T1]{fontenc}
\usepackage{ae,aecompl}

\usepackage{amsfonts,bm}
\usepackage{graphicx}
\usepackage[utf8]{inputenc}
\usepackage{adjustbox}   
\usepackage{subfig}
\usepackage{marvosym}

\newcommand  *{\diff}   {\mathop{}\!\mathrm{d}}
\renewcommand*{\vec}[1] {\boldsymbol{#1}}
\newcommand  *{\uvec}[1]{\hat{\vec{#1}}}
\newcommand  *{\s}[1]   {\mathsf{#1}}
\newcommand  *{\mat}[1] {\vec{\s{#1}}}
\newcommand  *{\Ups}    {\Upsilon}
\newcommand  *{\The}    {\Theta}
\newcommand  *{\I}      {\mathrm{i}}

\newcommand  *{\Exp}[1] {\mathrm{e}^{\textstyle #1}}
\newcommand  *{\phn}    {\phantom{-}}
\newcommand  *{\pfrac}[2] {\left(\frac{#1}{#2}\right)}
\newcommand   {\mbhm}{M_{\text{BH}}}
\newcommand   {\mbh}{$\text{M}_{\text{BH}}$}
\newcommand   {\mbul}{M$_{\text{bul}}$}
\newcommand   {\mbulm}{\text{M}_{\text{bul}}}

\newcommand   {\rsoi}{\text{r}_\text{SOI}}
\newcommand   {\ml}{$\Upsilon$}
\newcommand   {\mlm}{\Upsilon}
\newcommand   {\ellip}{$\varepsilon$}

\newcommand{\va}{$\left(\theta, \phi, \psi\right)$}

\title[The Black Hole of NGC~708]{Triaxial Schwarzschild Models of NGC~708: a 10-billion solar mass black hole in a low dispersion galaxy with a Kroupa IMF}

\author[S.~de Nicola et al.]{%
Stefano de Nicola,$^{\!\!2,1}$\thanks{E-mail: denicola@mpe.mpg.de}
Jens Thomas,$^{\!\!1,2}$
Roberto P. Saglia,$^{\!\!1,2}$
Jan Snigula,$^{\!\!1}$
Matthias Kluge,$^{\!\!1,2}$
\newauthor
and Ralf Bender $^{\!\!2,1}$
\\ \\
$^{1}$ Max-Planck Institute for Extraterrestrial Physics, Giessenbachstrasse 1, D-85748, Garching (Germany) \\
$^{2}$ Universit{\"a}ts-Sternwarte Muenchen, Scheinerstrasse 1, D-81679, Munich, Germany\\
}

\date{Accepted 2024 March 15. Received 2024 March 12; in original form 2024 January 19}

\pubyear{2023}

\begin{document}
\label{firstpage}
\pagerange{\pageref{firstpage}--\pageref{lastpage}}
\maketitle

\begin{abstract}
We report the discovery of a $(1.0 \pm 0.28) \times 10^{10}$ M$_\odot$ Supermassive Black Hole (SMBH) at the centre of NGC 708, the Brightest Cluster Galaxy of Abell 262. Such high BH masses are very rare and allow to investigate BH - host galaxy scaling relations at the high mass end, which in turn provide hints about the (co)evolution of such systems. NGC~708 is found to be an outlier in all the canonical scaling relations except for those linking the BH mass to the core properties. The galaxy mass-to-light ratio points to a Kroupa IMF rather than Salpeter, with this finding confirmed using photometry in two different bands. 
We perform this analysis using our novel triaxial Schwarzschild code to integrate orbits in a 5-dimensional space, using a semi-parametric deprojected light density to build the potential and non-parametric Line-of-Sight Velocity Distributions (LOSVDs) derived from long-slit spectra recently acquired at Large Binocular Telescope (LBT) to exploit the full information in the kinematic. We find that the galaxy geometry changes as a function of the radius going from prolate, nearly spherical in the central regions to triaxial at large radii, highlighting the need to go beyond constant shape profiles. 
Our analysis is only the second of its kind and will systematically be used in the future to hunt Supermassive Black Holes in giant ellipticals.

\end{abstract}

\begin{keywords}
	galaxies: elliptical and lenticular, cD --
	galaxies: kinematics and dynamics --
	galaxies: structure
\end{keywords}    



\section{Introduction}
\label{Sec.introduction}
NGC~708 is the Brightest Cluster Galaxy (BCG) of the Abell 262 cluster. It is located at a distance of 68.48 Mpc and has an absolute magnitude $M_{g'} =  -22.89$ \citep{Matthias20}. The galaxy is commonly classified as cD (see e.g. \citealt{Wegner12}). According to this classification, the galaxy is surrounded by a diffuse stellar envelope, whose origin is believed to lie in ex-situ stellar accretion \citep{Cooper15, Pillepich18}, and whose evolution is tightly linked to the whole cluster rather than to the BCG itself \citep{Gonzalez05}. This stellar envelope probably traces the cluster potential and is typically referred to as intra-cluster light (ICL) \citep{Matthias20}. \citet{Matthias21} attempt to dissect the BCG light from the ICL in several ways, finding that an accurate photometric decomposition cannot be obtained. A possible solution to the problem could be the comparison of SB profiles of BCGs with those of ordinary ETGs \citep{Matthias23}. \citet{Matthias20} report as best-fit parameters for the SB profile of this galaxy a single Sersic profile with Sersic index $n = 2.96 \pm 0.11$ and effective radius r$_e = 54.26^{+1.32}_{-1.29}$ kpc. 


HST images of the galaxy show the presence of a prominent, edge-on dust lane, extending $\sim$3" in the southern direction and $\sim$6" in the northern direction \citep{Wegner12}. Its origin probably lies in a merger, a commonly observed phenomenon in BCGs \citep{Lauer14}. In fact, BCGs lie at the center of potential wells in galaxy clusters, thus being in the ideal position to accrete material, and to experience several merging events. Another possible indication of a merger is given by the wiggles observed in the SB, ellipticity $\varepsilon$ and PA profiles (see Sec.~\ref{Ssec.iso_features}). \\
Perhaps the most interesting clue about merger(s) that this galaxy has experienced can be found in its light-deficient central core, which extends out to $\sim$2.2" ($\sim$0.73 kpc, see Sec.~\ref{Ssec.corerad}). The most plausible formation mechanism is core scouring: stars on radial orbits, which come closest to the galaxy centre, are ejected via gravitational slingshot by a shrinking Supermassive Black Hole (SMBH) binary \citep{Ebisuzaki91, Faber97, Merritt06, Milosavljevic01, Jens14, Jens16, Kianusch19} that formed after a major (dry) merger. The ejection of stars on radial orbits also generates a tangential anisotropy in the core, a commonly observed phenomenon in core-galaxies \citep{Gebhardt03, VDB10, Jens14, Jens16}. Scaling relations linking the black hole mass \mbh\,to the core size \citep{Rusli13b, Rob16}, the missing light with respect to an inward extrapolation of the SB profile \citep{Kormendy09}, and the SB of the central core itself \citep{Kianusch19} have been discovered. For NGC~708, the core size predicts an \mbh\,value of $\geq 10^{10}\,M_\odot$. Hunting SMBHs with masses in this range will help in filling the high-mass end of the BH-host scaling relations: as it can be seen from Fig. 11 in \citet{Kianusch19}, \mbh\,estimates in this range are currently almost completely missing. \\
Another interesting issue lies in the stellar Initial Mass Function (IMF) of massive galaxies. Although there are massive ETGs following a lightweight IMF \citep{Jens15, Jens16}, several studies \citep{Jens11, Cappellari12, Tortora14} suggest that the most massive ETGs may follow a bottom-heavy (e.g. \citealt{Salpeter55}) IMF. To test this idea, mass-to-light ratios \ml\,calculated by dynamically modeling these galaxies are compared to estimations from stellar population analysis \citep{Thomas03, Maraston05, Maraston11, Conroy17, Parikh18} assuming a \citet{Kroupa01} IMF, finding systematically overestimated \ml\,with respect to Kroupa IMF, even if this could also signal Dark Matter (DM) tracing the stars. In \citet{Wegner12}, \ml\,estimates using both SSP ($\mlm_\text{SSP}$) and dynamical models ($\mlm_*)$ for NGC~708 are published, finding evidence for a bottom-heavy IMF. However, \citet{Wegner12} did not include a central BH in their axisymmetric dynamical modeling. \\
In order to obtain reliable estimates of the mass parameters, it is essential to robustly recover the intrinsic shape of a galaxy. BCGs have been shown to be extremely triaxial objects \citep{dN22}, even more than ordinary early-type galaxies \citep{Vincent2005}. For NGC~708 \citet{dN22} showed by deprojecting the surface brightness profile of the galaxy that, for the best-fit orientations, the galaxy is close to being spherical in the central regions, but becomes triaxial at large radii: the triaxiality parameter $T = (1 - p^2) / (1 - q^2)$, where $p = b/a$ and $q = c/a$ are the intrinsic axis ratios with $a > b > c$, reaches the maximum value of 0.5 at $r \sim 3$ kpc, before going down to 0.3 at larger radii. In this case, an axisymmetric approximation can possibly yield biased black hole mass \mbh\, estimates \citep{VDB10} as well as \ml\, estimates biased by 50\% \citep{Jens07}.  \\  
In two recent works, \citet{dN22b} and \citet{Bianca23a} tested a new sophisticated triaxial machinery which combines the semi-parametric deprojection algorithm SHAPE3D \citep{dN20}, the non-parametric code for kinematics extraction WINGFIT \citep{Kianusch23} and the triaxial Schwarzschild code SMART \citep{Bianca21}, along with the novel model selection technique of \citet{Mathias21} and \citet{Jens22} which avoids $\chi^2$-biases linked to the different number of Degrees-of-Freedom (DOFs) in each model. The authors apply this to an $N$-body simulation \citep{Rantala18} which reproduces the formation of a typical core-galaxy, recovering the correct galaxy shape and orbit distribution with an accuracy $\Delta p, \Delta q, \Delta \beta < 0.1$, where $\beta$ is the anisotropy. In a companion paper \citep{Bianca23a} the same simulation is used to show that also \mbh\,and the mass-to-light ratio \ml\, can be recovered within 5-10\%. \\
The main goal of this paper is to apply our triaxial machinery to NGC~708 to investigate its intrinsic properties and in particular the mass of its SMBH. To this extent, we also make use of new LBT kinematical data and K-band photometry acquired at the Wendelstein observatory. The paper is structured as follows. Sec.~\ref{Sec.photo} describes the photometric data and the deprojections, while Sec.~\ref{Sec.kin} describes the spectroscopic data and kinematics extraction. Sec.~\ref{Sec.dynmod} presents the results of the dynamical modeling, which are discussed in Sec.~\ref{Sec.discussion}. Finally, we draw our conclusions in Sec.~\ref{Sec.conclusions}. 

\begin{figure*}

\subfloat[\label{Fig.A262_hst}]{\includegraphics[width=.3\textwidth]{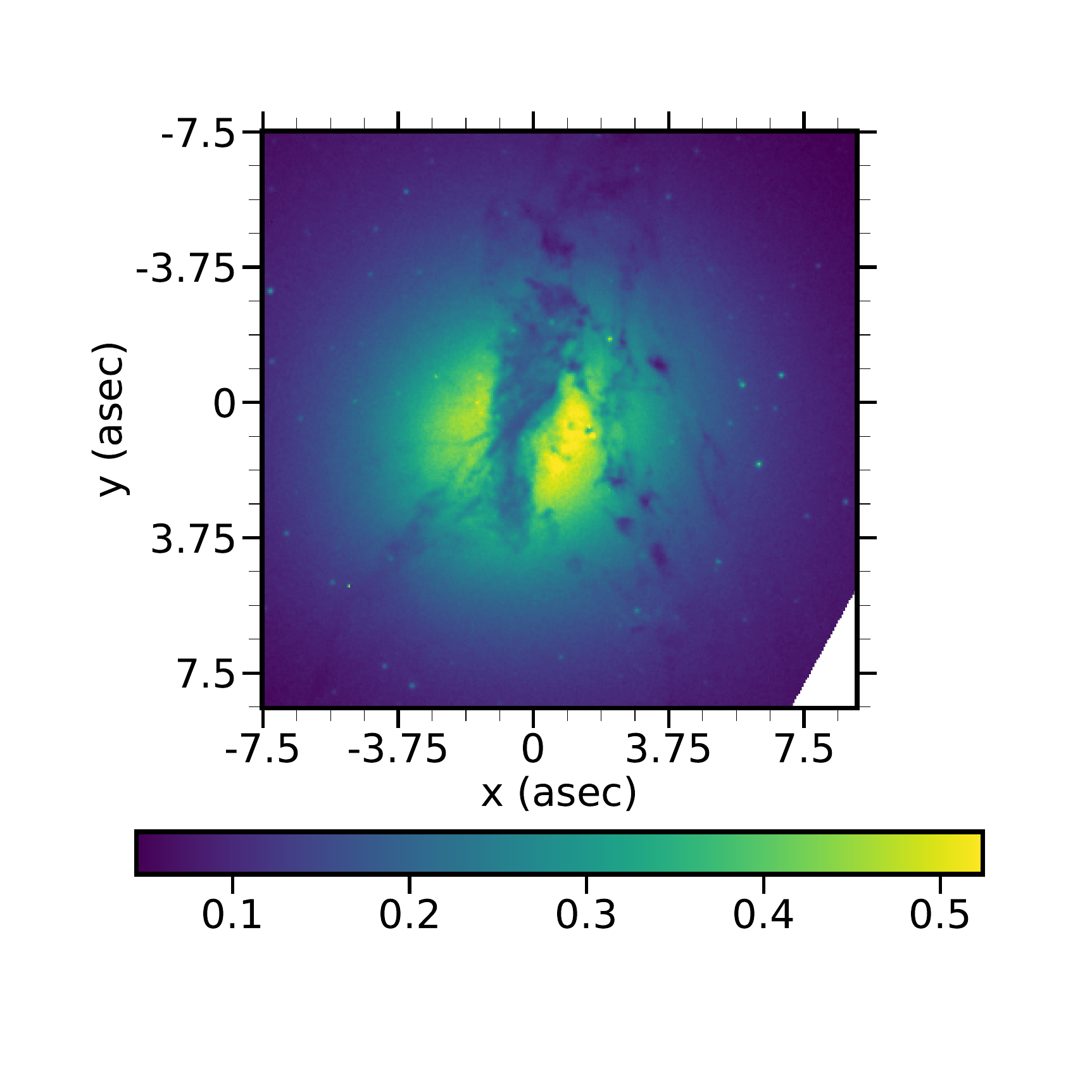}}     \hfil
\subfloat[\label{Fig.A262_hst_large}]{\includegraphics[width=.3\textwidth]{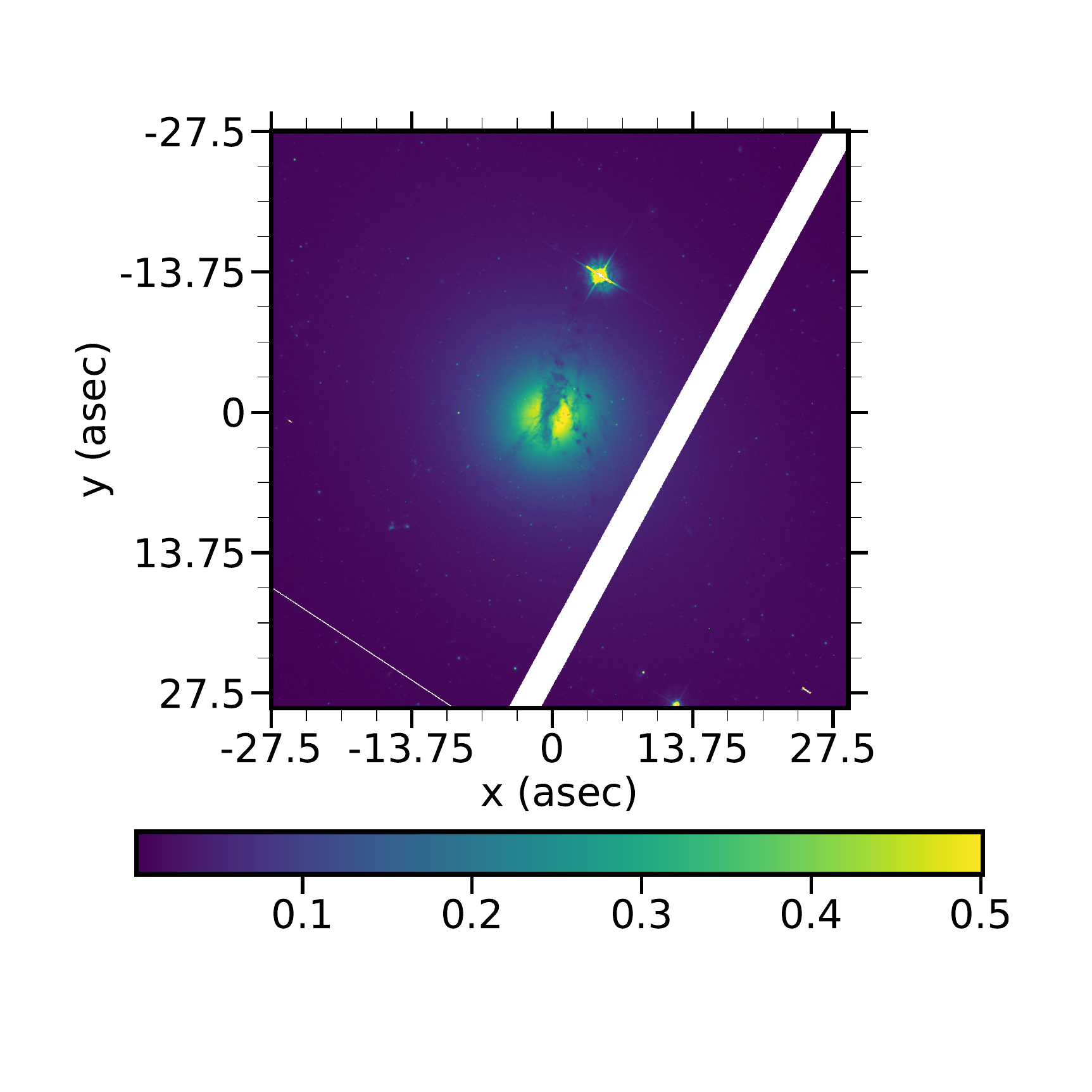}}     \hfil
\subfloat[\label{Fig.A262_hst_mask}]{\includegraphics[width=.3\textwidth]{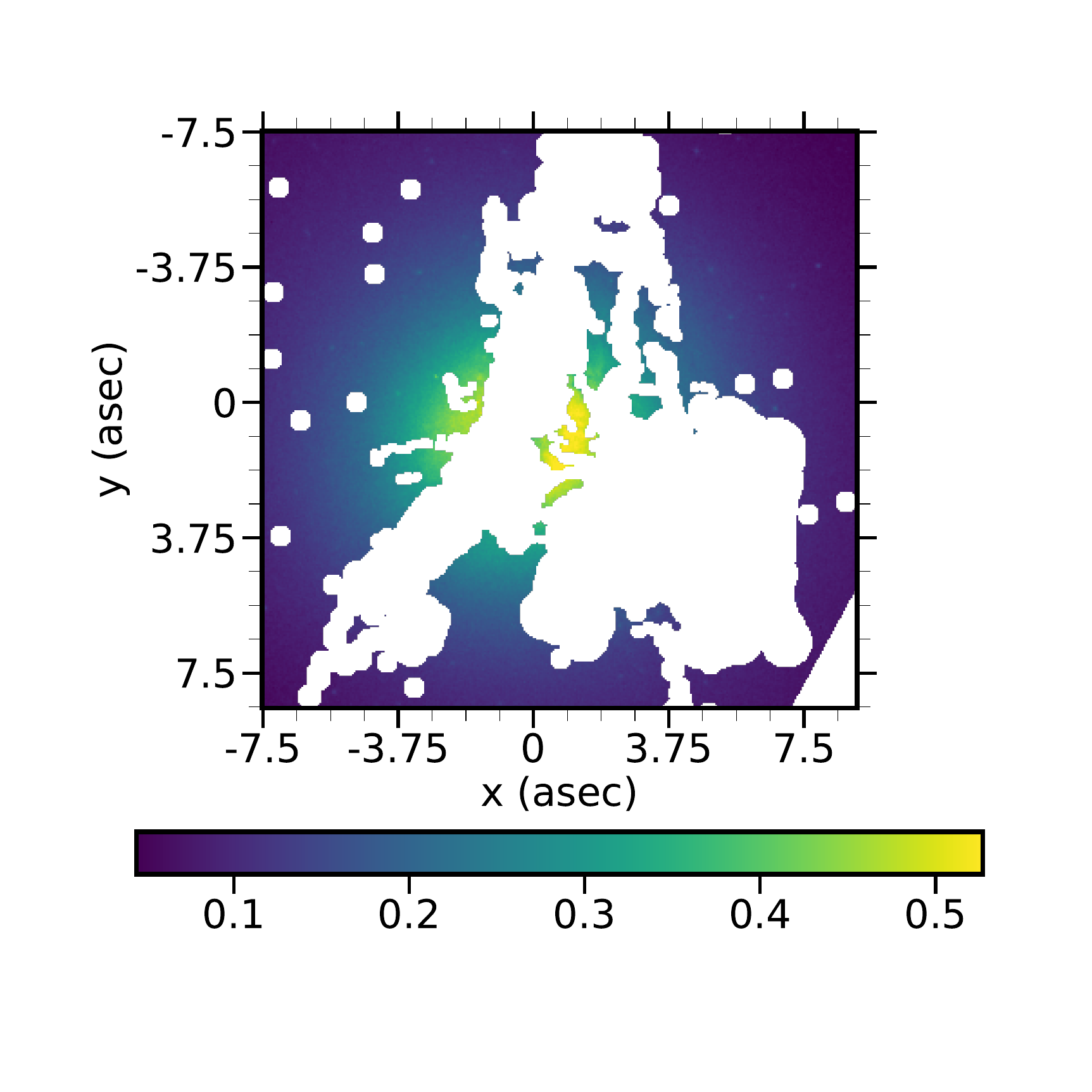}}

\subfloat[\label{Fig.A262_WST}]{\includegraphics[width=.3\textwidth]{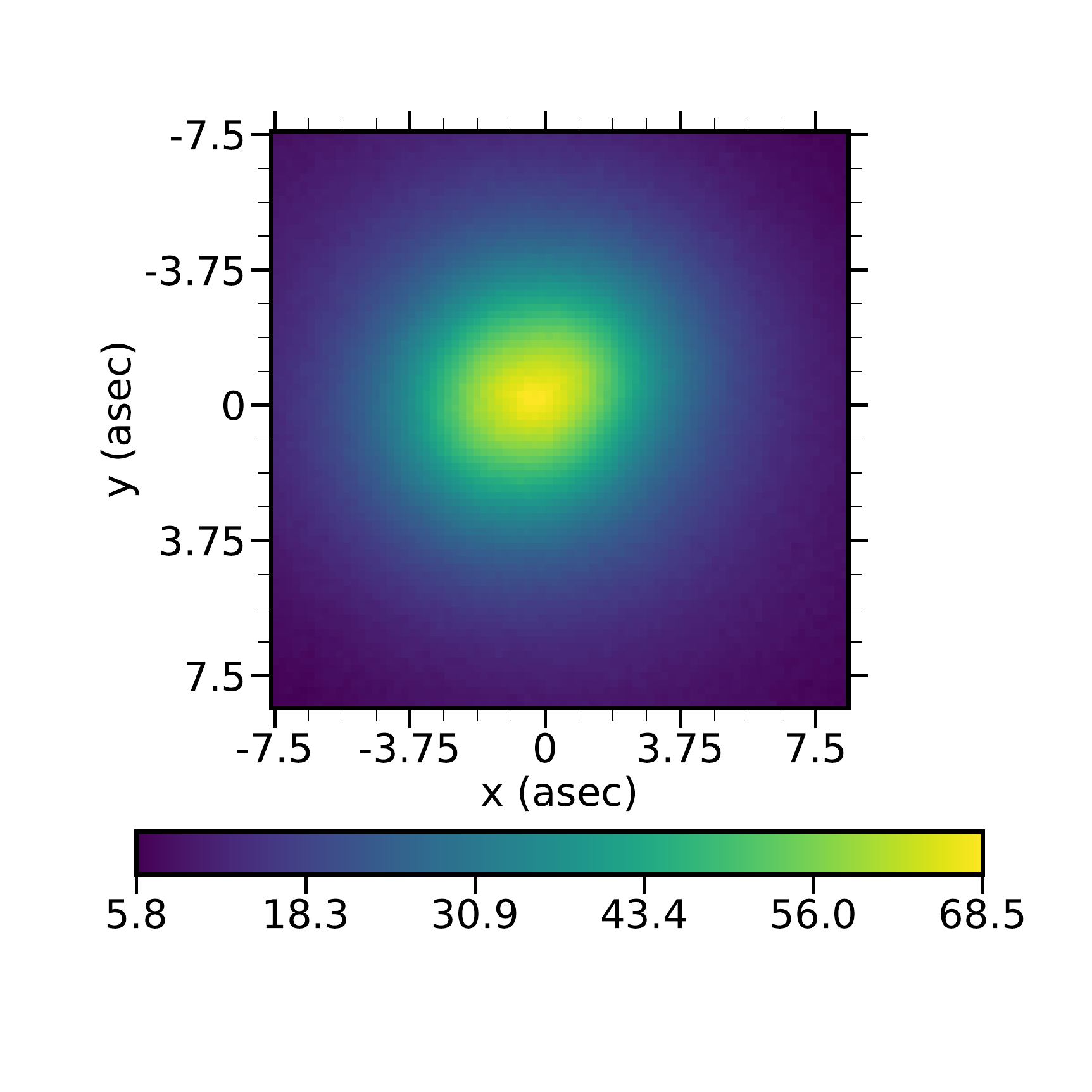}}   \hfil
\subfloat[\label{Fig.A262_WST_large}]{\includegraphics[width=.3\textwidth]{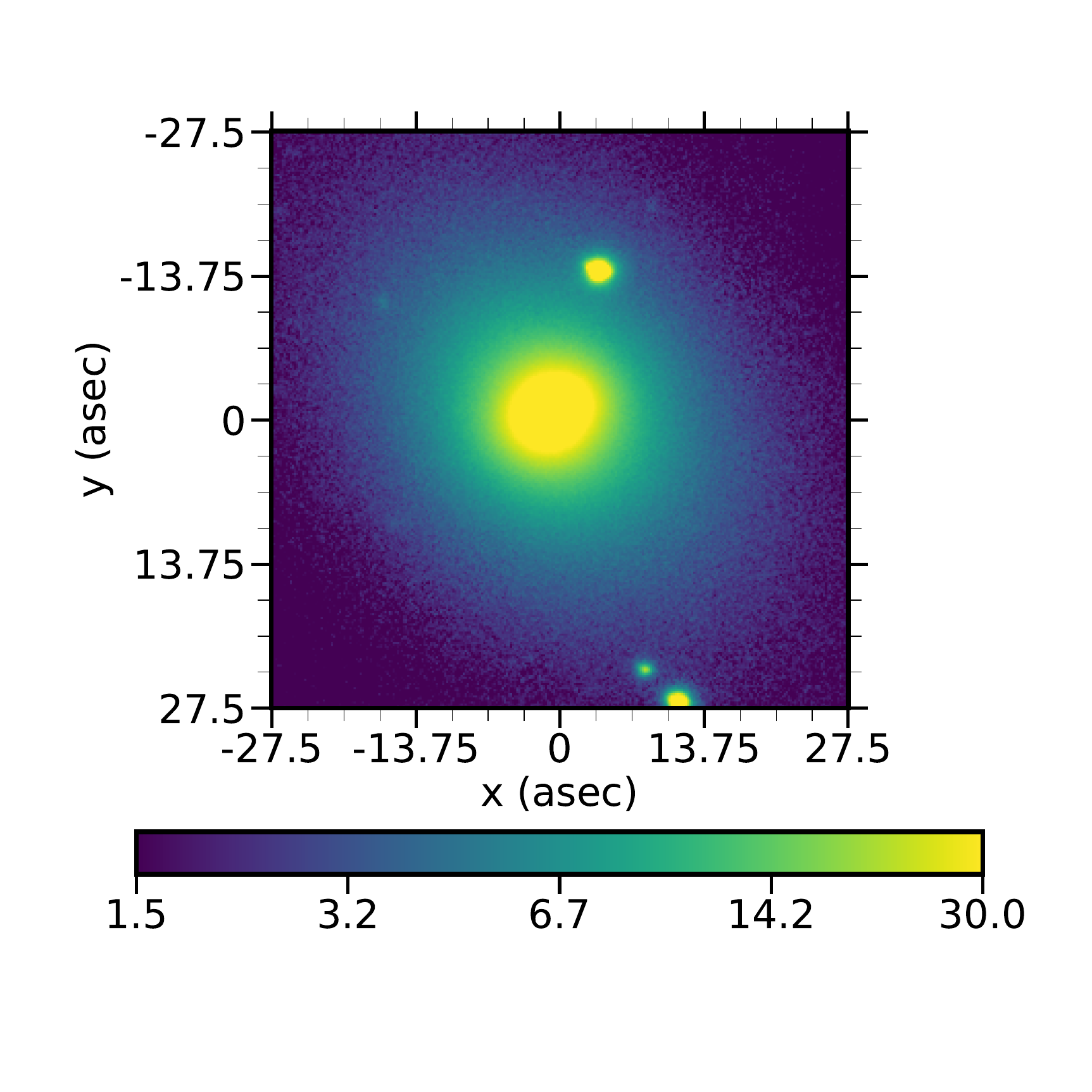}}   \hfil
\subfloat[\label{Fig.A262_WST_res}]{\includegraphics[width=.3\textwidth]{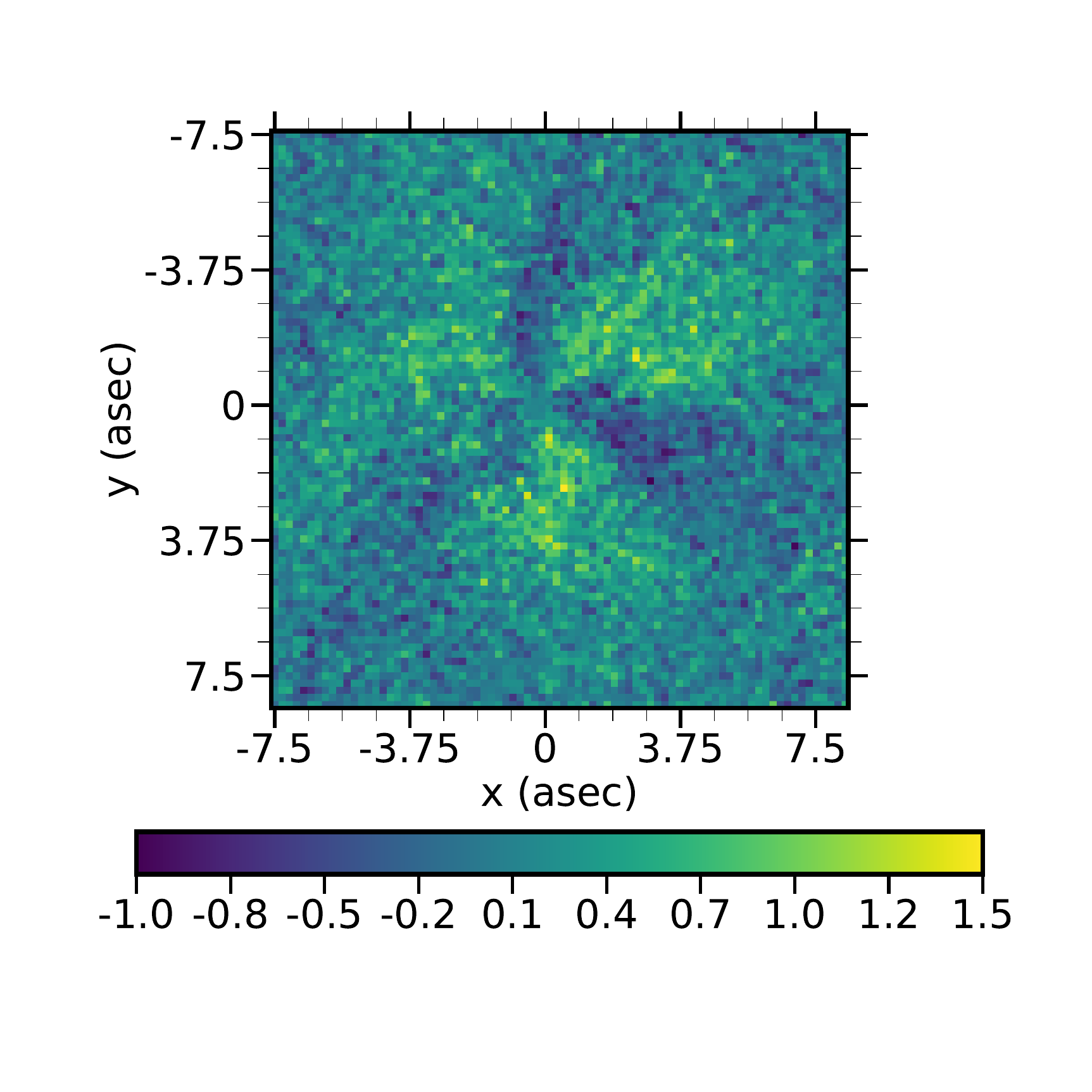}}

    \caption{HST/Wendelstein images of NGC~708. Images are color-coded according to the flux, which has been scaled for better visualization except for Figs.~\ref{Fig.A262_WST} \&~\ref{Fig.A262_WST_res} to show that the dust residuals are small. \textit{Top-left:} HST-image, with the prominent dust lane clearly visible. \textit{Top-middle} HST image with wider FOV. \textit{Top-right}: masked HST-image used to generate the isophotes shown in left panel of Fig.~\ref{Fig.photometry}. \textit{Bottom-left:} Wendelstein Ks-band image of NGC~708, no dust appearently visibile. \textit{Bottom-middle:} Wendelstein Ks-image with wider FOV. Here the isophote twist is evident. \textit{Bottom-right:} Ks-band residuals obtained by subtracting the fitted isophotes to the K-band image itself. A small residual due to dust is visible. In all images North is up and East to the left. }
    \label{Fig.A262}
\end{figure*}

\section{Photometry} \label{Sec.photo}

The photometry used in this work comes from three different image sources. The first is a $g'$-band image obtained with the Fraunhofer Telescope at the Wendelstein observatory \citep{Hopp10, LangBardl16} using the Wendelstein Wide Field Imager
(WWFI, \citealt{Kosyra14}), with typical seeing of FWHM = $(1.2 \pm 0.2)$" and pixel size of 0.2". The total field of view (FOV) of $27.6 \times 28.9$' allows to image the galaxy outskirts\footnote{The camera itself consists of 4 CCDs, each one having dimensions $4096 \times 4109$ pixels.}, where the galaxy light mixes up with the ICL. The galaxy was imaged following a 52-step dither pattern; corrections for bias, flat-field, cosmic rays, background, and bright stars were applied. Technical details can be found in Sec. 3 of \citet{Matthias20}. 
A second image comes from high-resolution HST observations (see top panels in Fig.~\ref{Fig.A262}), with typical resolution of 0.1". These were carried out using the Wide Field Planetary Camera 2 (WFPC2) using the filter F622W (GO program 10884; P.I. Wegner). The camera consists of a grid with dimensions $800 \times 800$ pixels, with pixelsize 0.0455", yielding a FOV of $800 \times 800$" \citep{Wegner12}. 
The galaxy belongs to a subset of 170 Brightest Cluster Galaxies (BCGs), observed down to a SB$_\text{g'}$ $\sim$ 30 mag asec$^{-2}$ \citep{Matthias20}. This allows to reach the galaxy regions where the interaction with the ICL becomes visible in the SB profile itself \citep{Matthias21}, as well as for a comparison between the galaxy intrinsic shape and the simulated DM halos \citep{dN22}. Complementarily, the HST images give us the necessary resolution in the central regions to resolve the black hole sphere-of-influence (SOI), where the potential is dominated by the black hole itself. This is crucial to derive reliable \mbh\,estimates. Finally, we also use a Ks-band Wendelstein image (see bottom panels in Fig.~\ref{Fig.A262}) to assess how much the presence of dust in the central region affects our analysis. \\
Because the two optical image sources are in different colour bands, we select the radii where we have data from both observation sets and combine them together. This is done by first interpolating the HST photometry at Wendelstein radii and then minimizing the differences between the two data sets. The resulting scaling factor is used to convert HST data to the $g$'-band. After the conversion has been made, we take HST data at $r < 15$", and WWFI data at $r > 15$". In this way, we have a very high resolution at the centre, which is what we need for a robust estimate of \mbh, and extending out to large radii ($\geq 30$ kpc). The resulting photometry is used to perform the triaxial deprojection described in Sec.~\ref{Ssec.depro}.

\subsection{Isophote features} \label{Ssec.iso_features}
\begin{figure*}
\centering

\subfloat[\label{Fig.iso}]{\includegraphics[width=.5\linewidth]{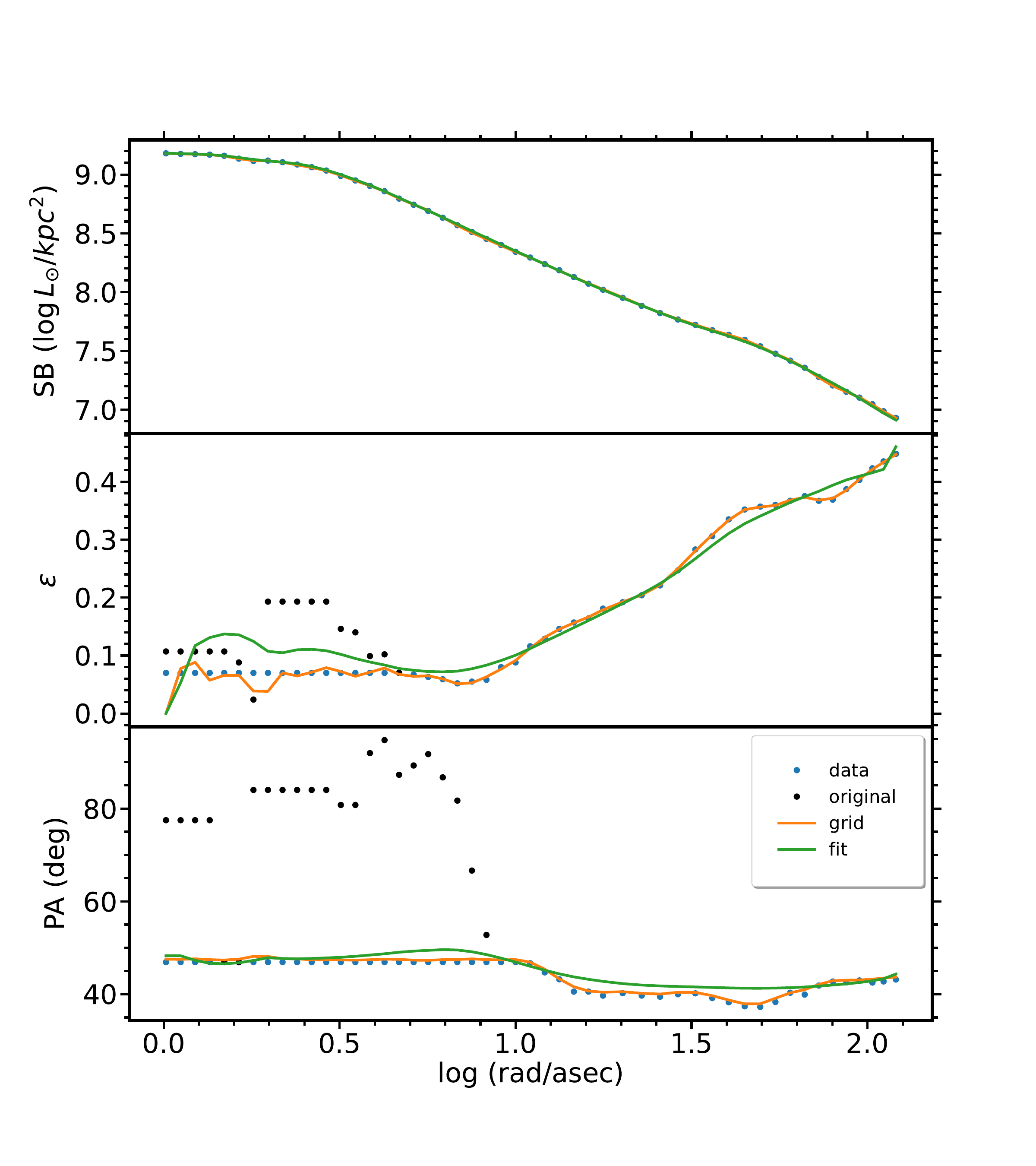}}
\subfloat[\label{Fig.css}]{\includegraphics[width=.5\linewidth]{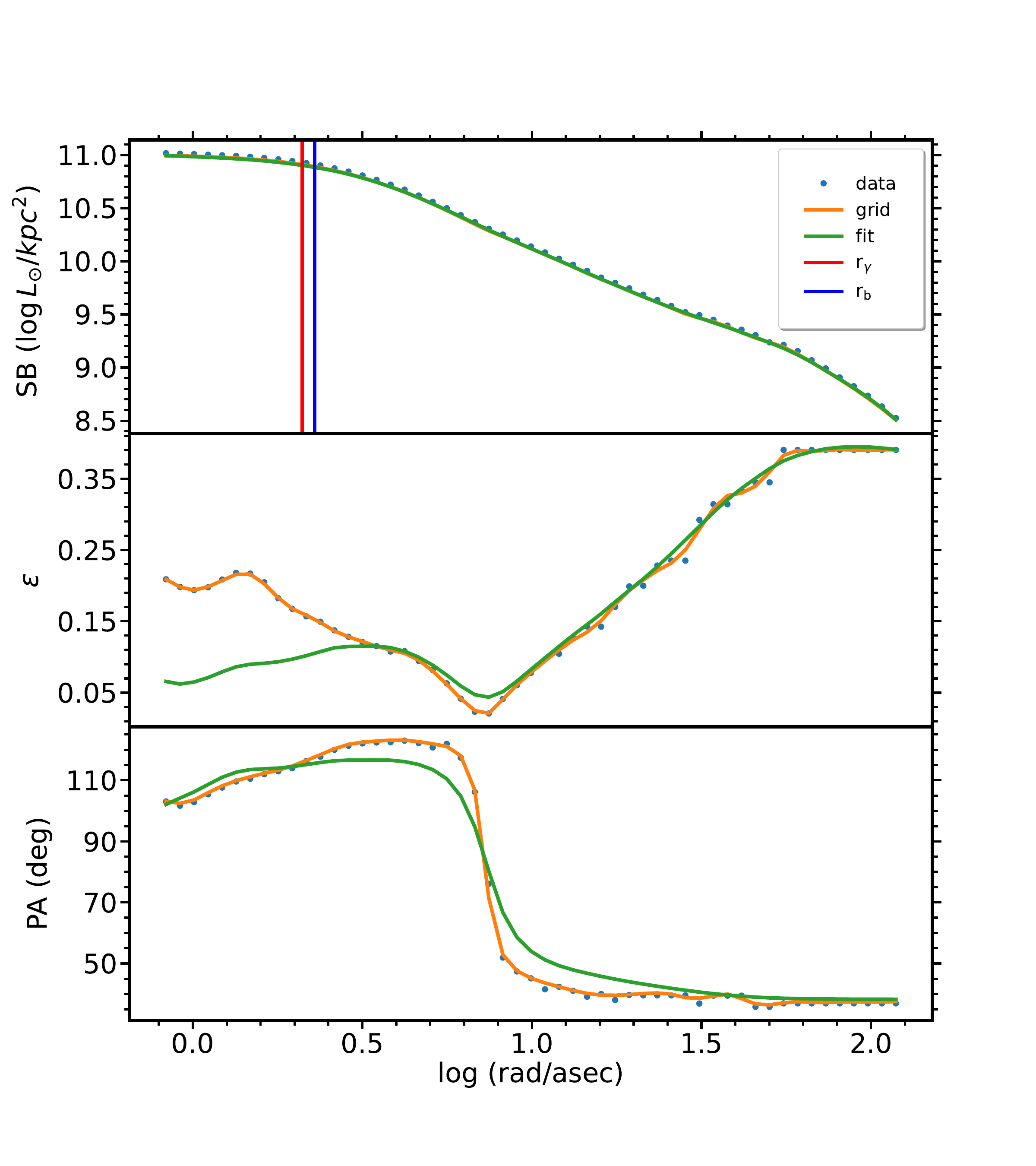}}

    \caption{Isophotes (from top to bottom: SB, ellipticity, PA) of the galaxy NGC~708. The blue points represent the observed photometry, while the orange lines are computed by placing the SB on the grid and performing isophotal fits. The green lines come from the projection of the deprojected SB. \textit{Right:} Same but using Ks-band Wendelstein deconvolved isophotes. In the top-right panel we show the position of cusp-radius r$_\gamma$ and break radius r$_\text{b}$, which we use to estimate the core size.}
    \label{Fig.photometry}
\end{figure*}

We use the method of \citet{Bender87} (for the g'-band image) and \citet{Matthias23, Matthias23b} (for the Ks-band image) to extract the isophote parameters from the galaxy image. The two methods yield very similar results. The variables we are interested in are the SB, the ellipticity $\varepsilon$, the position angle (PA) and the Fourier coefficients $a_n$, $b_n$ which quantify the deviations of the isophotes from the best-fit ellipses. This is done \textit{separately} for the two optical image sources before combining the isophotes as described above. The resulting g'- and Ks-band isophotes are shown in Fig.~\ref{Fig.photometry}. A few key-points to mention are:

\begin{itemize}
    \item There are wiggles in both the SB and the $\varepsilon$. These are probably signs of a recent merger;
    \item The $\varepsilon$ and PA profiles show huge bumps at small radii due to the dust lane. This also generates wiggles in the SB profile at small radii;
    \item With the exception of the innermost 10", all variables follow similar trends as those observed in ordinary ETGs \citep{Bender88, Bender92, Kormendy96}: \ellip\,is low at small radii before increasing at large radii, there is only a weak twist $< 10^\circ$ and the Fourier coefficients remain $\leq$ 1.5;
    \item Given the dust lane, it is difficult to tell whether the galaxy is indeed round in the central regions as it is usually the case for BCGs \citep{dN22}. Interestingly, \ellip\,goes towards zero already at 10", i.e. farther out than the core size (see Sec.~\ref{Ssec.corerad}).
    \item The ellipticity and PA profiles obtained from the Ks-band photometry also show large gradients. Indeed, after subtracting the best-fit elliptical model to the galaxy image, the residuals do show traces of dust, even if the flux loss due to dust is only $\sim$1.4\% in the Ks-band image (Figs.~\ref{Fig.A262_WST} \&~\ref{Fig.A262_WST_res}).
    
\noindent In practice, we are dealing with a (possibly still ongoing) merger with dust at the galaxy centre. To prevent dust from biasing our analysis, we set \ellip\,and the PA to constant values in the innermost regions (see left panel of Fig.~\ref{Fig.photometry}), while using the Ks-band image of the galaxy to discuss the impact of this assumption on the BH mass estimate. 

\subsubsection{ICL contamination}
 The light profile of NGC~708 is probably a superposition of the actual BCG light and an ICL component. 
According to \citet{Matthias21}, NGC~708 can be well described by a single Sersic component and, with the exception of the innermost regions, the ellipticity and PA profiles are typical of a massive ETG. For instance, the galaxy becomes flatter with increasing radius and displays a mild twist. Hence, the regions that we include in the dynamical modeling show no sign of being affected by a structurally distinct ICL component. Nevertheless, even in BCGs like NGC~708 without obvious signs of structural changes in the light profile, there is a potential contribution of the ICL. \citet{Matthias21} discuss several selection criteria that consistently yield typical ICL fractions around $\sim 50$ per cent. Among them a SB$_\text{g}'$ cut placed at 27 mag/asec$^2$: assuming that all the light fainter than this threshold belongs to the ICL on average results in 34\% of all diffuse light centered on the BCG belonging to the ICL. The outermost isophote used for our models has SB$_\text{g'}$ $<$ 25 mag/asec$^2$, well brighter than this threshold. This confirms again that our models are unlikely to be affected by the dynamics of the ICL. Hence, even if the ICL is contaminating the SB profile, any effect on the dynamical models is expected to be small and not to produce biases in our analysis. We may overestimate the luminosity of the actual BCG by up to a factor of $\sim 2$ though since we do not subtract any potential ICL contribution to the galaxy's light profile.
\end{itemize}





\subsection{Core radius} \label{Ssec.corerad}
Like many other massive ellipticals NGC~708 shows a light-deficient core with respect to an inward extrapolation of a single Sersic profile. Here, we focus on the core size using the (nearly) dust-free, deconvolved $K$-band Wendelstein photometry of the galaxy, whose FWHM $\lesssim$ 1.2" allows us to resolve the central core. The core size can be described both in terms of the cusp-radius r$_\gamma$, defined as the radius where dlog$\Sigma$/dlogr = -1/2, where $\Sigma = 10^{-0.4 \times \text{SB}}$, and of the break radius r$_\text{b}$, obtained by fitting a (double) core-Sersic law \citep{Graham03, Trujillo04} to the SB profile. Adopting the first definition, we measure a core of 2.10", corresponding to 0.70 kpc. The second method shows that the SB profile is better fitted by two components: a core-Sersic law, with Sersic index n$_1 = 1.84$ and effective radius R$_{\text{e,1}} = 0.50$ kpc, and a second simple Sersic profile with n$_2 = 1.59$ and R$_{\text{e,2}} = 2.59$ kpc. The size of the break radius is 2.29" - corresponding to 0.76 kpc - in line with the cusp-radius estimate.
However, given that the galaxy might also host an homogeneous exponential dust disk, we exploit the HST WFPC3 f110W image (GO program 14219; P.I. Blakeslee) and correct for dust following \citet{Bender15} (see also App. A of \citealt{Nowak08}) using both the HST f814W (GO program 5910; P.I. Lauer) and the HST f555W WFPC2 (GO program 7281; P.I. Fanti) images. Then, we measure the core radius from the two f110W dust-corrected images using both the cusp-radius r$_\gamma$ and a core-Sersic fit to the SB profile to determine r$_\text{b}$, reporting the results in Tab.~\ref{Tab.core_radius}. The procedure is summarized in App.~\ref{App.dust_correction}. As we can see, regardless of the approach and/or image we choose, the result is stable around 0.70 kpc: according to the \mbh-core size relation \citep{Jens16}, this would imply a central BH with mass $> 1.0 \times 10^{10} M_\odot$.  Moreover, because the size of the sphere of influence of SMBHs in core galaxies is similar to the core size \citep{Jens16} we expect the sphere of influence to be well resolved by our kinematic data.

\begin{table}
    \centering
    \begin{tabular}{c c c c}
       Image & Method & Core radius (") & Core radius (kpc) \\ \hline \hline
       Wendelstein-Ks & Cusp & 2.10 & 0.70 \\ 
       & CS & 2.29 & 0.76 \\ \hline
        f110W / f814W & Cusp & 1.94 & 0.65 \\ 
       & CS & 2.10 & 0.70 \\ \hline
       f110W / f555W & Cusp & 2.08 & 0.69 \\ 
       & CS & 2.43 & 0.81 \\ \hline
       & Avg Cusp & 2.04 $\pm$ 0.09 & 0.68 $\pm$ 0.03 \\ 
       & Avg. CS & 2.27 $\pm$ 0.16 & 0.75 $\pm$ 0.05 \\ \hline \hline
    \end{tabular}

    \caption{Summary of the different core-radius estimates derived for NGC~708. \textit{Col. 1:} Image used, where f110W / f814W means dust-corrected f110W image using f814W (and same for f110W / f555W). \textit{Col. 2:} Method used - either cusp-radius or Core-Sersic (CS). \textit{Col. 3-4:} Results in arcseconds and kpc. A value of 0.70 kpc implies $\mbhm = 1.22 \times 10^{10} M_\odot$ using the relation of \citet{Jens16}.}
    \label{Tab.core_radius}
\end{table}

\subsection{Deprojection} \label{Ssec.depro}
One crucial ingredient needed when dynamically modeling a galaxy is the reconstructed triaxial light density $\rho$. To recover this, we deproject the SB profile using the software SHAPE3D extensively described in \citet{dN20}. The code is a semi-parametric method which assumes the galaxy to be stratified on concentric ellipsoids:

   \begin{equation}
        m^{2-\xi(x)} = x^{2-\xi(x)} + \left[\frac{y}{p(x)}\right]^{2-\xi(x)} + \left[\frac{z}{q(x)}\right]^{2-\xi(x)}.
        \label{eq.def_ellips}
    \end{equation}

\noindent The exponent $\xi$ can be used to generate disky ($\xi > 0$) or boxy ($\xi < 0$) bias. The four one-dimensional functions $p(x), q(x)$, $\xi(x)$, along with the density on the x-axis $\rho_x (x)$, specify $\rho$ at each point of the grid: the density values which are not along the $x$-axis are updated through a log-linear interpolation/extrapolation. Finally, the code fits $p(x), q(x)$, $\xi(x)$ and $\rho_x (x)$, non-parametrically using a simulated annealing algorithm \citep{Metropolis53} and uses a one-dimensional radial smoothing to penalize against unsmooth solutions. \\
The code recovers the best-fit $\rho$ by minimizing RMS$=\sqrt{\langle \left(\ln (I_{\text{obs}}/I_{\text{fit}})\right)^2 \rangle}$, where $I_\text{fit}$ is the projected surface luminosity corresponding to $\rho$ for a given orientation. It allows to significantly restrict the range of viewing angles compatible with the observed photometry \citep{dN20, dN22} and to sample different solutions at the same orientation to explore the degeneracy intrinsically present in the deprojection \citep{Rybicki87, Gerhard96, GB96, VDB97, Magorrian99, dN20}. \\
For the deprojection, an intrinsic frame of reference $O(x,y,z)$ aligned with the galaxy principal axes is used, such that $x, y, z$ are on the major, intermediate and minor axis, respectively. The three viewing angles that are needed to determine the orientation of a triaxial body are the two angles $\left(\theta, \phi\right)$, which specify the position of the line-of-sight (LOS), and a third misalignment angle $\psi$, defined as the angle between the projected $z$-axis on the plane of the sky and the $x$'-axis, measured counterclockwise. \\
We place the SB on a circular 40 $\times$ 10 grid, with innermost radius at $\sim$1.02" and outermost radius at $\sim$120.4", corresponding to 0.34 and 40 kpc, respectively. These limits are set at the points where, when doing isophotal fits, one or more parameters need to be set to a constant value to get a reliable estimate of the SB values (see \citealt{Matthias20} for details). In order to check that the SB placed on the grid represents the observed isophotes well, we determine the isophote parameters using the SB on the grid and compare these to the observations, obtaining RMS$_{\log \Sigma} = 0.006$, RMS$_\varepsilon = 0.008$ and RMS$_{\text{PA}} = 0.602^\circ$. For the intrinsic density $\rho$ we choose an ellipsoidal 60 $\times$ 11 $\times$ 11 grid. The flattenings $P, Q$ of this grid are determined for each set of viewing angles calculating the expected values for a perfect ellipsoid (see App. A of \citealt{deZeeuwFranx1989} or eqs. 2-3 of \citealt{Cappellari02}) and averaging these, unless the deprojection is along the principal axes. \\
We sample the viewing angles in 10$^\circ$ step. Because of the assumption of triaxiality, $\theta, \phi$ only need to be sampled in $\left[0, 90\right]^\circ$, while $\psi$ in $\left[0, 180\right]^\circ$. As shown in \citet{dN20} this does not guarantee the canonical $1 \geq p \geq q$ relation. Therefore, for every selected $\rho$ (see below) we repeat the deprojection using the set of viewing angles such that the new deprojection is equivalent to the old one but the inequality $1 \geq p \geq q$ holds (see also Sec. 3.2 of \citealt{dN22}). Finally, every time the solution is along (or within 10$^\circ$ of) the principal angles, we test more deprojections at the same orientation to account for the degeneracy arising in these unfavourable cases (see \citealt{dN20}). For example, along the $y$-axis, we let the code recover $q(r)$ while for $p(r)$ we use different initial guesses, given that in this case the recovery of $p(r)$ is hampered by the position of the LOS. \\
From all deprojections we select those for which RMS $\leq 1.2 \times \text{RMS}_\text{min}$, where $\text{RMS}_\text{min} = 0.027$ is the best-fit RMS. The viewing angles that provide the best deprojection within the RMS interval are \va\,= (70,20,130)$^\circ$. The resulting light densities are those we use for the dynamical modeling of the galaxy. Finally, as a check of the goodness of the deprojections we compute, for the best-fit case, the RMS between the parameters of the fitted isophotes and those of the recovered ones, finding $\text{RMS}_{\log \text{SB}} = 0.017$, $\text{RMS}_\varepsilon = 0.027$ and $\text{RMS}_{\text{PA}} = 1.84^\circ$. The isophotal fits to the best-fit SB, superimposed to the observed photometry, are shown in Fig.~\ref{Fig.iso}. \\
Finally, we repeat this exercise with the Ks-band photometry without masking anything and identify the best-fit viewing angles $\left(\theta, \phi, \psi\right) = (77, 38, 141)^\circ$ for this case. These are slightly different from the g'-band deprojection given the twist in the central regions, which prevents the canonical $1 > p > q$ relation from holding at all radii. Fig.~\ref{Fig.css} shows the recovered photometry. Here we see that the rise in $\varepsilon$ cannot be entirely reproduced, as the recovered profile hits a ceiling at 0.1 close to the galaxy centre, similar to the deprojected ellipticity profile in g'-band. On the other hand, the twist is very well reproduced, meaning that it could be explained by triaxiality and we do not necessarily need to assume that the galaxy has a substructure in the centre.



\begin{figure}
\centering
\includegraphics[width=.7\linewidth]{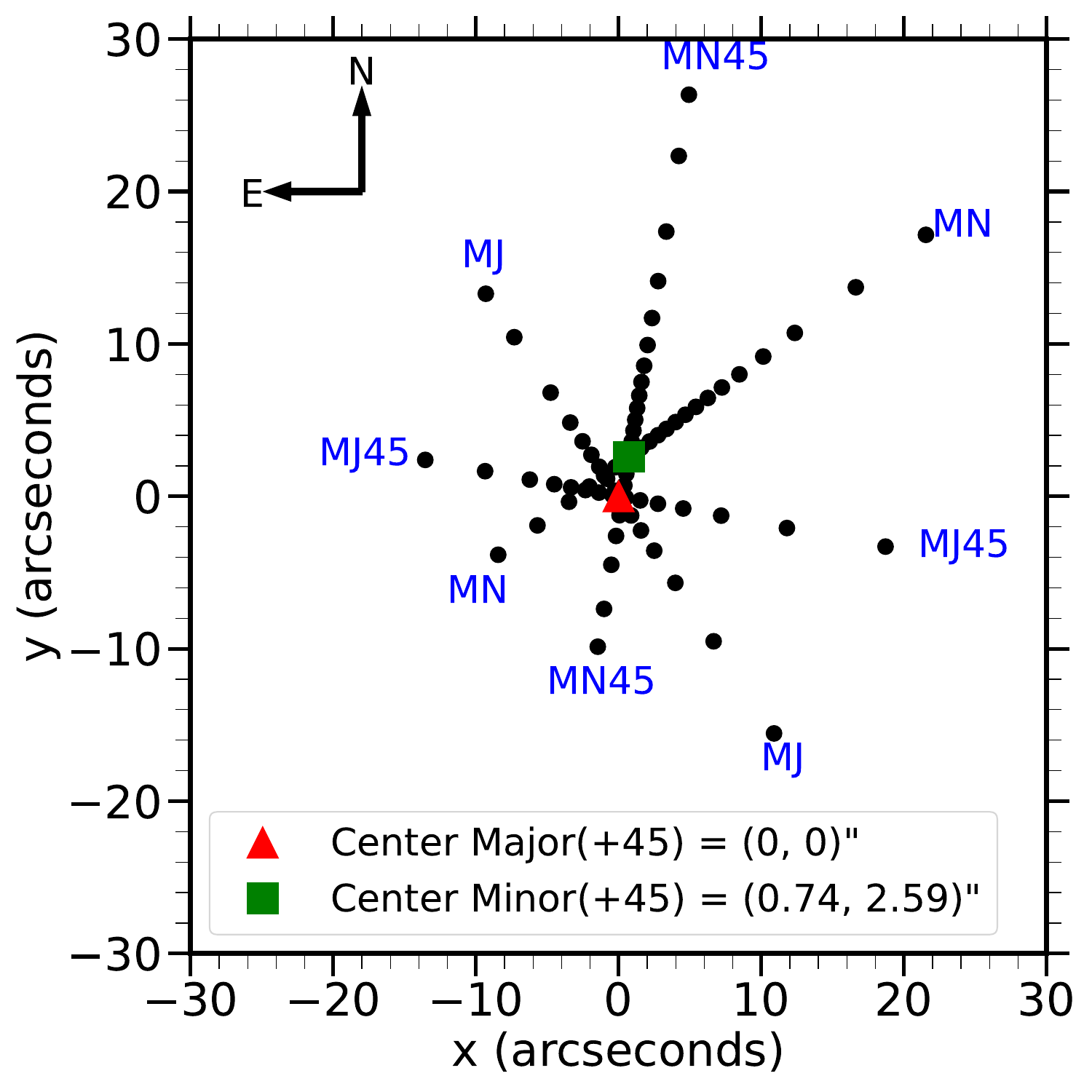}

    \caption{Illustration of the 4 slits configuration used throughout this work, whose centers are plotted as black dots. The "true" center of the galaxy is labeled with a red circle, while the green triangle shows where MINOR and MINOR+45 were centered.}
    \label{Fig.slits}
\end{figure}

\section{Spectroscopy} \label{Sec.kin}
In this section we present the spectroscopic observations of NGC~708 and the extraction of the kinematics. Apart from providing the parameters to be fitted when performing the dynamical model, the kinematics is also important for a variety of reasons: it can provide evidence for triaxiality if the galaxy shows minor-axis rotation \citep{Contopoulos56, Schechter79, Davies86, Davies88}, the velocity dispersion $\sigma$ can be used to get a prediction of \mbh\,using scaling relations \citep{Ferrarese00, Gebhardt00, McConnell13, Rob16}, while knowledge of the exact shape of the LOSVD is crucial to break the mass-anisotropy degeneracy which can lead to biased \mbh\ \citep{Binney82}. The few existing triaxial \mbh\ measurements in massive galaxies mostly used parametric LOSVDs  \citep{vdM93,Gerhard93} while we here use more flexible non-parametric LOSVDs.

\subsection{MODS Observations} \label{Ssec.MODS}
\begin{figure}
\centering

\includegraphics[width=.8\linewidth]{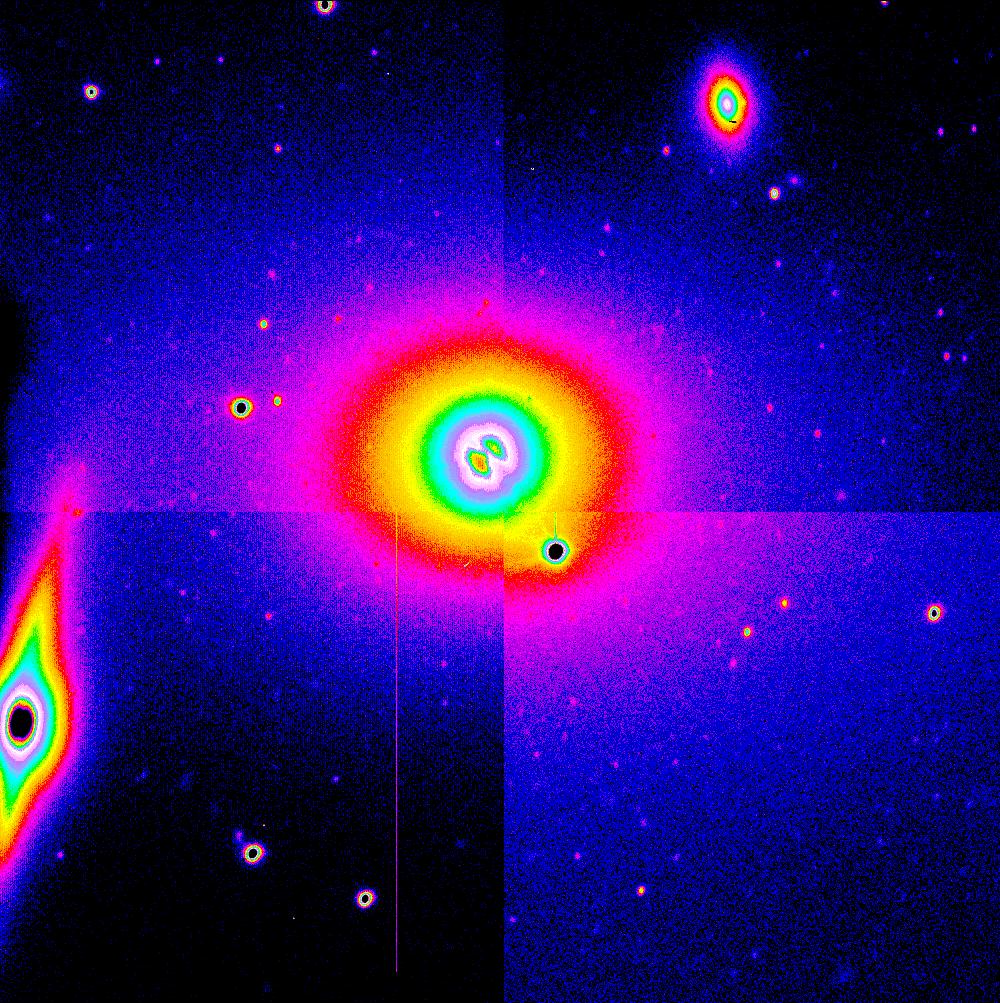}

    \caption{LBT acquisition image of NGC~708 before starting the spectroscopy. In the center of the galaxy there is a prominent "double center": this effect is generated by the dust lane.}
    \label{Fig.double_center}
\end{figure}

Long-slit spectroscopic data for NGC~708 were already published by \citet{Wegner12}. These data come from observations carried out at the 2.4-m Hiltner telescope of the MDM Observatory at Kitt Peak. We re-observed the galaxy at the Large Binocular Array (LBT) observatory, using the Multi-Object Double Spectrograph (MODS, \citealt{Pogge10}). Its binocular configuration (MODS1 - MODS2) allows to place two slits at two different orientations in the plane of the sky. The observations were carried out in two runs, the first one in October 2019 (observers Jan Snigula and Stefano de Nicola) and the second one in October 2020 (remote observers Jan Snigula and Roberto Saglia), with PI Roberto Saglia.
MODS has a field of view of $6 \times 6$', works in the range [3200-10000] \AA\,and has spectral resolution $\lambda / \Delta\lambda = 5000/3.2 \sim 1500$ with slit width 0.8" in the blue. All science images were corrected for bias, dark, flat fields and wavelength calibrated. Moreover, we acquired one sky image after each setup to allow for background subtraction. The pixel scale of each image is $\sim$ 0.12 arcsec/pixel.\\
The galaxy was observed using two different configurations. First the two slits, each one with width of 0.8", were placed along the galaxy projected principal axes on the plane of the sky (MAJOR and MINOR). The second configuration was obtained by rotating the slits by 45$^\circ$ (MAJOR+45, MINOR+45), as shown in Fig.~\ref{Fig.slits}. In both runs we took data in the range 3200-8450 \AA, splitting the spectra in a blue ($\lambda \in$ [3200-5700] \AA) and a red ($\lambda \in$ [5700-8450] \AA) part. The typical seeing is $\sim$ 1.4" (FWHM). The relevant pieces of information of the four setups for the blue part of the spectrum are reported in Tab.~\ref{Tab.slits}. \\
Because of the dust lane, the galaxy shows two apparent nuclei (see Fig.~\ref{Fig.double_center}). While MAJOR and MAJOR+45 were centered on the brightest peak, MINOR and MINOR+45 were centered on the fainter one. In a NW coordinate system centered on the brightest peak, the fainter has coordinates (0.7369, 2.5913) arcseconds.  \\
In order to achieve a nearly constant signal-to-noise of $\sim$40 per pixel, for each one of the four setups we spatially binned the data (see Tab.~\ref{Tab.slits}) along the slit . In each spatial bin we extracted the spectrum and measured the kinematics.

\begin{table}
    \centering
    \begin{tabular}{c c c c c}
       Setup  & PA ($^\circ$) & Center (") & Number of bins & Exp. time (')\\ \hline \hline
       MAJOR    & 35 & (0,0) & 16 & 110\\
       MINOR   & 125 & (0.7369, 2.5913) & 22 & 100 \\
       MAJOR+45  & 80 & (0,0) & 15 & 110\\
       MINOR+45 & 170 & (0.7369, 2.5913) & 23 & 100\\ 
         \hline
    \end{tabular}
    \caption{Technical details about the four slit configurations that we used to observe NGC~708. \textit{Col. 1:} Setup name. \textit{Col. 2:} PA of the slit, measured from North to East. \textit{Col. 3:} Center of the slit coordinates (in arcseconds) in a NW frame of reference centered on the brightest peak. \textit{Col. 4:} Number of spatial bins each slit is divided into. The details refer to the blue part of the spectrum since we did not use the red part in our fits (see text). \textit{Col. 5:} Total exposure time in minutes adding up the spectra of the two runs together. All galaxies were observed for one hour in 2020, while MAJOR and MAJOR+45 were observed 10' longer in 2019.}
    \label{Tab.slits}
\end{table}

\begin{figure*}
\centering
\subfloat{\includegraphics[height=.3\textheight]{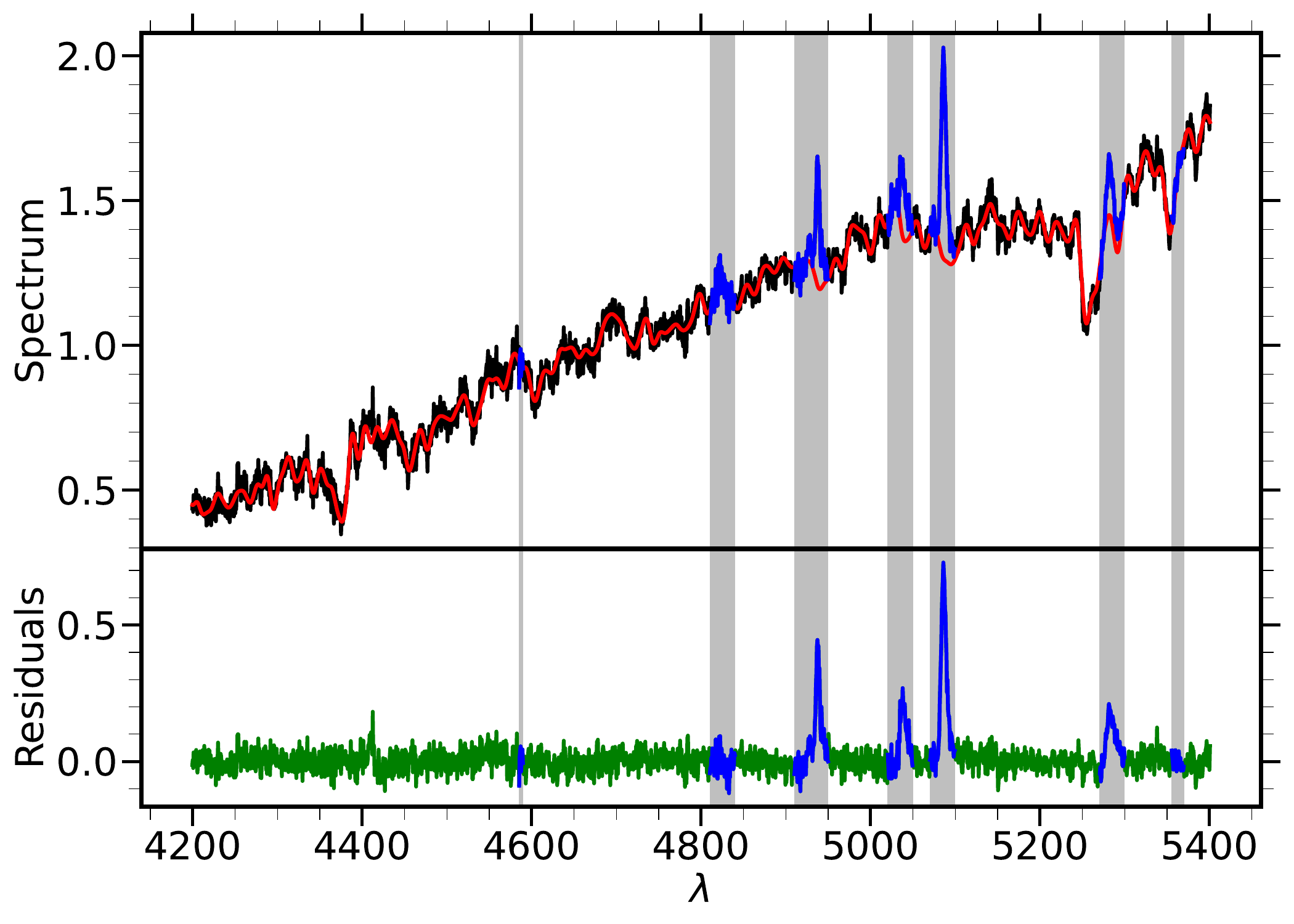}}
\subfloat{\includegraphics[height=.3\textheight]{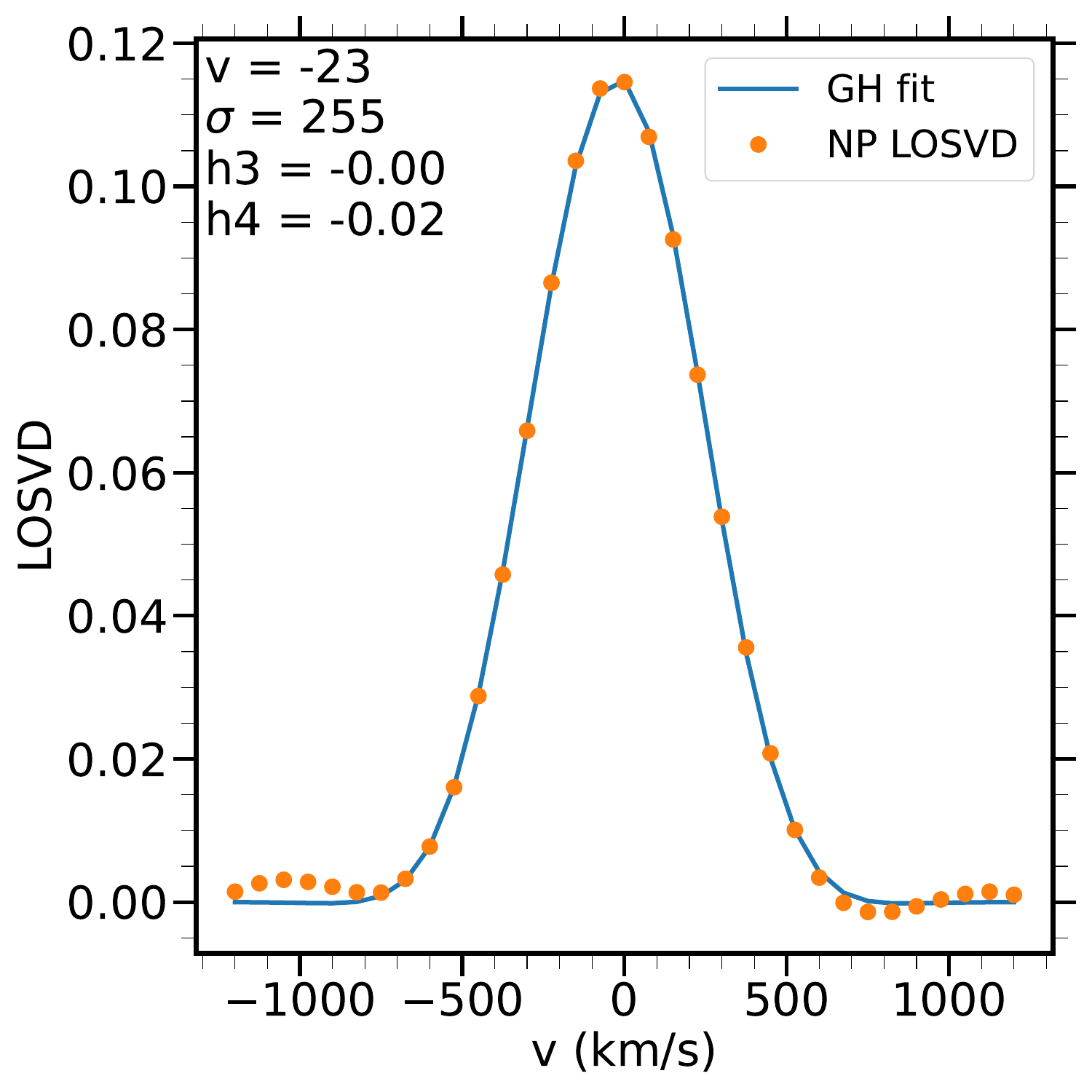}}

    \caption{\textit{Left:} Example of fitted spectrum by WINGFIT. The top panel shows the observed (black) and the fitted (red) spectrum, while the parts of the spectrum highlighted in blue and surrounded by grey rectangles are masked and omitted from the fit. Finally, the bottom panel shows the fit residuals. \textit{Right:} Corresponding non-parametric LOSVD recovered by WINGFIT (orange points) along with the best-fit parametric LOSVD (solid blue line), whose coefficients are determined fitting a 4-th order Gauss-Hermite polynomial to the recovered LOSVD. The bin belongs to the MAJOR slit (Tab.~\ref{Tab.slits}) and its coordinates are (-0.8415, 1.2017) in a NW frame of reference.}
    \label{Fig.kin_wingfit}
\end{figure*}

\subsection{Extracting the kinematics}
The extracted spectra show prominent emission lines: in the blue part, [OIII] 5007 \AA, [NI] 5199 \AA, H$\beta$ are clearly visible. The red part of the spectrum is dominated by emission lines such as H$\alpha$, [NII] 6583 \AA\,and [SII] 6730 \AA.  \\
To extract the kinematic variables from the spectrum we use the non-parametric LOSVD fitting routine WINGFIT (Thomas et al., in prep). The code exploits the full information contained in the LOSVD, allowing for accurate reconstruction of the wing profiles, a crucial ingredient to accurately recover \mbh. In fact, the well-known mass-anisotropy degeneracy \citep{Binney82} can bias the measurement given that a radial anisotropy profile can also generate an increase in the velocity dispersion. The information contained in the wings can help lifting the degeneracy. The code uses a novel model selection technique to optimize the smoothing of the LOSVDs \citep{Jens22}. It is based on a generalization of the classical Akaike Information Criterion (AIC, \citealt{Akaike74}) and uses the concept of effective degrees of freedom \citep{Mathias21} to take into account the different number of DOFs each model has. Instead of minimizing the $\chi^2$, the optimization is done in

\begin{equation}
\text{AIC}_\text{p} = \chi^2 + 2*\text{m}_\text{eff}.
    \label{eq.meff}
\end{equation}

\begin{figure*}

\subfloat{\includegraphics[width=.5\linewidth]{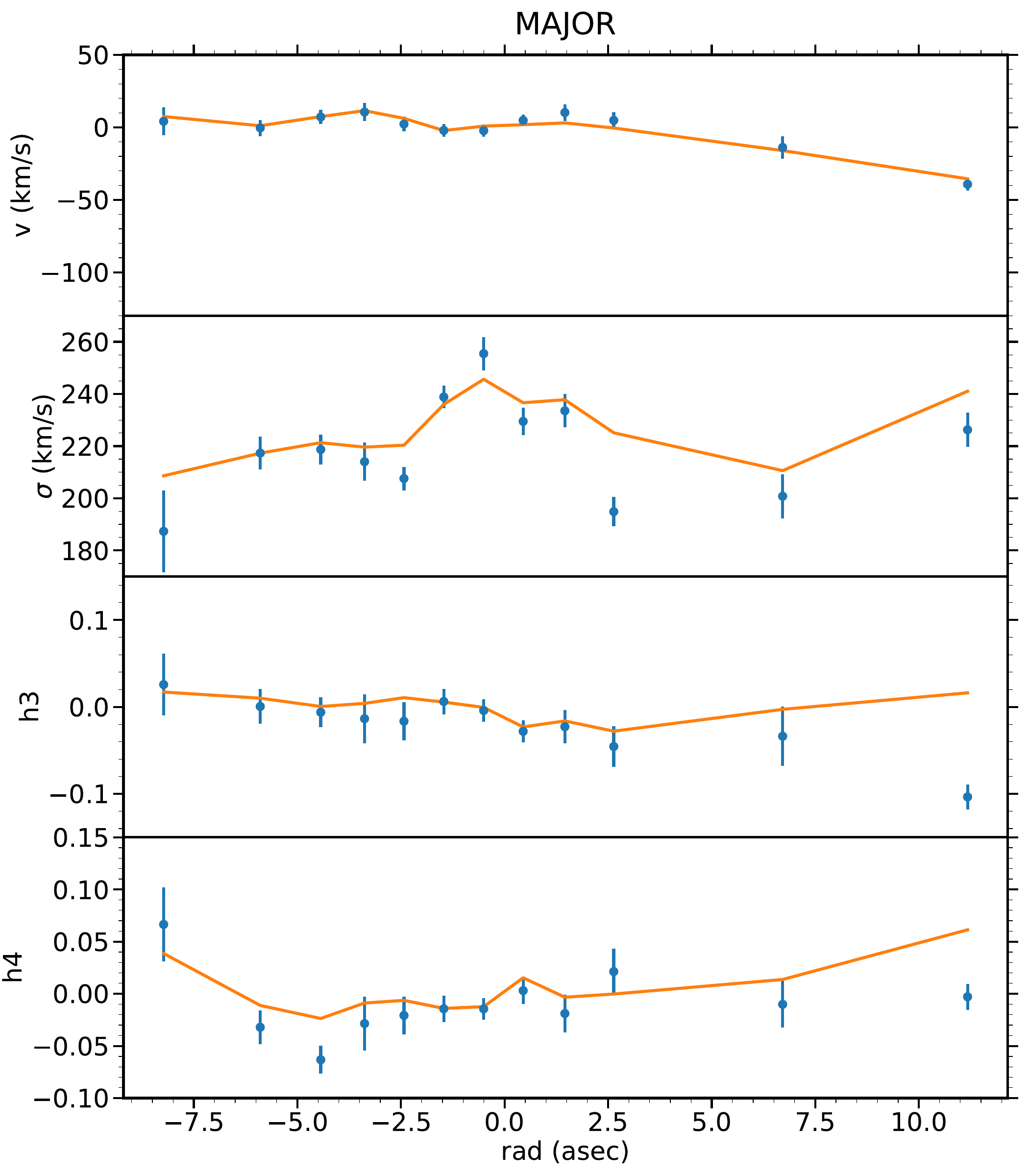}}
\subfloat{\includegraphics[width=.5\linewidth]{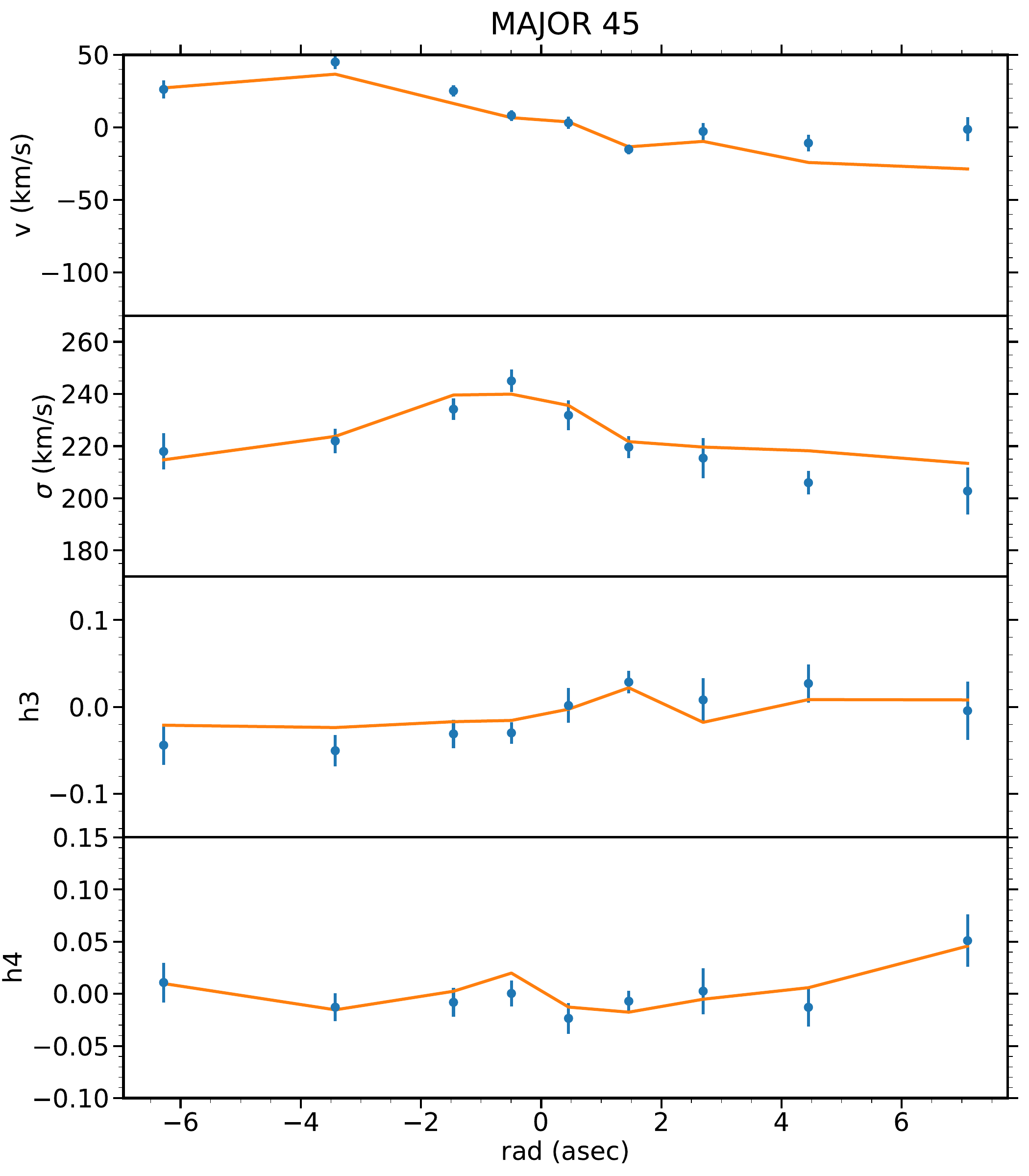}}

\subfloat{\includegraphics[width=.5\linewidth]{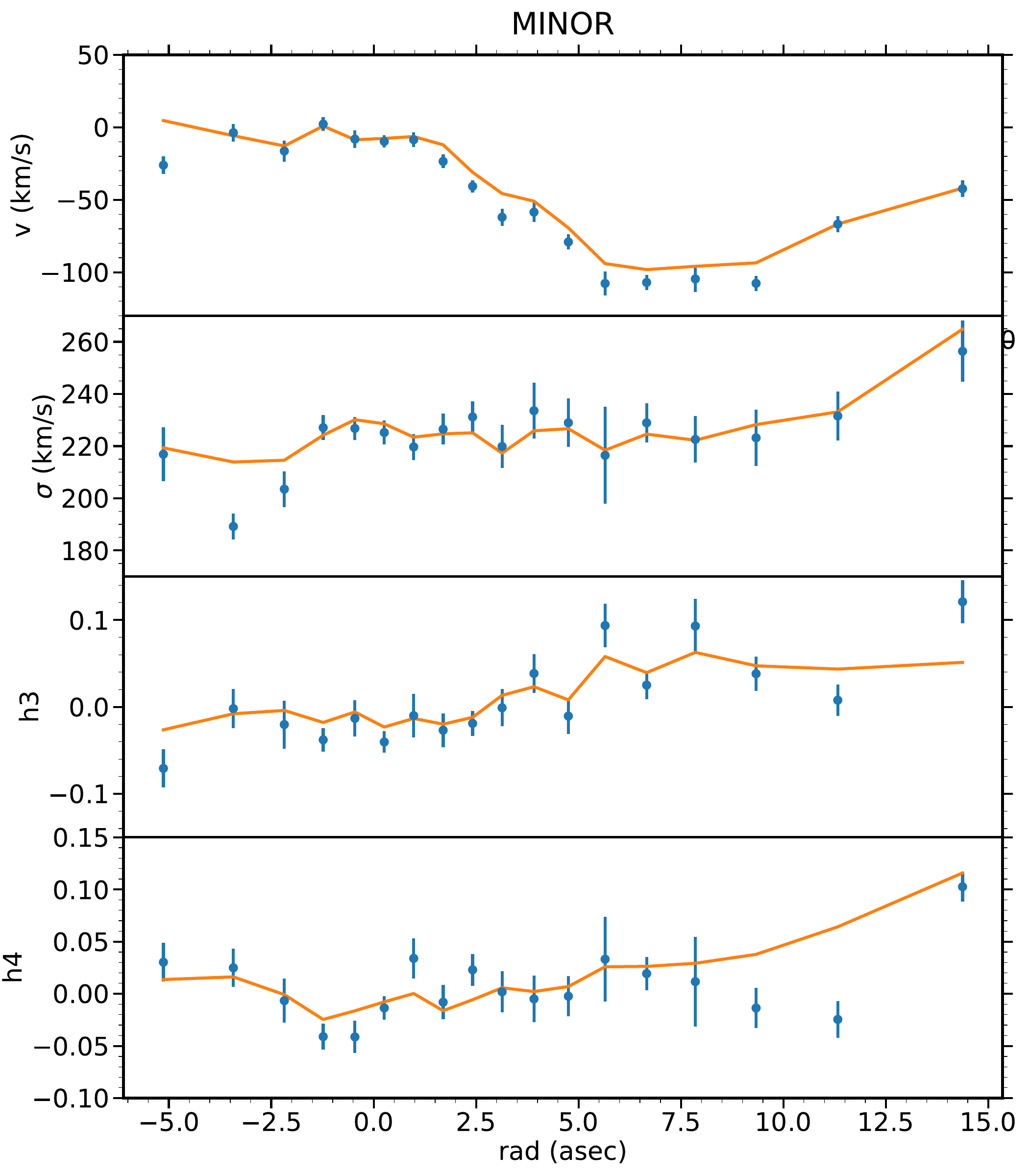}}
\subfloat{\includegraphics[width=.5\linewidth]{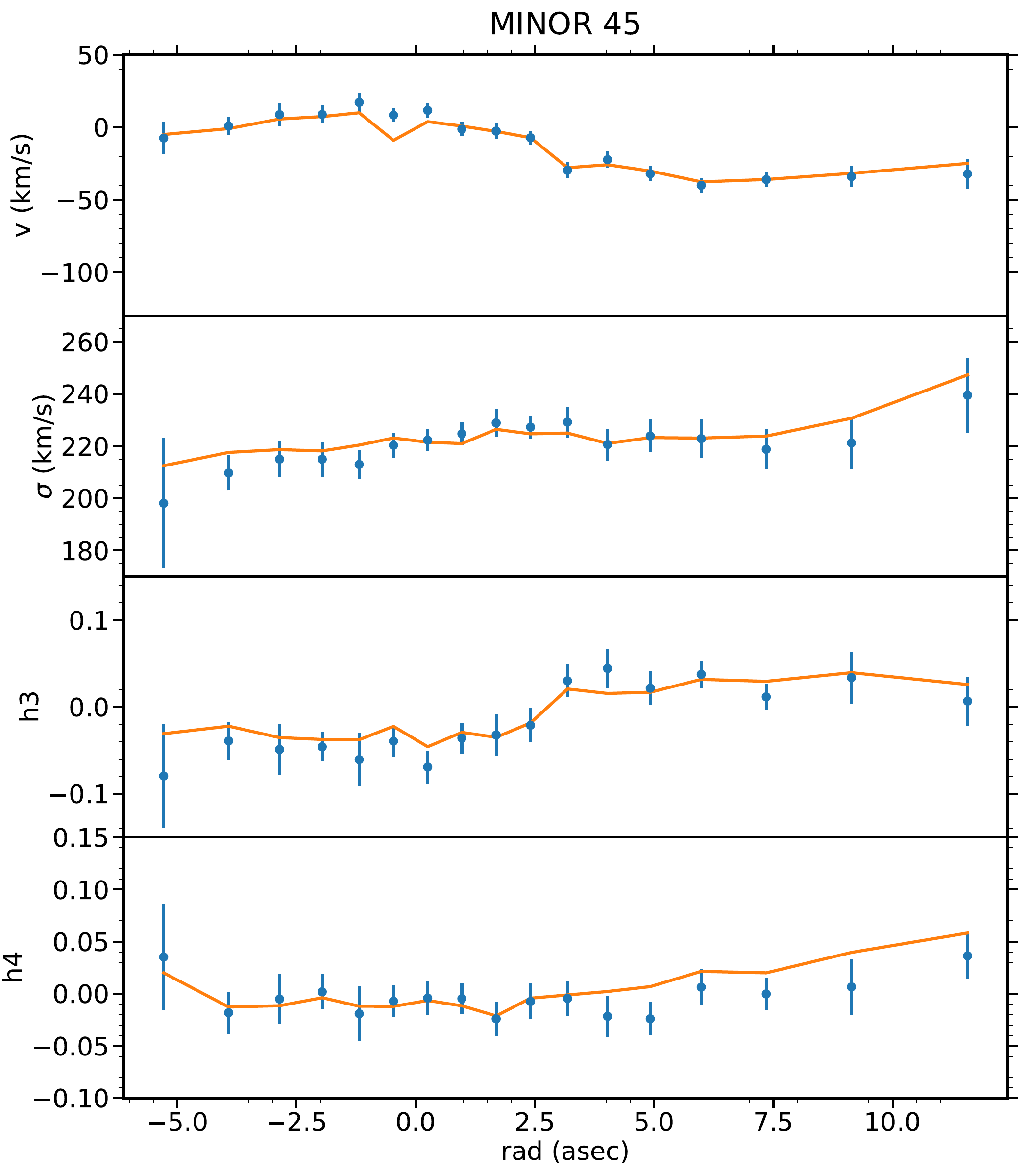}}



    \caption{Kinematics of NGC~708 along the four different slits (see Tab.~\ref{Tab.slits}). Blue dots are the measured kinematics using WINGIFT, while orange lines show the result from the best-fit dynamical modeling using g'-band photometry. In particular, the minor-axis rotation is very well reconstructed. Radii values grow from East to West.}
    \label{Fig.kinematics_WS}
\end{figure*}

\noindent In practice, eq.~\ref{eq.meff} allows to find a model which fits the data well but is also smooth. \\
WINGFIT is inefficient when the number of templates is $>$ 50. To preselect a pool of template stars we launch a preliminary fit using the parametric algorithm PPXF \citep{Cappellari04, Cappellari17}. The spectrum is fitted using all stars found in the MILES stellar library \citep{FB11}. Given the higher resolution of the library (FWHM $\sim$ 2.5 \AA), the stellar spectra are broadened to match the resolution of the observations (FWHM $\sim$ 3.2 \AA). We fit the blue spectrum in the range [4200-5400] \AA. This allows us to fit the most prominent absorption features and exclude too noisy regions. We do not use the red part of the spectrum, which is dominated by emission lines and does not allow for a wavelength range large enough to obtain a reliable fit. \\
In PPXF, emission lines can be fitted along with the absorption features with separate templates, yielding two LOSVDs, one for the absorption and one for the emission lines. We tested two approaches obtaining comparable results: we first tried the fit as described above, and then repeated the procedure fitting only the absorption features, masking the emission lines, obtaining similar results. One important caveat lies in the multiplicative polynomials. As shown in \citet{Kianusch23}, their usage can generate artificially enhanced wings. Therefore, we limited ourselves to 4-th order multiplicative polynomials and avoided to use additive polynomials at all. \\
This procedure is repeated for every spatial bin. PPXF assigns weights to each template that it used to fit the galaxy spectrum. We take the best $\sim$15-20 templates for each bin and pass these to WINGFIT to reconstruct the fully non-parametric LOSVD. An example of fitted spectrum along with its corresponding LOSVD is shown in Fig.~\ref{Fig.kin_wingfit}.\\    
The recession velocity of the galaxy barycentre needs to be subtracted from the actual estimate 
to set the zero-point of the systemic velocity at the galaxy centre. As first guess, we try $v = c\ln(1+z)$, where $z$ is the galaxy redshift. The fits are then repeated once more to fine-tune the velocity. \\
The resulting kinematics, parametrised in terms of standard Gauss-Hermite coefficients for illustration purposes, is shown in Fig.~\ref{Fig.kinematics_WS} as blue dots. This is in good agreement with the published values of \citet{Wegner12}. For the Schwarzschild fits, the LOSVDs are binned into $N_\text{vel} = 25$ velocity bins. We discard bins with unrealistic Gauss-Hermite (GH) coefficients or spatial bins with too low S/N, retaining a total of 56 bins. We see that along the MINOR there is a velocity variation $\Delta v \sim$100 km s$^{-1}$, which is a clear indication of triaxiality. The fact that the rotation is only retrograde depends on the asymmetric position of the MINOR slit (see Sec.~\ref{Ssec.MODS}). The low h$_4$ values indicate that the wings are not very strong, while the slightly negative h$_3$ might indicate a small residual template mismatch \citep{Bender94,Kianusch23}. Finally, the velocity dispersion $\sigma$ hits a ceiling at $\sim$250 km s$^{-1}$ in the central regions. This value is low compared to what is typically observed in large ETGs. Given that the core size of this galaxy predicts a black hole with mass $\sim 10^{10} M_\odot$, while using the $\mbhm-\sigma$ relation of \citet{Rob16}, we infer a black hole mass smaller than $10^9$ M$_\odot$. Thus, the galaxy is a potential catastrophic outlier in the $\mbhm-\sigma$ relation, similar to NGC1600 \citep{Jens16}.

\section{Dynamical modeling} \label{Sec.dynmod}

We now put together the results of the previous two sections and turn to the dynamical modeling of NGC~708 under the assumption of triaxiality. The galaxy has already been modeled assuming it to be axisymmetric (see \citealt{Wegner12}): the photometry does not show unambiguous signs of triaxiality. Nevertheless, several of our deprojections are strongly triaxial (see also \citealt{dN22}), and even if the galaxy may be axisymmetric, triaxial solutions cannot be excluded\footnote{An example is a triaxial galaxy with constant axis ratios as a function of the distance from the galaxy centre. In this case a twist cannot be observed.}. Moreover, the rotation along the minor axis at least excludes an oblate axisymmetric shape.

\subsection{Our code}  \label{Ssec.ourcode}

To compute the dynamical models of NGC~708 we use our recently developed triaxial Schwarzschild code SMART \citep{Bianca21}. SMART can deal with any deprojection or DM halo - parametric or non-parametric - and exploits a 5D orbital sampling space, allowing to characterize every orbit family (tubes, box orbits, Keplerian orbits) which may be found in a triaxial potential \citep{Poon01}. Moreover, it uses the model selection technique described above \citep{Mathias21, Jens22} to optimize the smoothing individually in each trial mass model. \\
Specifically, SMART computes the potential by integrating Poisson's equation. The density is constructed as 

\begin{equation}
\rho_\text{TOT} = \mbhm \times \delta (r) + \Gamma \times \rho_* + \rho_\text{DM}
    \label{eq.dens_SMART}
\end{equation}

\noindent where $\mbhm \times \delta (r)$ is the point-like contribution from the central SMBH, $\Gamma \times \rho_*$ is the stellar contribution - obtained by multiplying the deprojected light density\footnote{Thus, in our code the light density is not fitted, but treated as a constraint.} with the mass-to-light ratio. While $\Gamma$ is indeed the parameter which SMART fits, for reasons that will become clear later we follow \citet{Kianusch24} and introduce the parameter \ml, computed at the minimum of $\mlm (r) = [\rho (r)  + \rho_\text{DM} (r)] / \rho_*$, where $\rho$ is the stellar mass density, obtained as $\rho_* \times \Gamma$. Finally, $\rho_\text{DM}$ is the DM density, for which we assume a triaxial \citet{Zhao1996} model 

\begin{equation}
  \rho_\text{DM} (r) = \frac{\rho_0}{\text{p}_\text{DM} \cdot \text{q}_\text{DM} \cdot \left(\frac{r}{r_0}\right)^{\gamma} \left[1 + \left(\frac{r}{r_0}\right)^{1/\alpha} \right]^{(\beta - \gamma)/\alpha}} 
    \label{eq.Zhao_models}
\end{equation}

\noindent where $\text{p}_\text{DM}$ and $\text{q}_\text{DM}$ are the intrinsic flattenings of the DM halo, we fix $\alpha = 1$ and $r = \sqrt{x^2 + \left(y/\text{p}_\text{DM}\right)^2 + \left(z/\text{q}_\text{DM}\right)^2}$. \\
Once the potential has been calculated, a time-averaged orbit library is constructed, and time-averaged projections are computed taking seeing into account. The stellar weights are then updated to maximize the entropy-like quantity

\begin{equation}
\hat{S} = S - \alpha\chi^2
    \label{eq.Shannon}
\end{equation}

\noindent where $S$ is related to the Shannon entropy, $\alpha$ is a smoothing term and $\chi^2$ compares the differences between fitted and observed LOSVDs:

\begin{equation}
       \chi^2 = \sum_{i=1}^{N_{\text{losvd}}} \sum_{j=1}^{N_{\text{vel}}} \left(     \frac{\text{LOSVD}_{\text{model}}^{i,j} - \text{LOSVD}_{\text{data}}^{i,j} }{\Delta\text{LOSVD}_{\text{data}}^{i,j}} \right)^2.         
        \label{eq.chi2_Schw}
    \end{equation}
    
\noindent Here, the summations are carried over every $i$-th spatial bin and every $j$-th velocity bin (cfr. Sec.~\ref{Ssec.MODS}), while $\Delta\text{LOSVD}_{\text{data}}^{i,j}$ is the uncertainty on the $i$-th, $j$-th bin, yielded by WINGFIT. The smoothing optimization is defined in \citet{Jens22} and its implementation into SMART is described in \citet{Bianca23a}. \\
SMART was tested by \citet{Bianca21} using an $N$-body simulation \citep{Rantala18} which aims at reconstructing the formation and the evolution of massive core-galaxies. In particular, the simulation closely resembles NGC1600 \citep{Jens16}. When supplied with the 3D light and DM densities calculated from the particles, the code returned recoveries of \mbh, \ml\,and the normalization of the DM with unprecedented accuracy. Recently, \citet{Bianca23a} moved one step further and simulated the full modeling of a galaxy, deprojecting the SB profile using SHAPE3D and including the DM halo in the dynamical modeling, recovering all the mass parameters with an accuracy of 5-10\%. In a companion paper, \citet{dN22b} used the same approach to recover the intrinsic shape $p, q$ and the anisotropy $\beta$

\begin{equation}
   \beta = 1 - \frac{\sigma_\theta^2 + \sigma_\phi^2}{2 \sigma_r^2}
    \label{eq.beta}
\end{equation}

\noindent obtaining $\Delta p, \Delta q, \Delta\beta < 0.1$. \\
These results have been obtained using simulated high-resolution IFU kinematics. Given that the kinematics used in this work comes from long-slit data with seeing $\sim$ 1.4", we repeated the same exercise, this time fitting a kinematic data set with the same geometry as that of NGC~708. The results are shown in App.~\ref{App.Nbody}, where it can be seen that we still reach an accuracy of 10-20\% when recovering the mass parameters of the galaxy. Moreover, by fitting more than one mock kinematics, we can obtain a first-order estimate of the statistical uncertainties on the recovered mass parameters. 

\begin{table}
    \centering
    \begin{tabular}{c c c c}
       Variable & Range & No. Values & Step size\\ \hline \hline
       \mbh  &  [4 $\times 10^9$ - 1.2 $\times 10^{10}$ ] M$_\odot$ & 5 & 4 $\times 10^9$ M$_\odot$\\
        $\Gamma$   & [0.6 - 4.8] & 7 & 0.7\\
       $\log \rho_0$ & [7.8-8.2] & 7 & 0.0667 \\
       $\gamma$ & [0.0 - 1.0] & 6 & 0.2 \\ 
       p$_\text{DM}$ & [0.8 - 1.0] & 3 & 0.1  \\
       q$_\text{DM}$ & [0.8 - 1.0] & 3 & 0.1    \\
         \hline
    \end{tabular}
    \caption{Sampling the we used for our NOMAD runs. \textit{Col. 1:} Variable name.  \textit{Col. 2:} Sampled range (linear spacing). \textit{Col. 3:} Number of sampled values. \textit{Col. 4:} Step size.}
    \label{Tab.NOMAD}
\end{table}

\subsection{Modeling strategy}
The parameters needed to build the trial gravitational potential (defined in eq.~\ref{eq.dens_SMART}) are: \mbh, $\Gamma$, the halo normalization $\rho_0$, the inner and outer slopes $\gamma$ and $\beta$ and the break radius $r_0$. Moreover, since we use a triaxial halo, we also need to specify the two flattenings p$_\text{DM}$ and q$_\text{DM}$. Finally, since we have several deprojections, the three viewing angles \va\,also need to be varied, for a total of 11 unknowns. As a preliminary step, we fix the break radius of the DM halo $r_0 = 50$ kpc and the outer slope $\beta = 4.5$. The value of $\beta$ differs from the canonical value of 3 of the gNFW models: we assume that the halo progenitors can be modeled by a \citet{Hernquist90} sphere. Given that a large number of trial potentials must be tested, we use the software NOMAD \citep{Audet06, LeDigabel11, Audet17} to efficiently search for the global minimum by launching several models in parallel. \\
Our modeling strategy can be summarized as follows:

\begin{enumerate}
    \item We consider the g'-band deprojection launching a NOMAD run fixing the viewing angles at best-fit orientation \va\,= (70,20,130)$^\circ$. The other 6 parameters are (log-)linearly sampled as reported in Tab.~\ref{Tab.NOMAD}. With the exception of \mbh, which can be estimated using the BH-core size relation, we need to assume fiducial intervals. All the models are constructed by assuming a gaussian PSF with FWHM of 1.4", measured from our spectroscopic observations. Given that the size of the central core is expected to approximate the sphere-of-influence radius r$_\text{SOI}$ well and given that we model with a DM halo, we conclude that we can reliably estimate \mbh\,(see discussion in \citealt{Rusli2013}). We verify this assumption below.
    
    \item We fix the mass parameters to the values coming from step 1) and launch a second NOMAD run, this time \textit{only} fitting for the viewing angles. For all the orientation along (or close to) the principal axes, we sample several deprojections to account for the degeneracy (see Sec.~\ref{Sec.results}).
    
    \item We launch a third and final NOMAD run fixing the viewing angles at the best-fit orientation found in step 2), sampling the mass parameters as in step 1) to determine our final estimate of the best-fit model. 

    \item Given that here we model all bins together, in order to get an estimate of the error bars on our best-fit estimates we rely on the values we find in App.~\ref{App.Nbody} for the $N$-body simulation. In that case, we have a total of 9 configurations, so we take the standard deviation as (pessimistic) estimate of the error bars, scaled to the values we find for NGC~708.

    \item To evaluate the impact of the dust, we launch a further NOMAD run using the best-fit Ks-band deprojection at \va\,= (77,38,141)$^\circ$. Given that the photometry sets and the deprojections are similar in the outermost regions, we take only \mbh\,and $\Gamma$\,as free parameters while keeping the halo parameters fixed. As second test, we deproject the g'-photometry with masking at the best fit Ks-band viewing angles and model this with fixed halo parameters. In this way, we can verify how much the BH mass remains stable and compare $\Gamma$\, estimates in different bands.

\end{enumerate}

In \citet{dN22b} we have shown that by selecting deprojections using the approach described in Sec.~\ref{Ssec.depro} and computing the average shape profiles $p(r)$ and $q(r)$ among all these, the resulting profiles approximate the correct ones very well. The modeling strategy adopted here allows us to test this \textit{a posteriori} by comparing the deprojection corresponding to the best-fit viewing angles found in step 2) with $< p(r) >$ and $< q(r) >$, where the average is performed over all selected deprojections.

\begin{figure}
\centering

\includegraphics[width=.9\linewidth]{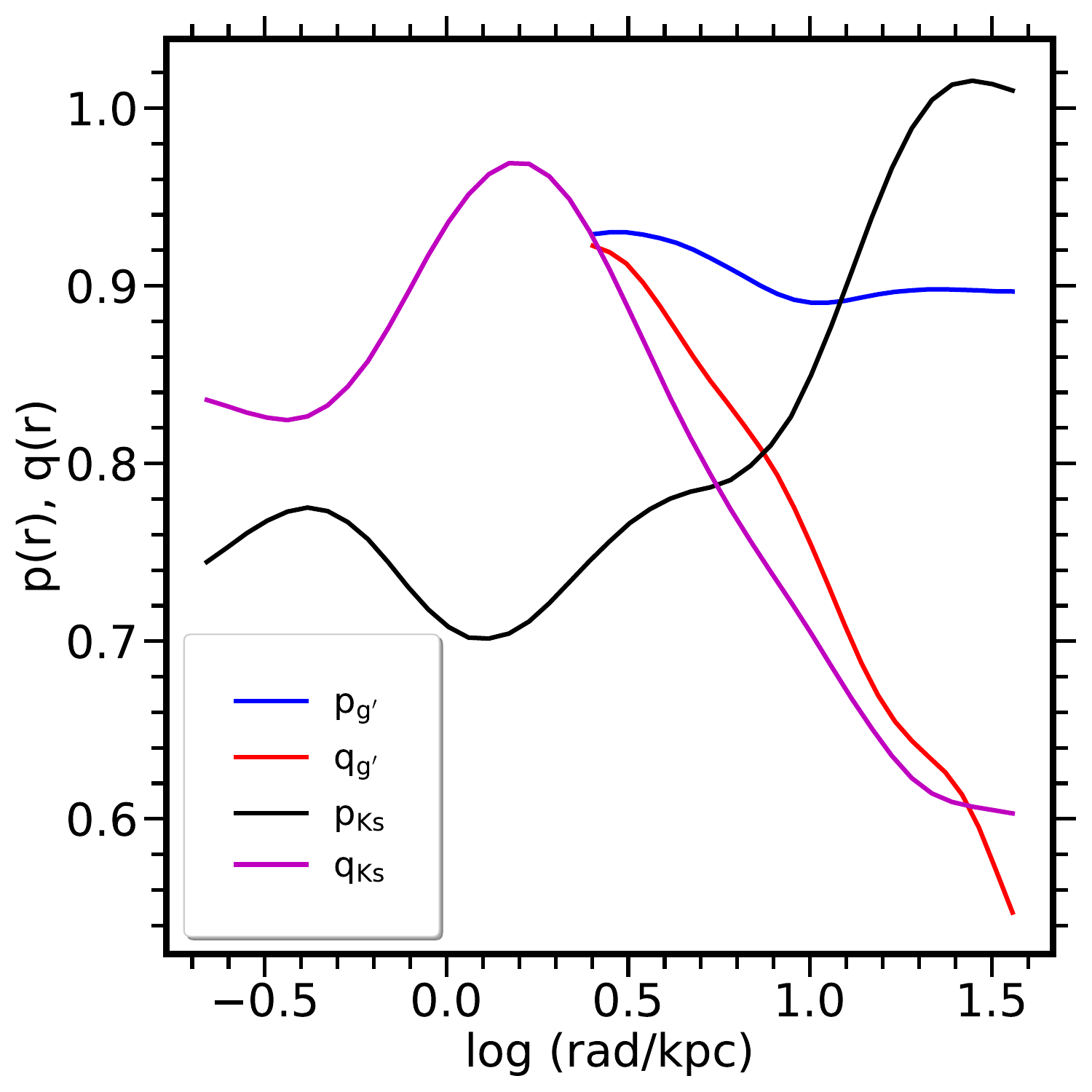}
    
    \caption{The recovered intrinsic shapes for the best-fit solutions, found at $\left(\theta, \phi, \psi\right) = (80,90,135)^\circ$ for the g'-band and at $\left(\theta, \phi, \psi\right) = (77,38,141)^\circ$ for the Ks-band. The innermost regions are not shown for the g'-band due to dust contamination; in the outermost region the galaxy becomes mildly triaxial.}
    \label{Fig.pqT_bf}
\end{figure}

\begin{figure*}

\subfloat{\includegraphics[width=.3\linewidth]{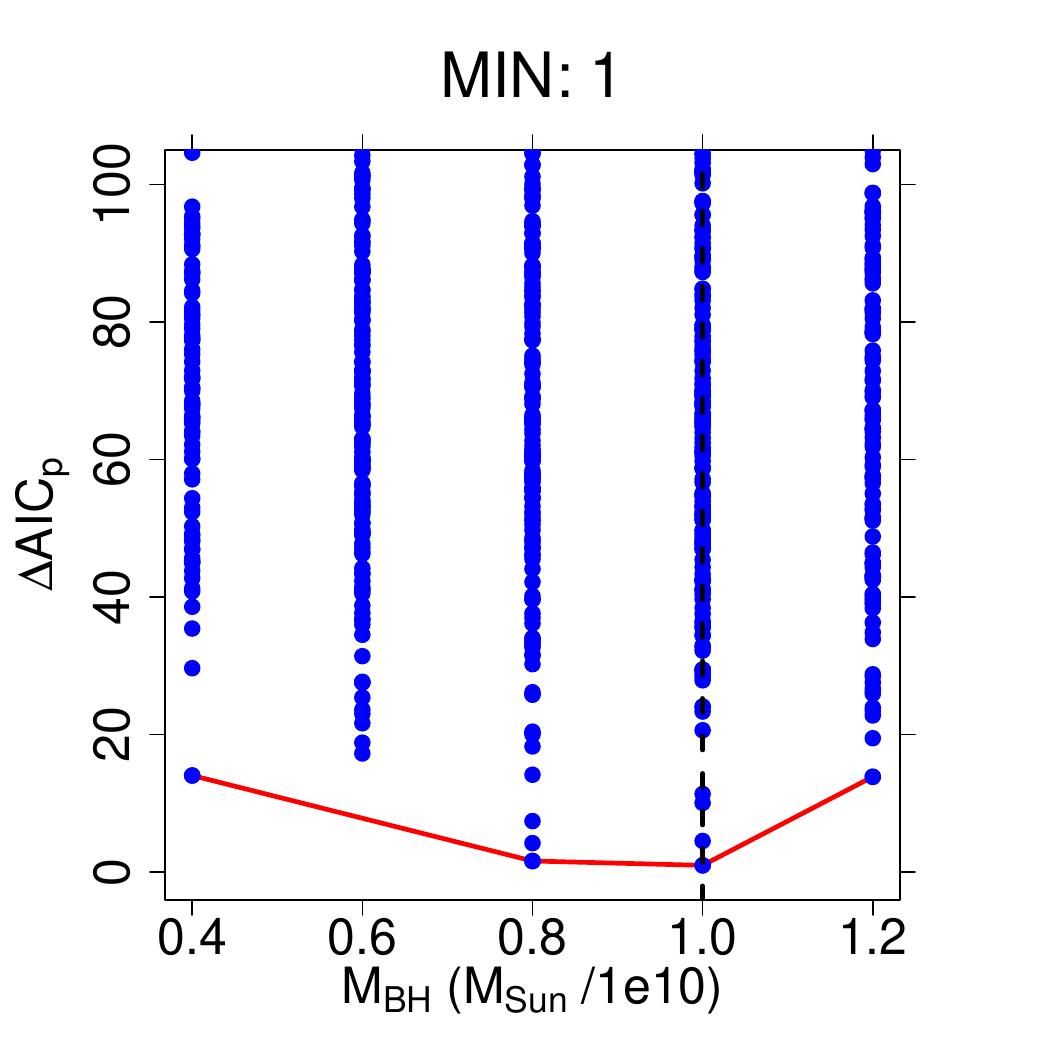}}
\subfloat{\includegraphics[width=.3\linewidth]{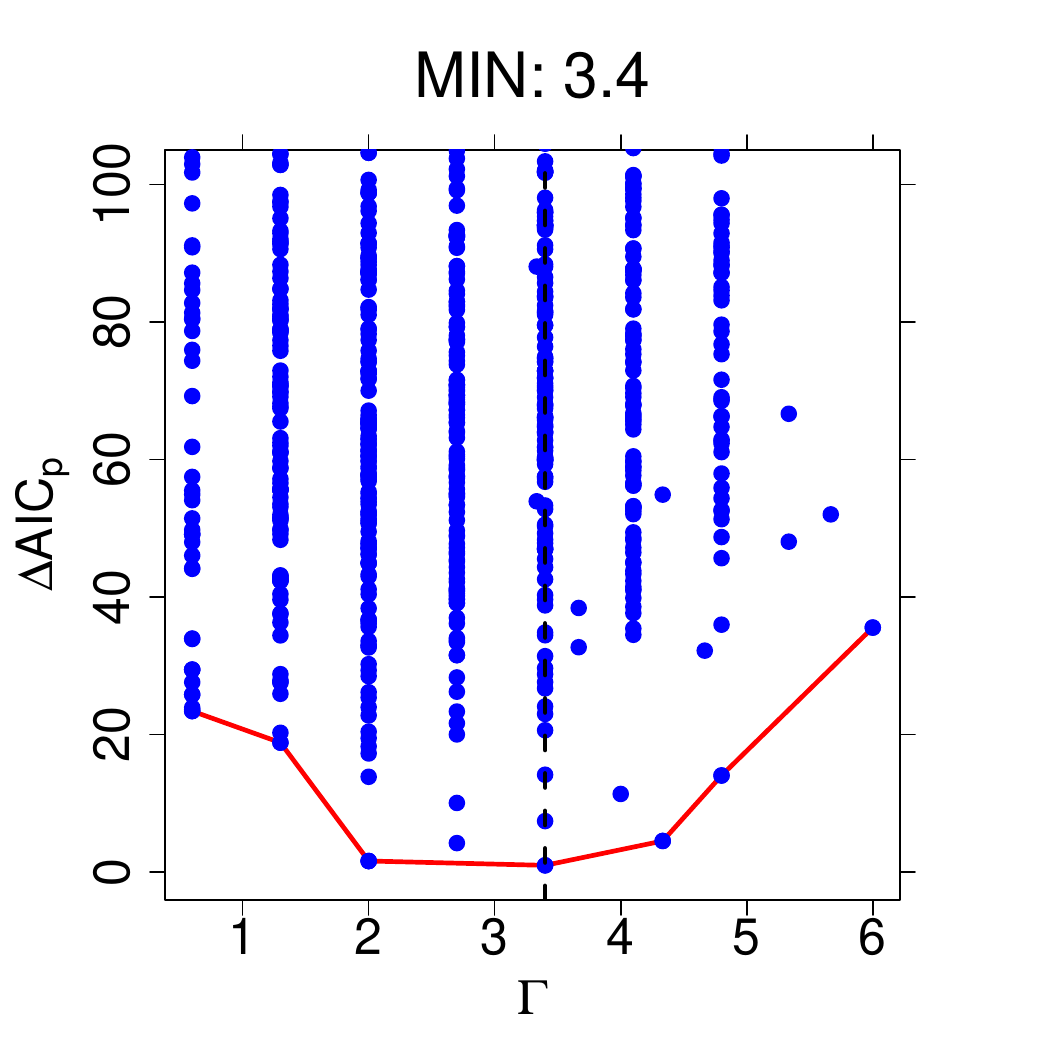}}
\subfloat{\includegraphics[width=.3\linewidth]{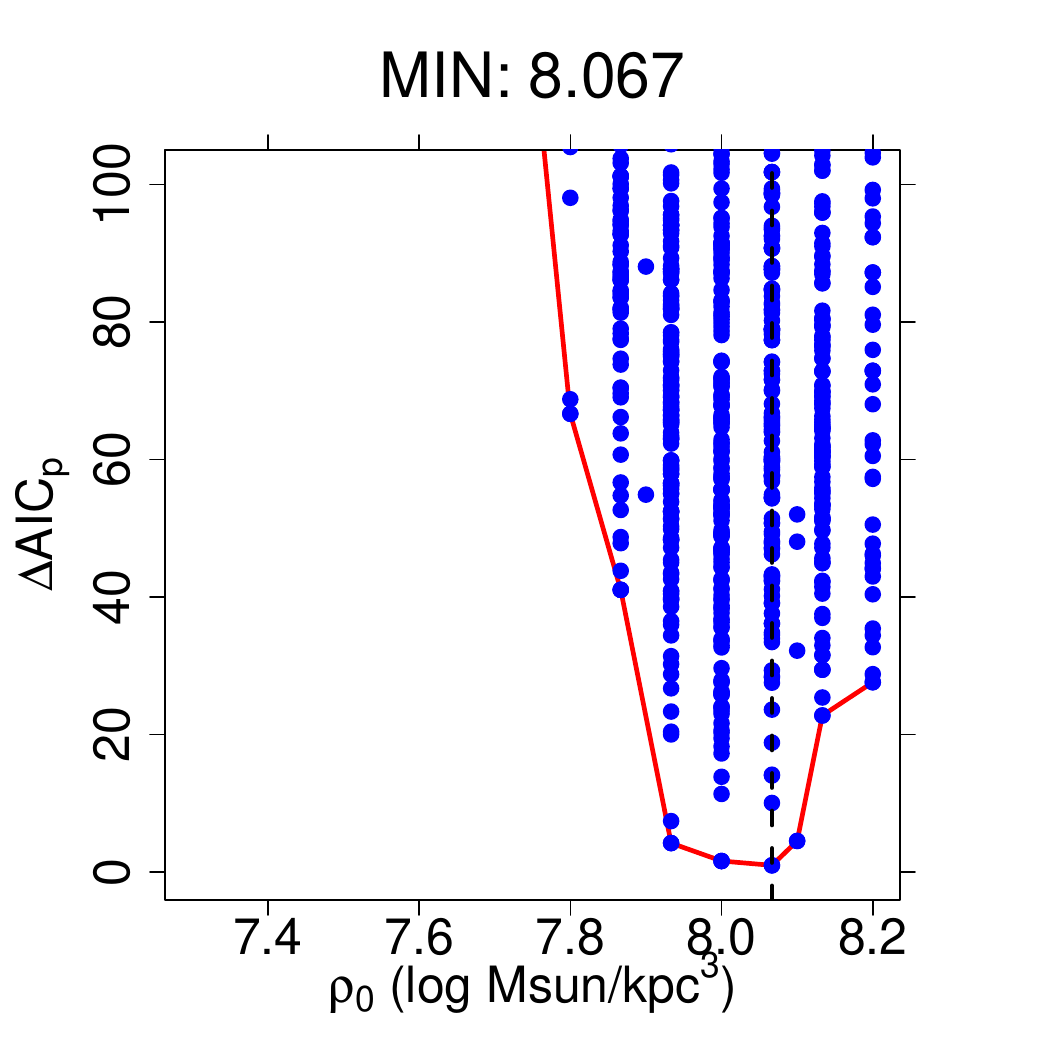}}

\subfloat{\includegraphics[width=.3\linewidth]{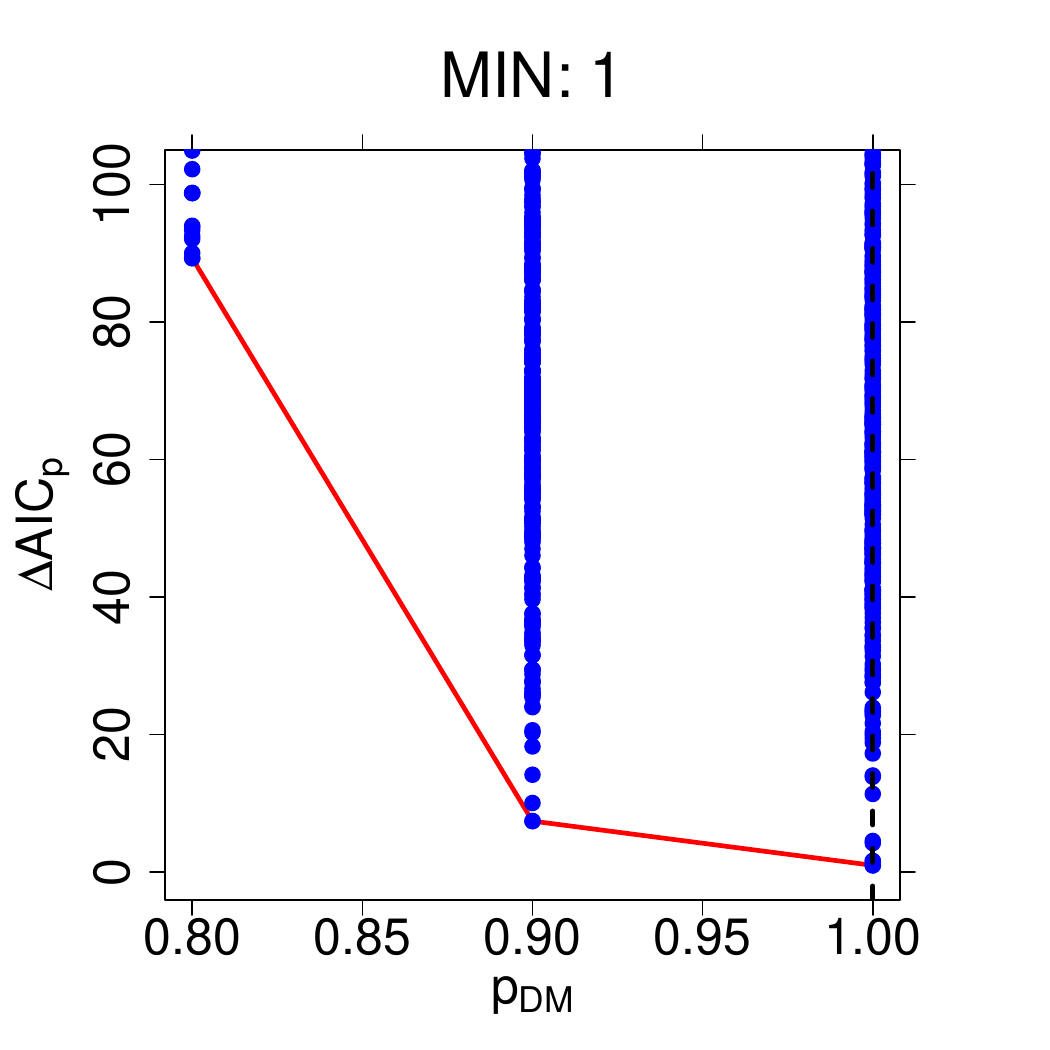}}
\subfloat{\includegraphics[width=.3\linewidth]{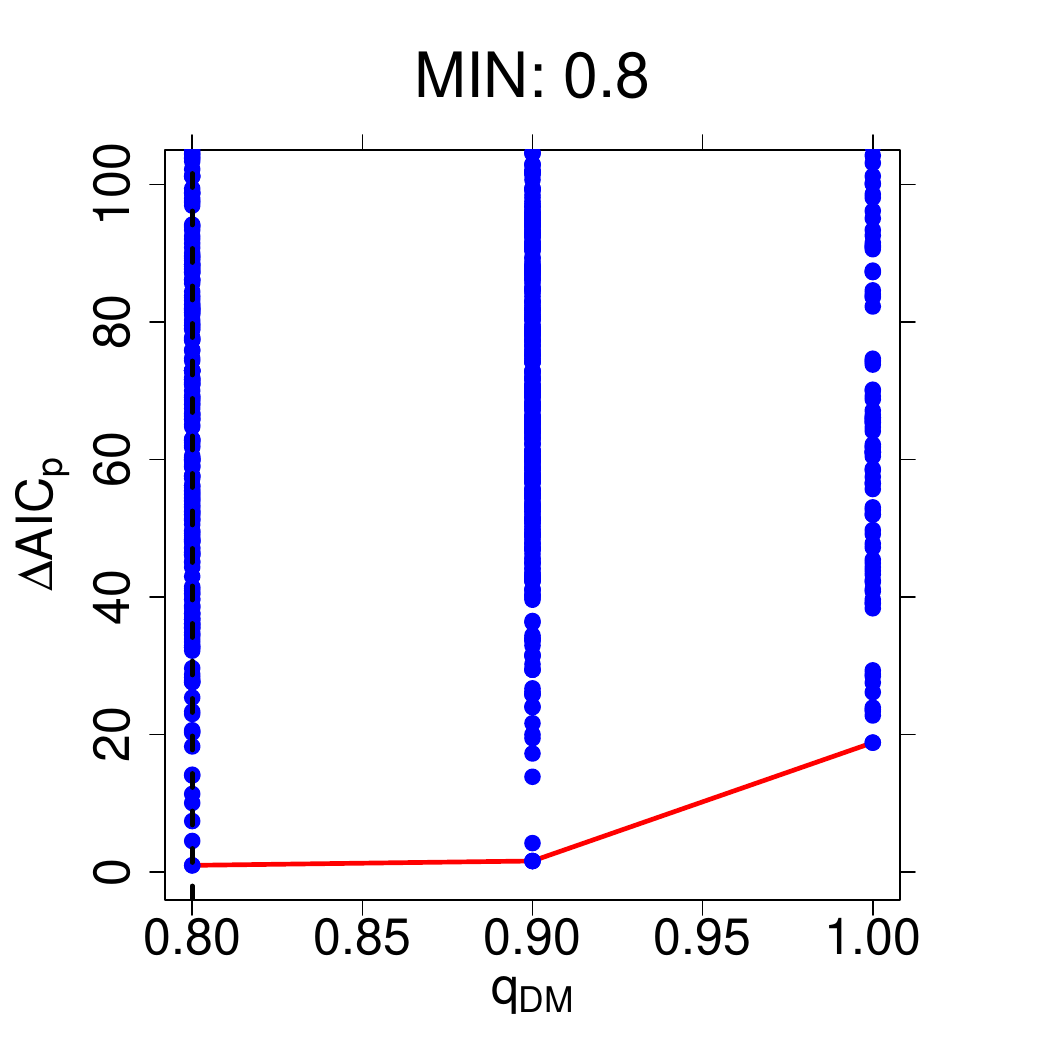}}
\subfloat{\includegraphics[width=.3\linewidth]{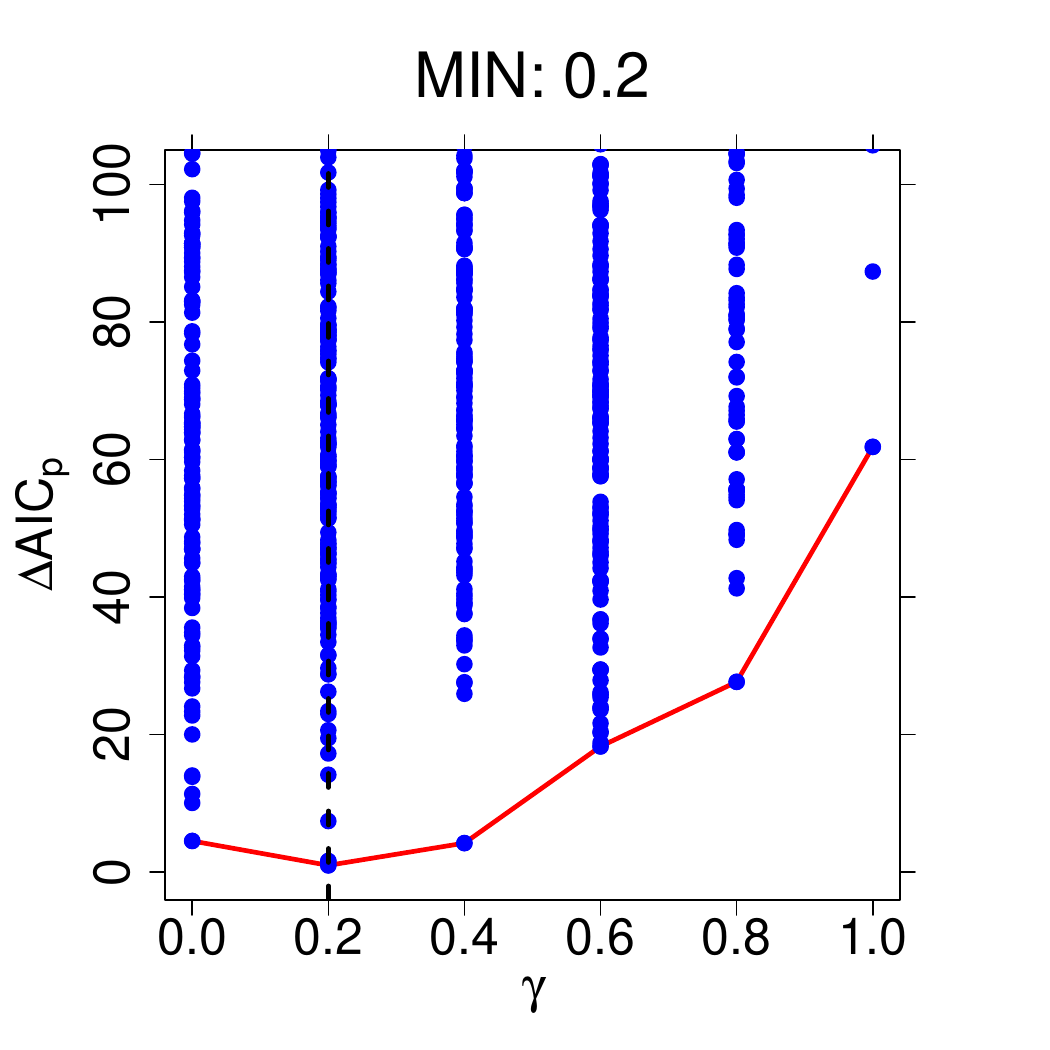}}

    \caption{AIC$_\text{p}$ values, calculated following eq.~\ref{eq.meff}, plotted against the 6 variables fitted in our final NOMAD run. The blue points are the individual models. The majority of them are not shown as they fal outside the plotted range. The black dashed line labels the best-fit value, while the red line follows the best-fit models for each tested value. \textit{Left to right, top to bottom:} \mbh, $\Gamma$, $\rho_0$, $\gamma$, $\text{p}_\text{DM}$, $\text{q}_\text{DM}$.}
    \label{Fig.Schw_results}
\end{figure*}

\begin{figure}
\includegraphics[width=\linewidth]{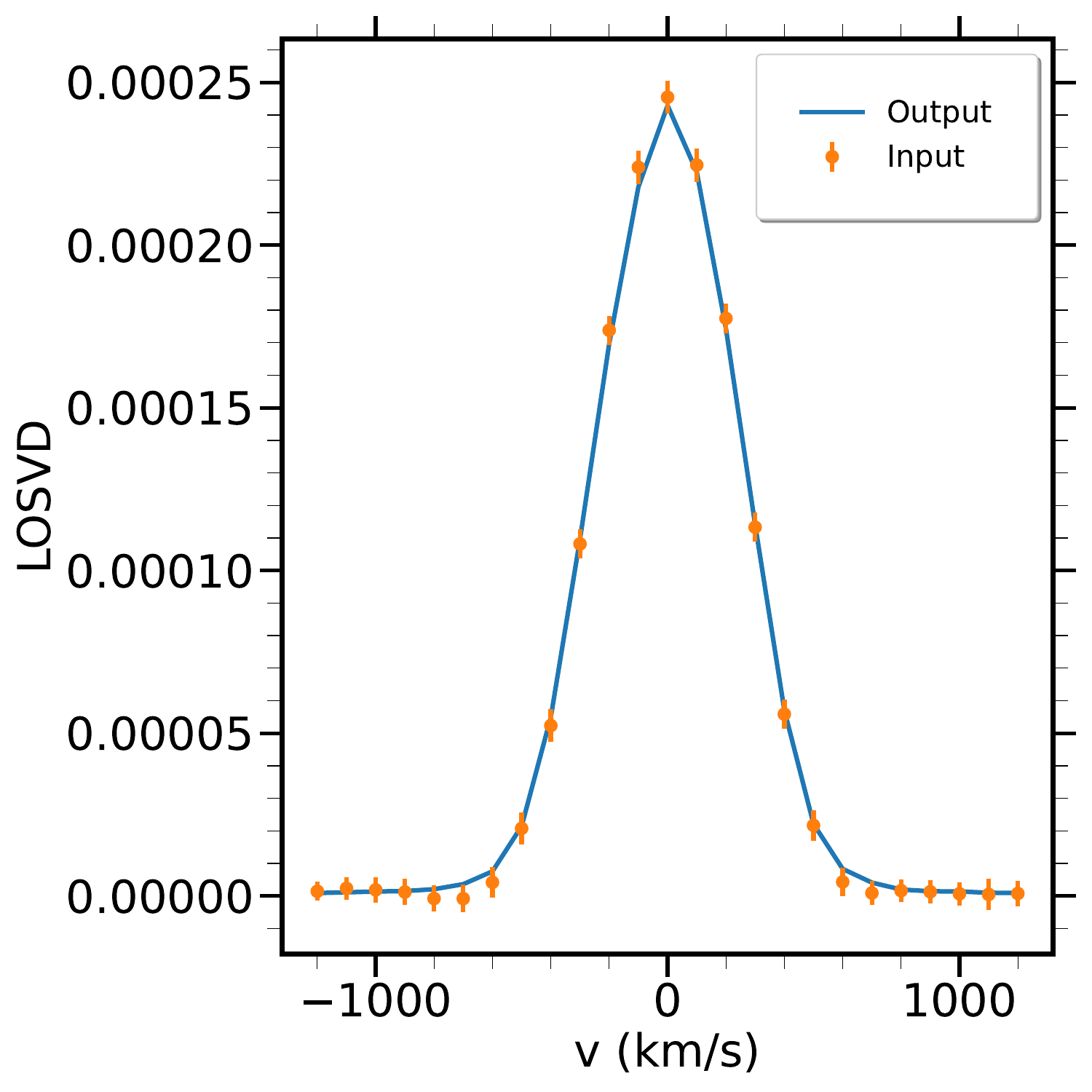}

    \caption{Example of SMART fit to a single LOSVD. The orange points are the input velocity bins, i.e. those provided by our code for kinematics extraction WINGFIT, while the light blue line is the recovered LOSVD. The bin is the same as that shown in Fig.~\ref{Fig.kin_wingfit}, i.e. located at (x,y) = (-0.8415, 1.2017) on the plane of the sky and belonging to the MAJOR slit (Tab.~\ref{Tab.slits}).}
    \label{Fig.kin_SMART}
\end{figure}


\subsection{Results} \label{Sec.results}
The shapes of the best-fit g'- and Ks-band densities are shown in Fig.~\ref{Fig.pqT_bf}, where we leave out the dust-contaminated central regions of the galaxy for the g'-band. In the optical, the density is prolate in the central regions and then becomes triaxial at larger radii. The viewing angles for this density coming out from the second NOMAD run are $\left(\theta, \phi, \psi\right) = (80,90,135)^\circ$: for this orientation, close to the $y$-axis, we sampled three different deprojections using as starting values $p(r) = 1.0, 0.95, 0.9$ at all radii\footnote{Note that also in this case the code \textit{does} fit $p(r)$. However, given the orientation, the fitted profile is expected to be close to the initial value.}. The last value provided the best-fit density. This agrees well with the findings of \citet{dN22}: the $p(r)$, $q(r)$, $T(r)$ intervals shown in Fig. 9 in that paper overlap well with those found in this work. It should be noted that the best-fit solution does not appear in \citet{dN22} because there only one deprojection\footnote{For the line-of-sight along the $y$-axis, we tested an almost oblate projection with $p(r)$ = 0.95 at all radii.} per orientation was tested. Instead, the recovered shape profile in Ks-band intersect with each other in the central regions, which is expected given the large isophote twist. It is reassuring that the two $q(r)$ profiles agree nicely with each other, while for $p(r)$ deviations of $\sim$ 0.1 are observed. Nevertheless, such scatter is expected (see \citealt{dN22b}). \\
The results of the third and final g'-band NOMAD run are shown in Fig.~\ref{Fig.Schw_results}, for a total of 1617 models. For each variable, we marginalize the 6-dimensional AIC$_\text{p}$ distribution to recover the six one-dimensional functions 
AIC$_\text{p} (\mbhm)$, AIC$_\text{p} (\Gamma)$, AIC$_\text{p} (\rho_0)$, AIC$_\text{p} (\gamma)$, AIC$_\text{p} (\text{p}_\text{DM}$ and AIC$_\text{p} (\text{q}_\text{DM})$. The fit to the kinematics is shown in Fig.~\ref{Fig.kinematics_WS} as orange lines. Here, we see that the code is able to recover all four kinematic variables well\footnote{Note the SMART fits the entire LOSVDs: Gauss-Hermite coefficients are only used for illustration.}; in particular, the minor-axis rotation is well reproduced. Moreover, in Fig.~\ref{Fig.kin_SMART} we show an example of a fit to an individual LOSVD, showing that SMART reproduces it very well: in this case, we have $\chi^2/\text{N}_{data} = 0.73$. \\
The most relevant result is the detection of a $(1.0 \pm 0.28) \times 10^{10}$ M$_\odot$ SMBH in the center of NGC~708. We calculate the size of its SOI in three ways: using the central velocity dispersion $\sigma_0$ as $\rsoi = G\mbhm / \sigma_0^2$ and using the stellar mass derived from our models as $M_\text{tot} (\rsoi) = 2\,\mbhm$ \citep{Merritt04} and as $M_\text{tot} (\rsoi) = \mbhm$ \citep{Jens16}. The \citet{Merritt04} definition is equivalent to the "velocity dispersion" one if the mass density profile of the galaxy can be described by a Singular Isothermal Sphere (SIS). Using the three definitions, we find $\rsoi = 0.73, 1.83$ and 1.21 kpc, respectively\footnote{These values correspond to 2.20, 5.51 and 3.64 arcseconds, respectively.}. The strong inconsistency points out that the SIS model does not work well for this galaxy.\\
The derived $\Gamma = 3.4 \pm 0.51$ is lower than the expected value for a Kroupa IMF: indeed, \citet{Wegner12} found $4.17 \pm 1.05$ in the Kron-Cousins $R$-band for a \citet{Kroupa01} IMF using SSP models \citep{Maraston05}. We discuss this further in Sec.~\ref{Ssec.starsgasDM}. \\
The DM halo has a normalization $\rho_0 = 10^{8.06}$ M$_\odot / \text{kpc}^3$ and, in the central regions, is less steep ($\gamma = 0.2$) than predicted by a NFW model ($\gamma = 1$). Finally, we note that DM halo is oblate ($\text{p}_\text{DM}$ = 1.0, $\text{q}_\text{DM}$ = 0.8). \\ 

\begin{figure*}

\subfloat{\includegraphics[width=.4\linewidth]{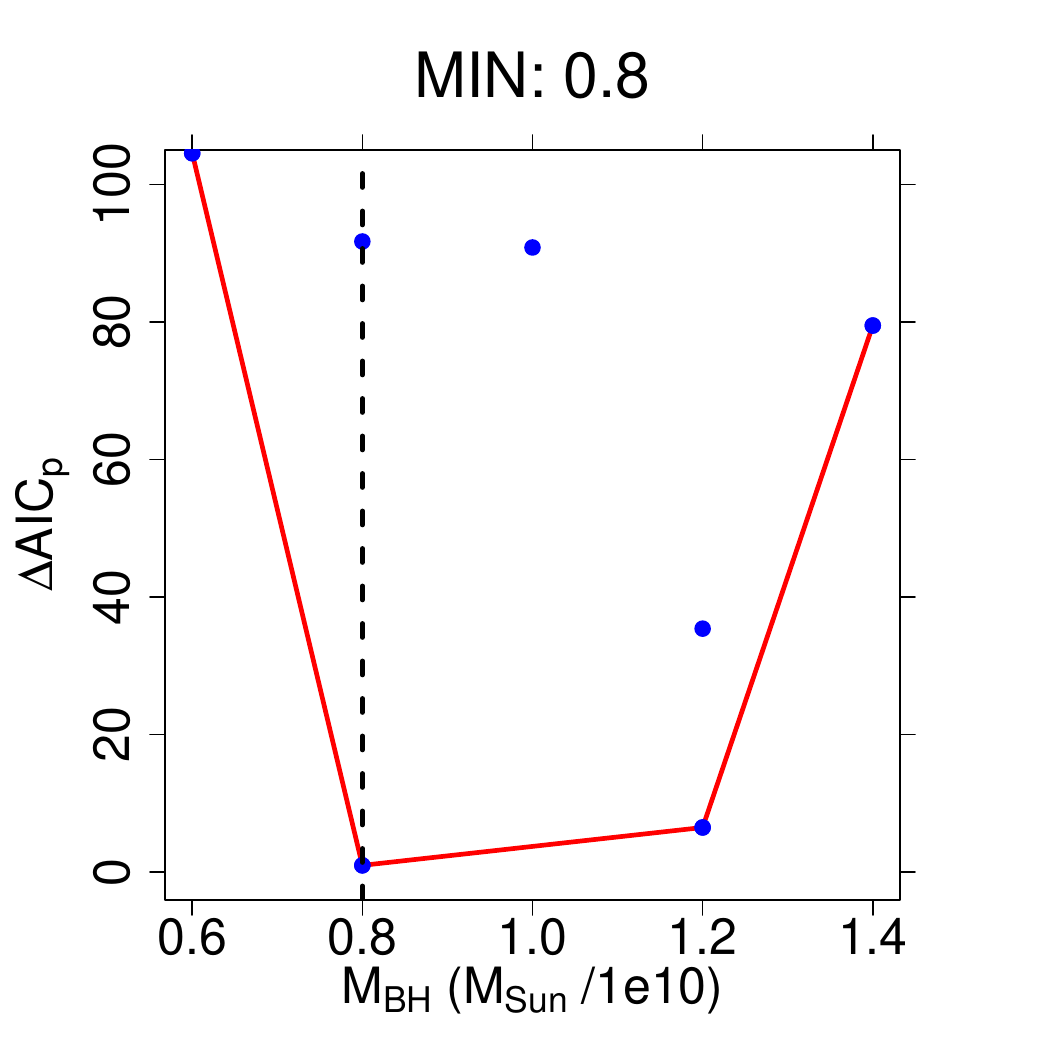}}
\subfloat{\includegraphics[width=.4\linewidth]{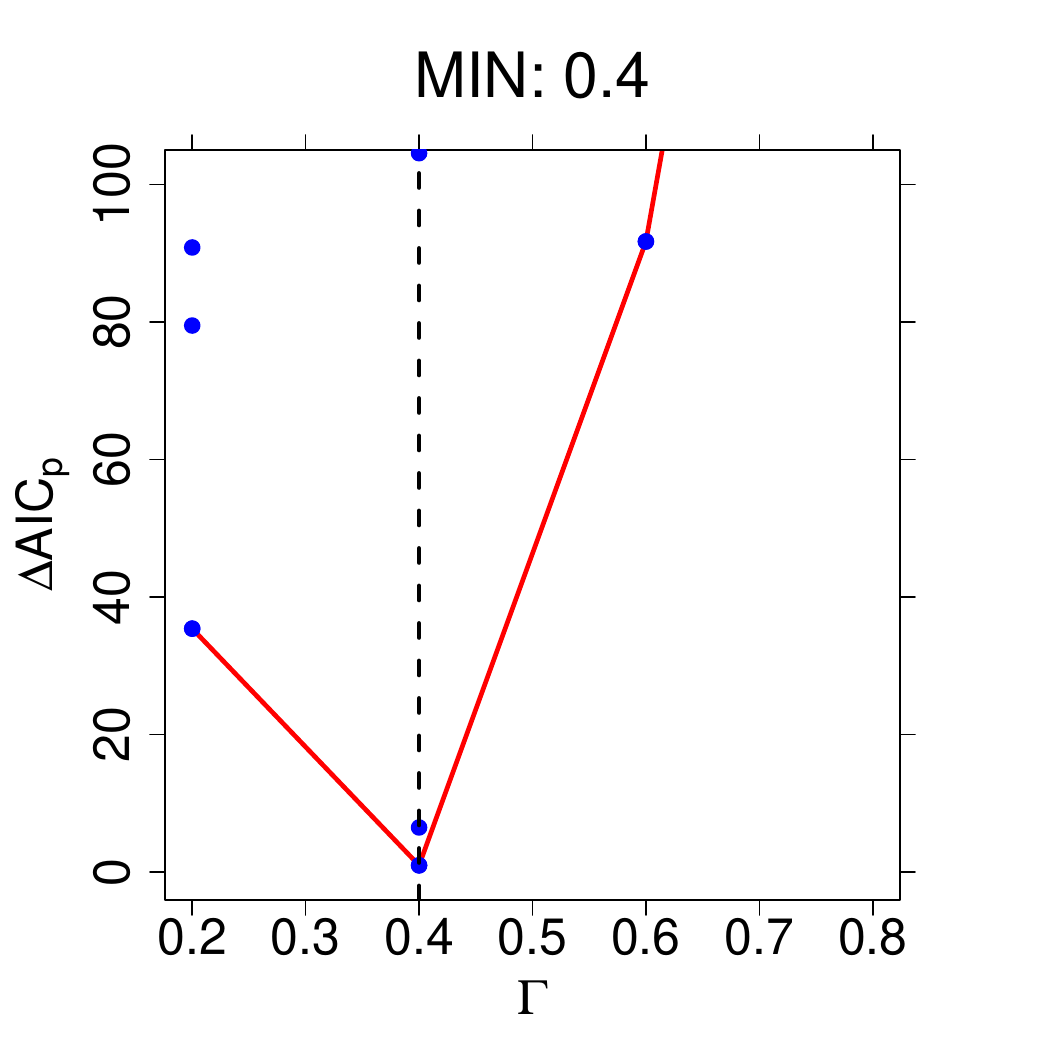}}

\subfloat{\includegraphics[width=.4\linewidth]{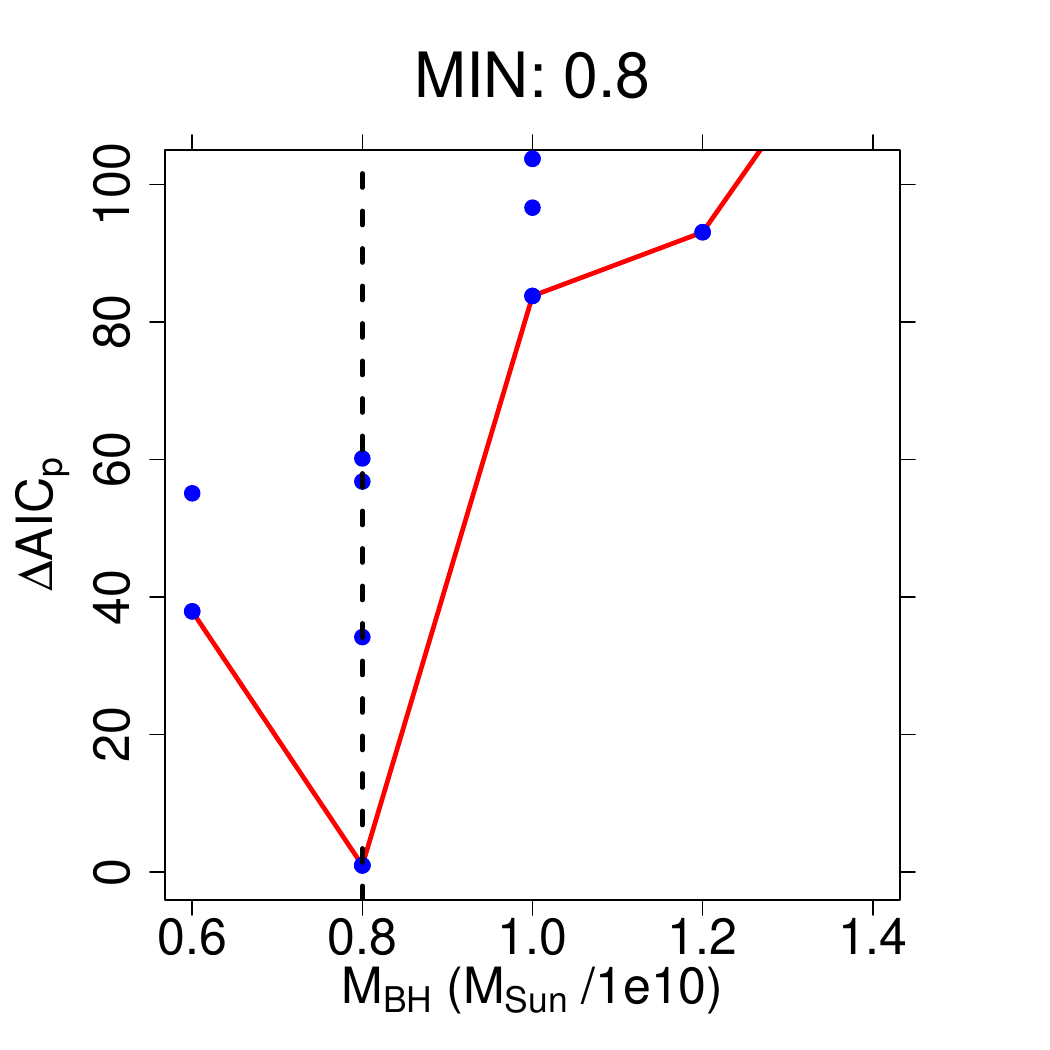}}
\subfloat{\includegraphics[width=.4\linewidth]{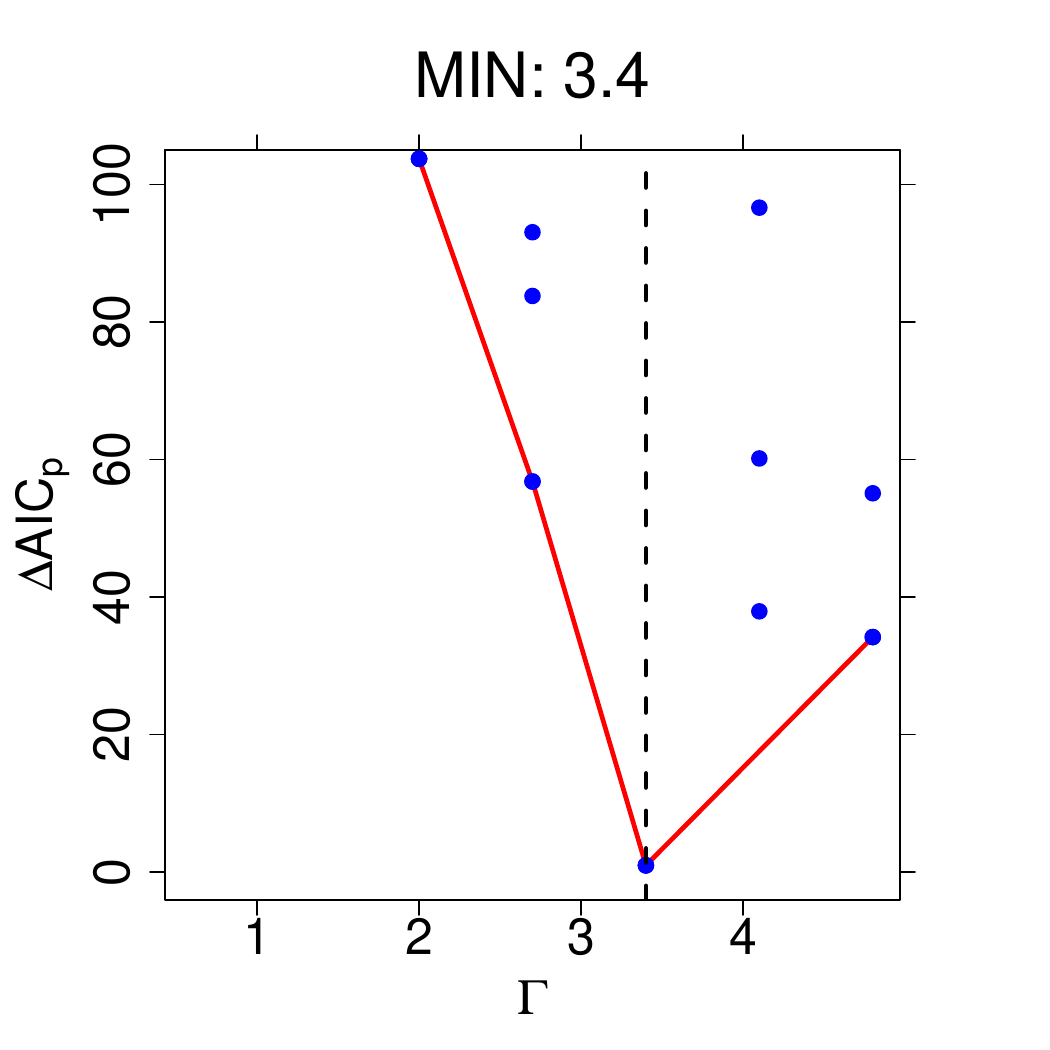}}

    \caption{Dynamical estimates of BH mass and M/L in Ks-band (top panels) and in g'-band (bottom panels) using the best-fit viewing angles found by deprojecting the Wendelstein Ks-band photometry. In both cases, the BH mass gets lower by 20\%, well within our error bars. Both M/L values are too low with respect to SSP values assuming a Kroupa IMF, but this can be explained assuming that the DM halo traces the stars.}
    \label{Fig.dyn_Ks}
\end{figure*}

Finally, the upper panels of Fig.~\ref{Fig.dyn_Ks} show the results of the NOMAD run, with fixed halo, using the best-fit Ks-band deprojection. Here the best-fit \mbh\,value is only 20\% smaller, which is well within our error bars. Furthermore, the value of $\Gamma_K$ = 0.4 is also lower than the prediction assuming a Kroupa IMF, as we also found for the g'-band deprojection. In the bottom panels of Fig.~\ref{Fig.dyn_Ks} we show what we get when modeling, again with fixed halo, the g'-band photometry at the best-fit Ks-band orientation: the results highlight that the black hole mass estimate also becomes 20\% smaller, while the $\Gamma$\,estimate remains the same. Given the uncertainty values which we estimated using the $N$-body simulation (Tab.~\ref{Tab.Nbody}), we conclude that our \mbh\,estimate is robust.

\begin{figure}

\includegraphics[width=\linewidth]{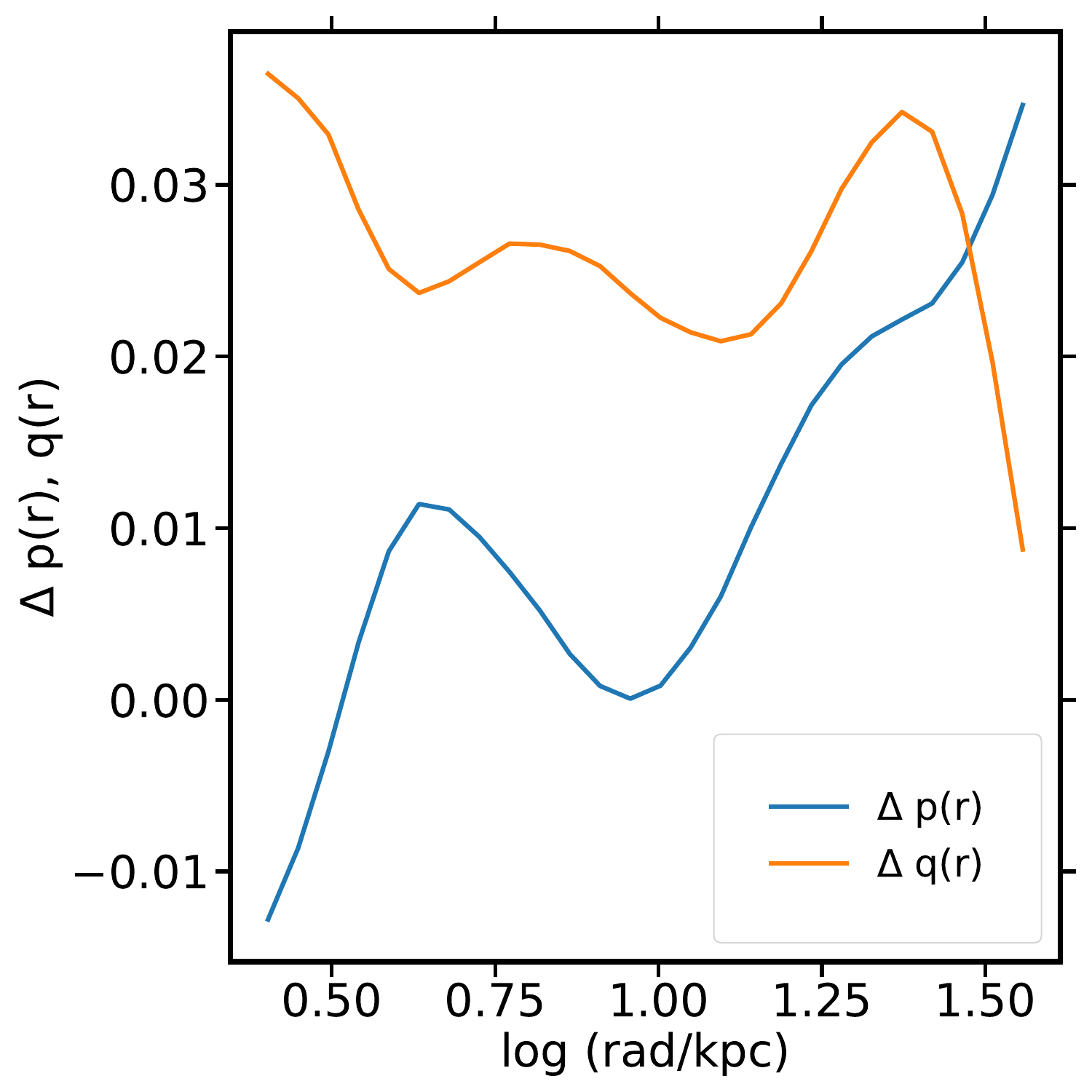}

    \caption{Differences between the intrinsic shape profiles $p(r)$ (blue) and $q(r)$ (orange) for the best-fit light density $\rho$ found by NOMAD and the average shape profiles $< p(r) >$ and $< q(r) >$ among all deprojections selected for the dynamical modeling. The very small residuals confirm the findings of \citet{dN22b}: only one deprojection is needed for the dynamical modeling.}
    \label{Fig.delp_delq}
\end{figure}



\section{Discussion} \label{Sec.discussion}

We have presented the full data analysis of NGC~708, BGC of A262. This is the second BCG which we modeled using our full non-parametric triaxial pipeline (SHAPE3D + WINGIFT + SMART): another work taking on dynamical modeling of a triaxial ETG is \citet{Bianca23b}. In what follows we discuss our results in more detail.

\subsection{Intrinsic shape} \label{Ssec.intr_shape}
As shown in \citet{dN22b}, the deprojection alone suffices if one wants to recover the correct intrinsic shape of the galaxy: we need to compute the average shape profiles among all deprojections\footnote{Intrinsic shape profiles which cannot be fitted because of deprojections performed at orientations along or close to the principal axes are not considered. For example, for the deprojection along the $y$-axis we do not consider the $p(r)$ profile.} which we select for the dynamical modeling and look for the deprojection which is closest to it. Instead, the dynamical modeling is needed to reduce the scatter of this estimate. \\
Here, we tested this approach with a real galaxy: after obtaining a first estimate of the mass parameters, we used these to estimate the best-fit viewing angles and thus the best-fit deprojection, using more than one density for the degenerate cases. The best-fit solution, found at $\left(\theta, \phi, \psi\right) = (80,90,135)^\circ$, is indeed remarkably close to the average $< p(r) >$, $< q(r) >$ profiles (see Fig.~\ref{Fig.delp_delq}). We notice that $p(r)$ mostly oscillates around the grid flattening $P = 0.9$, which is expected given the orientation close to the $y$-axis. Instead, in this case $q(r)$ is very close to the observed flattening $q' \equiv 1 - \varepsilon$: it starts decreasing at $r \sim 2.5$ kpc, as expected given the $\varepsilon$ profile shown in Fig.~\ref{Fig.photometry}, reaching $q(r) < 0.6$ at large radii. The corresponding triaxiality shows a monotonically decreasing profile, reaching the maximum triaxiality $T = 0.5$ at $\sim$10 kpc. Averaging over all radii, we find  $< T > = (1 - < p(r) >^2) / (1 - < q(r) >^2)$ = 0.455.


\begin{figure*}
\centering

\subfloat[\label{Fig.bhsigma}]{\includegraphics[width=.5\linewidth]{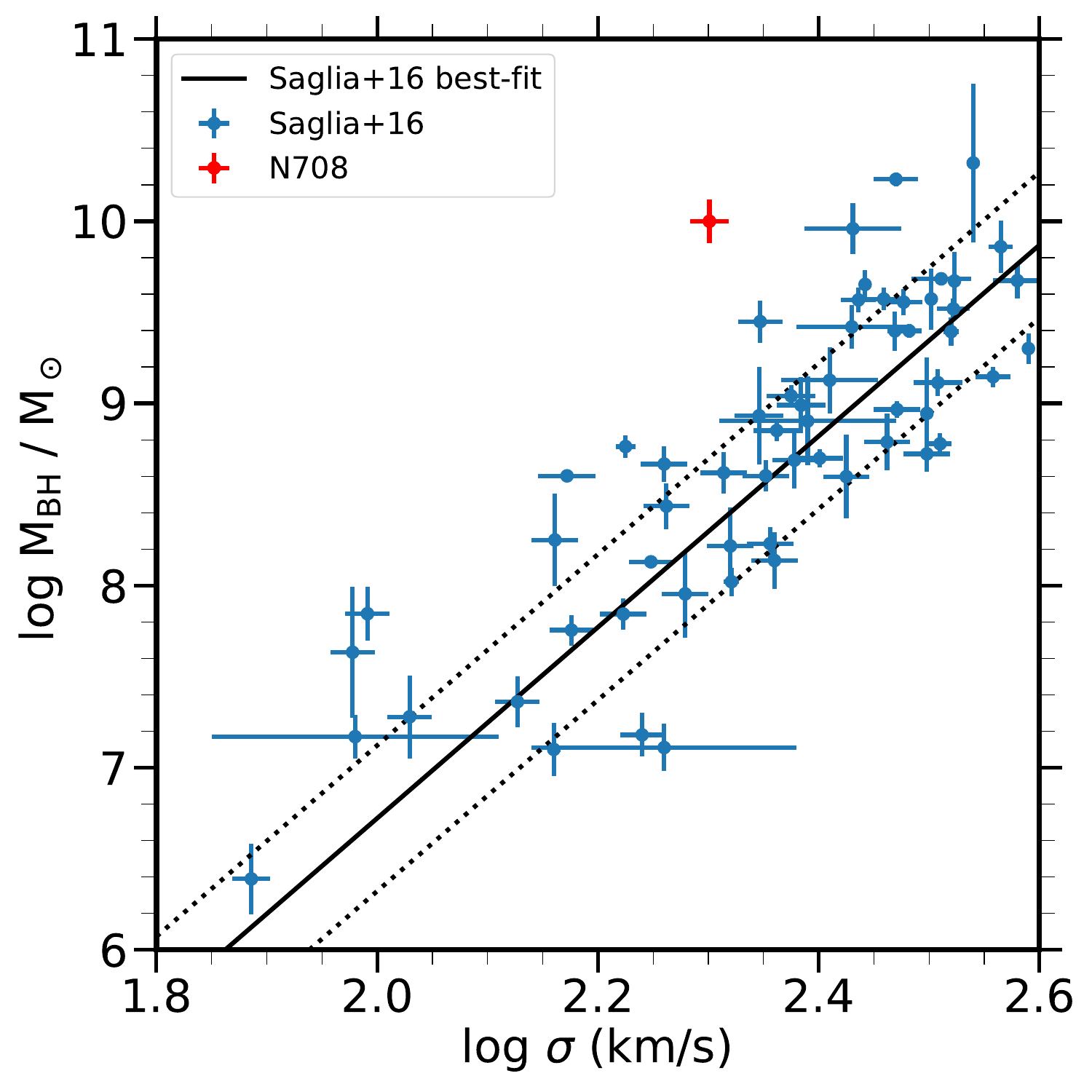}}
\subfloat[\label{Fig.bhrb}]{\includegraphics[width=.5\linewidth]{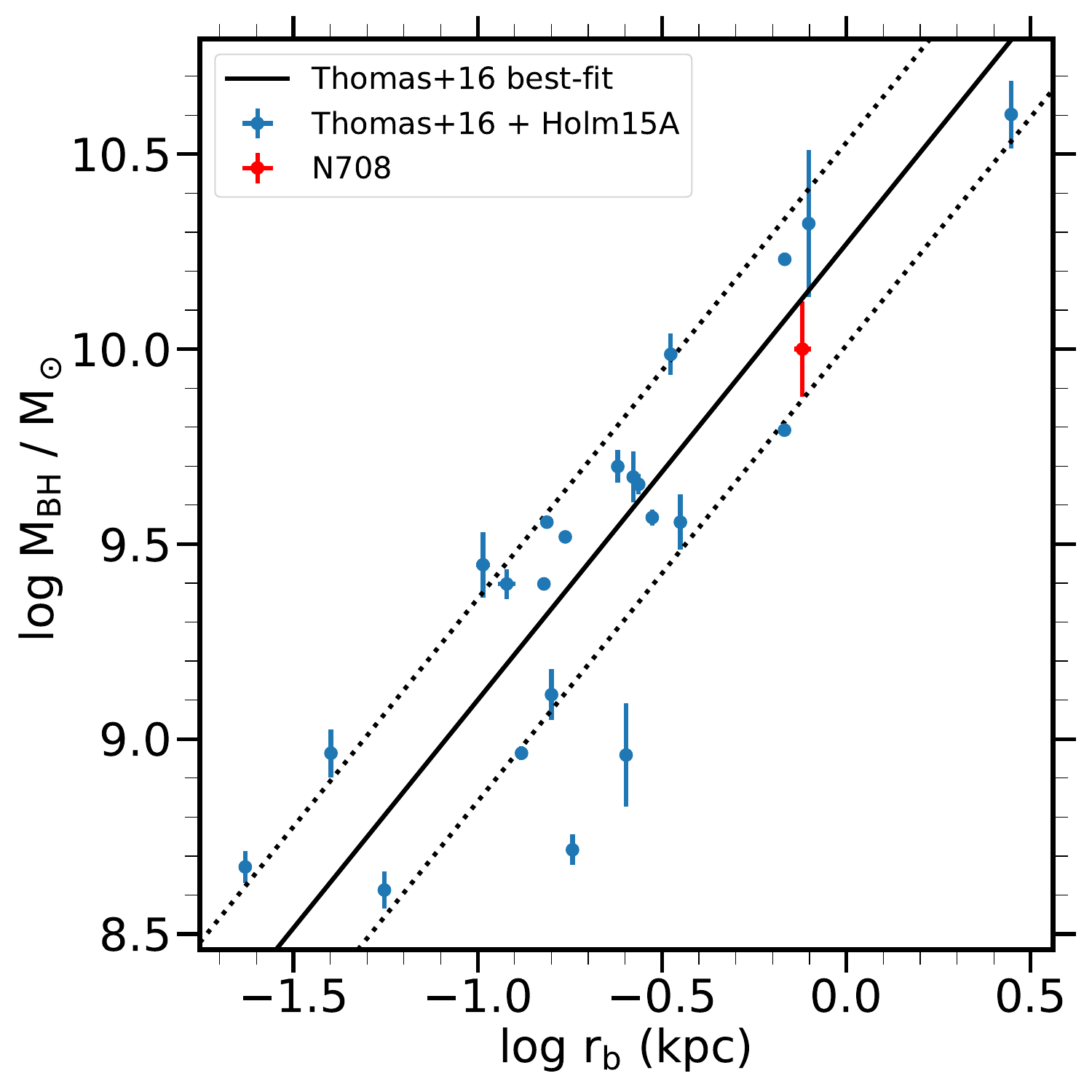}}

    \caption{\textit{Left:} \mbh-$\sigma$ relation for ETGs; the gray line is the best-fit line relation for ETGs from \citet{Rob16}. \textit{Right:} \mbh-r$_\text{b}$ relation with best=fit line from \citet{Jens16}. We also add the estimate of Holm15A \citep{Kianusch19}. Here the size of error bars on r$_\text{b}$ estimates is too small to be clearly visible in the plot. 
    In both plots NGC~708 is shown as a red dot, while the dotted lines enclose the intrinsic scatters of the linear relations.}
    \label{Fig.scaling}
\end{figure*}

\subsection{The Black Hole: scaling relations, anisotropy and orbital structure}
\begin{figure}
\centering

\includegraphics[width=.8\linewidth]{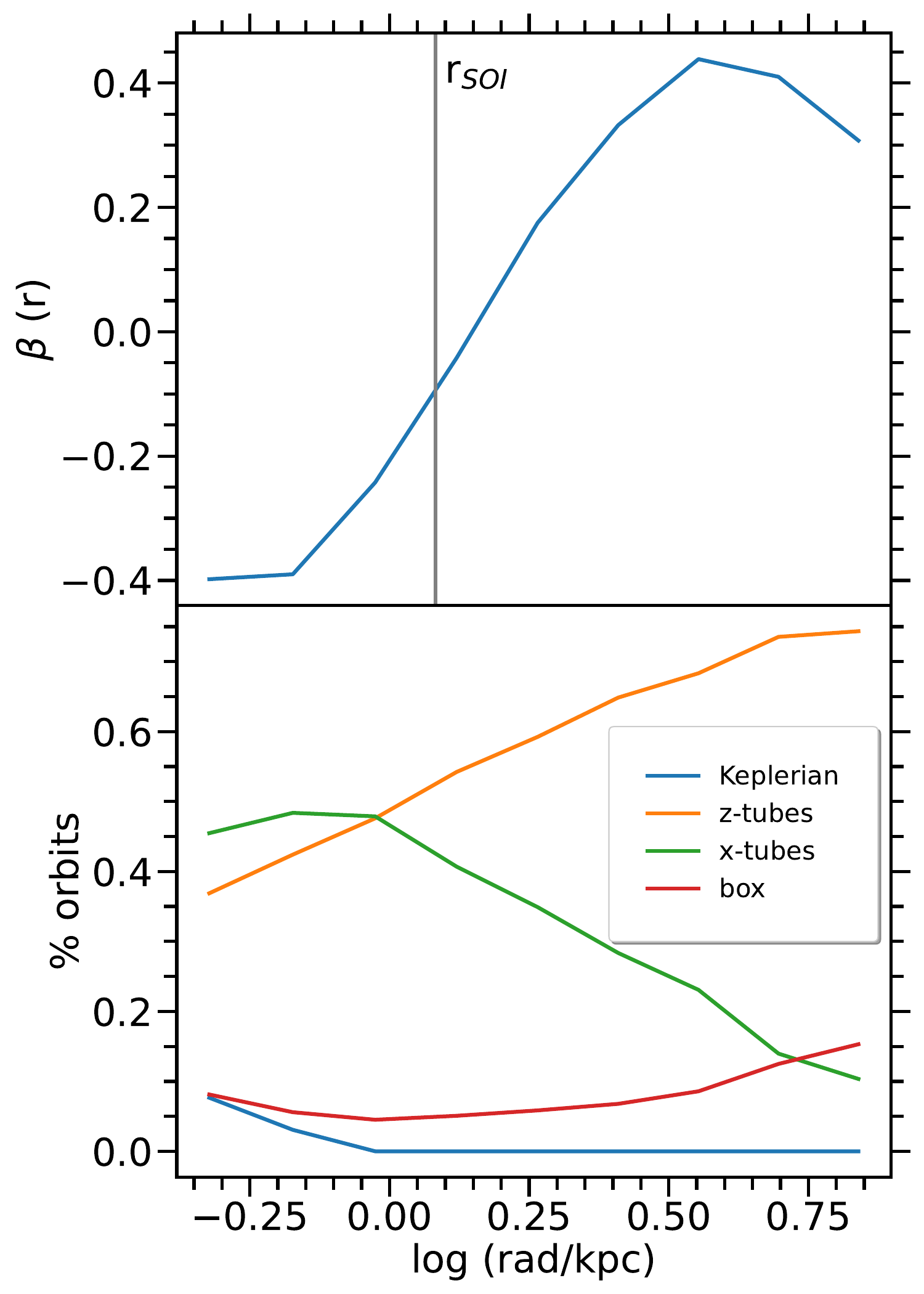}
    
    \caption{\textit{Top:} The recovered anisotropy of NGC~708 for the best-fit model. We find tangential anisotropy in the central regions. This is a fingerprint of BH core scouring and therefore provides evidence for past merger(s). The fact that the tangential bias is not so pronounced as in most core-galaxy might indicate that the progenitors of NGC~708 were core-galaxies themselves. At larger radii, the galaxy shows a small radial anisotropy. The core radius computed using the definition of \citet{Jens16} is also plotted. \textit{Bottom:} Orbit distribution for the best-fit model. $x$- and $z$-tubes dominate in the central region with tangential anisotropy, while $z$-tubes alone take over at large radii.}
    \label{Fig.beta}
\end{figure}

Our dynamical modeling delivers a BH mass of $(1.0 \pm 0.28) \times 10^{10}$ M$_\odot$. Even using the smallest $\rsoi$ value among the three we report in Sec.~\ref{Sec.results}, we would still have 18 kinematical bins inside the BH SOI: this, along with the fact that we included a DM halo in our models, confirms that our estimate is secure. Objects with masses in this range are rare: there is an almost empty region between the most massive SMBH dynamically detected \citep{Kianusch19} and SMBHs with masses $< 10^{10}$ M$_\odot$. Measurements in this region include NGC4889 ($\mbhm = 2.1 \times 10^{10}\,M_\odot$, \citealt{McConnell12}), NGC1600 ($\mbhm = 1.7 \times 10^{10}\,M_\odot$, \citealt{Jens16}) and the recently published combined mass ($\mbhm = 1.0 \times 10^{10}\,M_\odot$) of the two SMBHs of NGC5419 using triaxial models \citep{Bianca23b}. \\
At the very high-mass end, the scaling relation between SMBHs and the velocity dispersion of the host bulge $\sigma$ saturates. This is linked to the evolution history of these galaxy: massive ETGs accrete mass through gas-poor ("dry") mergers which do not significantly alter $\sigma$ \citep{Naab09} and, hence, generate \mbh\,values which are higher than the prediction of the canonical $\mbhm-\sigma$ relation, for which typically $\mbhm \propto \sigma^{5.2 \div 5.4}$ \citep{Rob16, VDB16}. This is expected: SMBHs correlate with the bulge parameters \citep{Rob16, dN19}, which are locked together through the Fundamental Plane (FP, \citealt{Djorgovski87}) and, for BCGs, different \citet{Faber76} (FJ) and FP relations with respect to ordinary ETGs are found \citep{Matthias20}. \\
Regardless of which $\mbhm-\sigma$ we assume, NGC~708 is an outlier (see Fig.~\ref{Fig.bhsigma}). The coefficients found by \citet{Rob16} omitting pseudobulges predict $\mbhm = 5.75 \times 10^{8}$ M$_\odot$, making the galaxy an outlier \textit{by almost a factor of 20}. We note that NGC~708 has a low velocity dispersion in comparison with most BCGs. This stays roughly constant at all radii, peaking at 250 km s$^{-1}$ in the most central bins. \\
The same conclusions apply by considering the \mbh-\mbul\,relation \citep{Magorrian98}. As estimate of \mbul\, we take the stellar mass profile, computed using the deprojected density multiplied by \ml, up to the largest radius used for the deprojection, yielding $\mbulm = 2.8 \times 10^{11} \text{M}_\odot$. Using the coefficients of \citet{Rob16}, again omitting pseudobulges, the galaxy is an outlier by a factor 10.5, while using the relation of \citet{Bogdan18} the galaxy is a factor 10.6 off. Thus, the galaxy is not only an extreme case among the galaxy population, but also among core galaxies only\footnote{Note that all scaling relations have intrinsic scatters. Using the (large) intrinsic scatter of the \mbh-\mbul\,relation $\varepsilon = 0.61$ from \citet{Bogdan18}, NGC~708 remains an outlier by a factor of 2.6.}. Note that since we did not try to subtract a potential ICL component from the light of NGC~708 our \mbul\ are upper limits and the offsets from the relations might be even larger. \\
The commonly proposed formation mechanism for these central cores is the gravitational slingshot caused by SMBHs lying at the center of the progenitors, and forming a binary after the merging process. This phenomenon ejects stars, causing a light deficit in the central regions (i.e. the core, \citealt{Ebisuzaki91}), and in dry mergers there is no gas that can replenish the center. Therefore, scaling relations linking \mbh\,to the core properties are theoretically expected and have indeed been observed: these include a correlation with the missing mass \citep{Kormendy09}, with the core size r$_\gamma$ \citep{Rusli13b, Jens16} and with the central surface brightness SB$_0$ of the core itself \citep{Kianusch19}. Indeed, the coefficients of \citet{Jens16} for the \mbh-r$_\gamma$ and those of \citet{Kianusch19} for the \mbh-SB relations\footnote{This relation was derived in the $V$-band, which is close to the $g'$-band photometry used in this work nonetheless.} predict a BH with mass $\sim 1.5 \times 10^{10}$, a much better prediction for the \mbh\,value found in this work (see also Fig.~\ref{Fig.bhrb}). \\
Another good predictor of \mbh\,for BCGs is the virial temperature kT (i.e. the total gravitating mass) of the host cluster \citep{Bogdan18}. This happens because the mass drives the accretion onto the SMBH itself. Indeed, using the kT value of $2.25 \pm 0.04$ keV for A262 \citep{Ilic15} yields $\mbhm = 6.5 \times 10^9$, also a reasonable estimate of the actual \mbh\,value. \\
The core scouring mechanism generates a tangential anisotropy ($\beta < 0$, eq.~\ref{eq.beta}) in the central regions, because stars on radial orbits come closer to the SMBH and are more likely to be ejected. In the top panel of Fig.~\ref{Fig.beta} we plot the anisotropy profile $\beta (r)$ for the best-fit model. Even if the profile shows the typical negative $\beta$, the less pronounced negative anisotropy with respect to other BCGs hints at a merger between two core-galaxies as formation scenario for NGC~708. In the future we will compare the derived $\beta (r)$-profile with simulations which reproduce the formation of cores in ETGs \citep{Rantala19, Frigo21}. \\
Instead, the bottom panel of Fig.~\ref{Fig.beta} shows the corresponding orbit distribution, where we see that tubes dominate over box orbits. In particular, while in the innermost regions with tangential anisotropy we find a similar amount of $x$- and $z$-tubes, orbits rotating around the minor axis take over at larger radii. This $z$-tubes dominance agrees with $p(r) \sim 0.9$, which indicates a galaxy shape close to an oblate spheroid, for which $x$-tubes cannot be observed at all.\\

\subsection{Stars, DM and gas} \label{Ssec.starsgasDM}
\begin{figure}
\centering

\includegraphics[width=.8\linewidth]{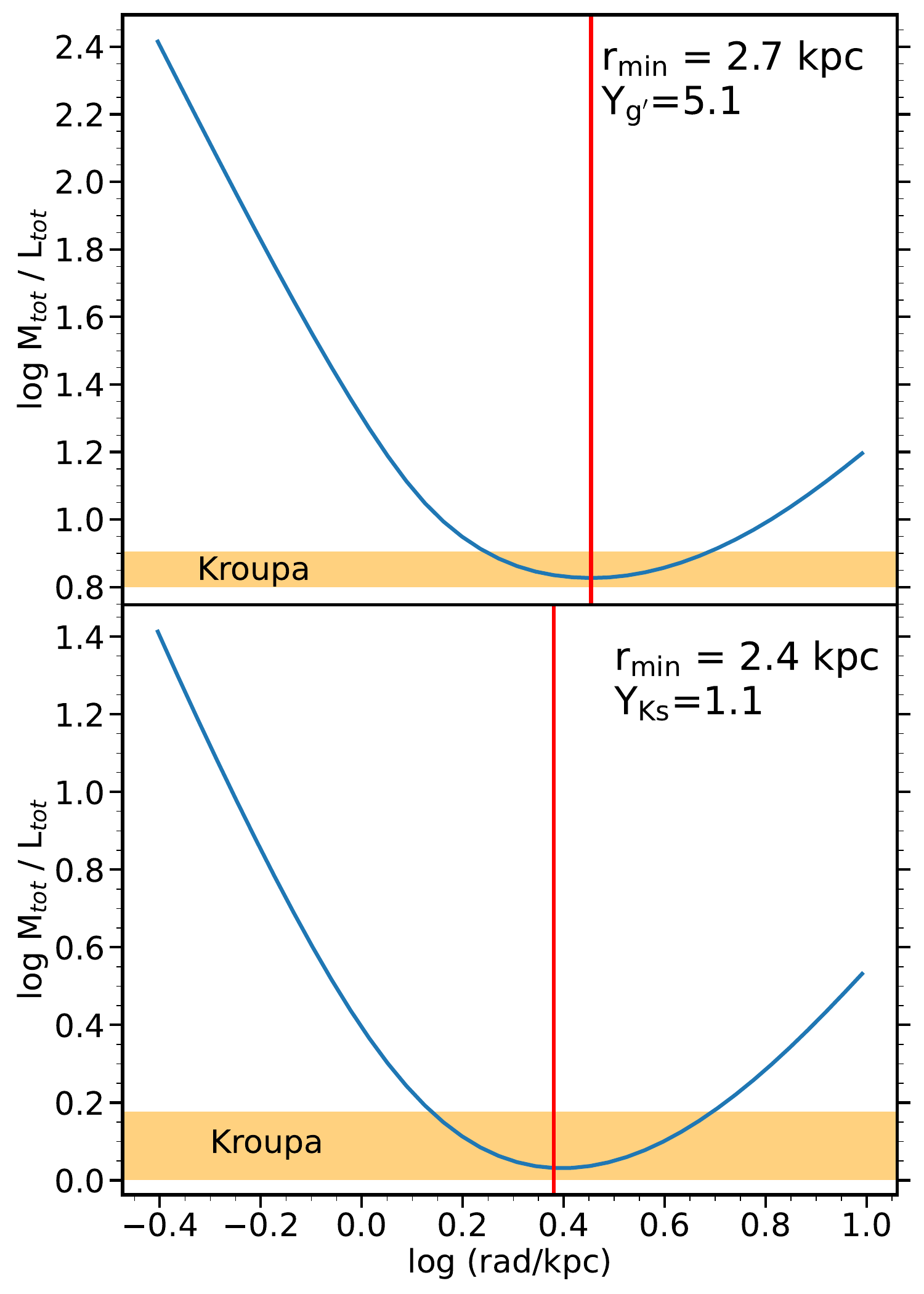}

    \caption{The total mass-to-light profile for the best-fit g'-band model (top) and Ks-band model (bottom) of NGC~708. The value at the minimum is what we use to derive an estimate of $\mlm_*$ under the assumption that the halo partially traces stars. The shaded regions show the expected values range under the assumption of Kroupa IMF in order to match the R-band value of \citet{Wegner12}.}
    \label{Fig.MtotL}
\end{figure}

\begin{figure}
\centering
\includegraphics[width=.8\linewidth]{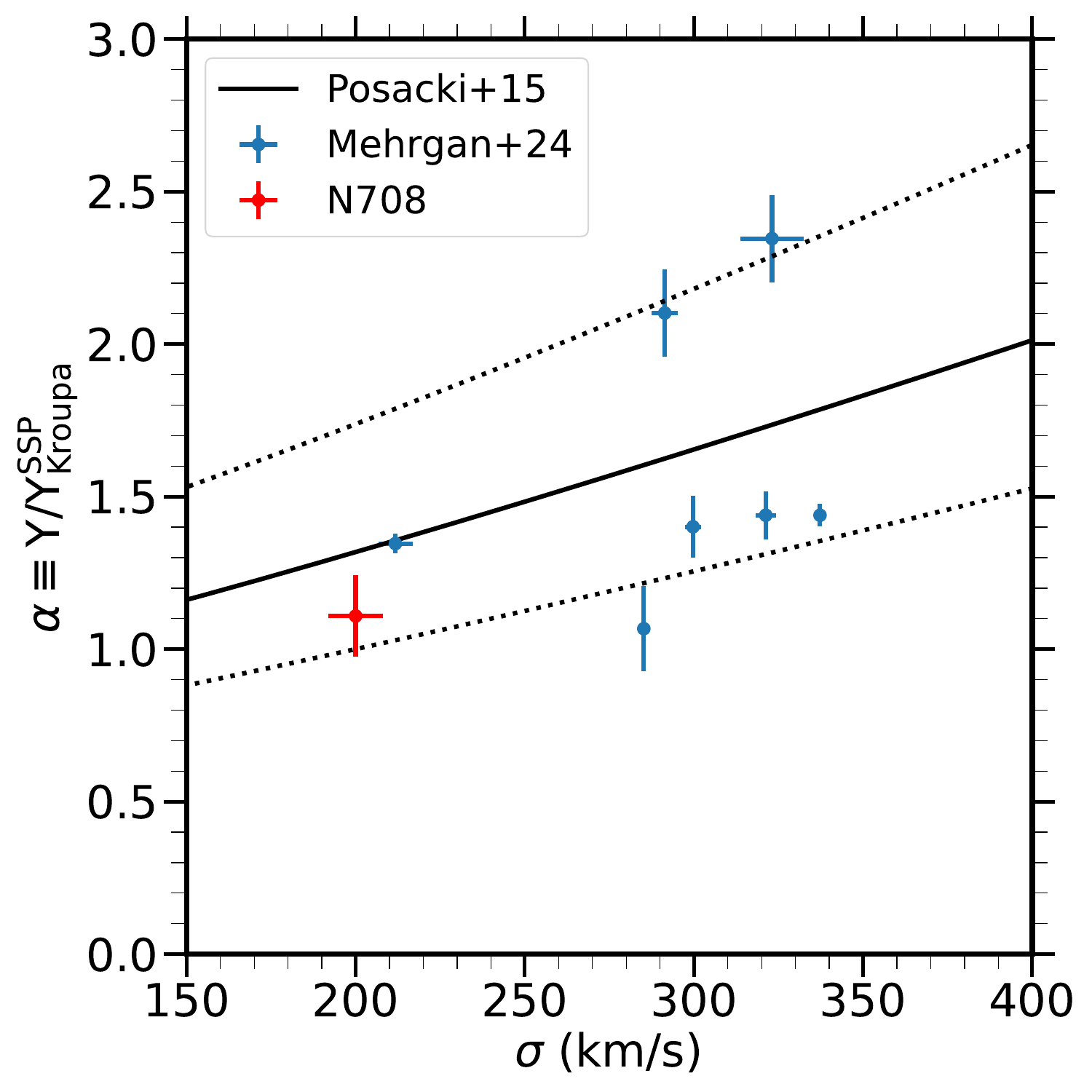}
\caption{Similar as in Fig.~\ref{Fig.scaling} for the $(\alpha, \sigma)$ relation evaluated at the radius r$_\mathrm{main}$ for the 9 ETGs of \citet{Kianusch24} with the best-fit quadratic relation and its scatter of \citet{Posacki15}. NGC~708 is shown as a red dot; despite having a Kroupa-like IMF, the dispersion is so low that the galaxy lies within the scatter of \citet{Posacki15}'s relation.}

    \label{Fig.alphasigma}
\end{figure}

Mass-to-light ratio estimates for NGC~708 have been published by \citet{Wegner12} by means of stellar population analysis. In that paper the authors use the SSP method from \citet{Maraston05} to derive the mass-to-light ratio assuming a Kroupa IMF. According to this value, we would need $\Gamma$ to be \textit{at least} 4.6 for a Kroupa IMF assuming an old stellar population with twice solar metallicity. In a recent study \citep{Kianusch24} the authors suggest to evaluate the mass-to-light ratio using the \textit{total} mass-to-light profile, i.e. by using the parameter \ml\,introduced in Sec.~\ref{Ssec.ourcode}. This is particularly relevant in this case because of the possible degeneracy between stars and DM: the dynamical mass-to-light ratio estimations assume that the whole mass comes from stars in the galaxy, while it may well be that the DM traces the stars and reduces M/L. Given that we do find a massive halo, we choose to use \ml\,rather than $\Gamma$ to estimate M/L, showing the total mass-to-light profile in the upper panel of Fig.~\ref{Fig.MtotL}. We see that the profile shows a clear minimum - signaling the region where stars dominate the potential - at 2.7 kpc. Evaluating \ml\,then yields a g'-band mass-to-light ratio of 5.1, in good agreement with SSP models assuming a Kroupa IMF and with the findings of \citet{Wegner12}. A second consistency check is given by the Ks-band mass-to-light ratio $\Upsilon_K$: as in g'-band, by evaluating $\Upsilon_K$ at the minimum of the $\text{M}_\text{tot} / \text{L}$ curve (found at 2.4 kpc, see bottom panel of Fig.~\ref{Fig.MtotL}, we find a value to 1.1, in good agreement with the estimates in R and g'-bands which would foresee $\Upsilon_K \in [1.0-1.5]$ for a Kroupa IMF and an old, metal-rich stellar population.  \\
Following \citet{Kianusch24} we define an IMF mismatch parameter with respect to a Kroupa IMF as $\alpha = \Upsilon / \Upsilon^\mathrm{SSP}_\mathrm{Kroupa}$, which is known to correlate with $\sigma$ (e.g. eq. 6 of \citealt{Posacki15}), and show in Fig.~\ref{Fig.alphasigma} the $(\alpha, \sigma)$ relation
for the 9 ETGs with MUSE kinematics from \citet{Kianusch24}
and NGC~708 and the best-fit parabola of \citet{Posacki15} obtained using SLACS + ATLAS$^\text{3D}$ data. Both $\alpha$ and $\sigma$ are evaluated at the radius where the $\text{M}_\text{tot} / \text{L}$ profile for that particular galaxy has its minimum. This is called r$_\mathrm{main}$ in \citet{Kianusch24}. While most ETGs fall below the best-fit curve, indicating that also at high dispersion IMFs are more Kroupa-like\footnote{\citet{Kianusch24} found evidence for mass-to-light gradients in 8 of the 9 galaxies they analyzed, reporting central regions with Salpeter-like IMF. Nevertheless, these gradients are so concentrated that already beyond 2 kpc the IMF becomes Kroupa-like.}, the dispersion of NGC~708 is so low that the galaxy does fall within the scatter of the \citet{Posacki15} relation.\\
Finally, from the best-fit potential yielded by SMART we can reconstruct the rotation curve and compare it to the results of \citet{Wegner12}. While we did find a deprojected velocity profile agreeing with the circular velocity measured from our models, the inclination of the gas disk needed to match our results does not place it on any of the principal planes (x,y), (x,z), (y,z), stressing the need of IFU data to better clarify this issue.

\section{Summary and conclusions} \label{Sec.conclusions}
We have obtained dynamical models of NGC~708, BCG of A262, combining high-resolution HST images with deep-photometry data taken with the WWFI at the Wendelstein observatory and using long-slit spectroscopy acquired with the MODS instrument at LBT. The analysis is done using a fully non-parametric triaxial pipeline, combining our deprojection code SHAPE3D, our code for kinematics extraction WINGFIT and, finally, our triaxial Schwarzschild code SMART. \\
The galaxy shows several interesting features. We detect a SMBH with mass $(1.0 \pm 0.28) \times 10^{10}$ M$_\odot$, one of the few measurements in this mass range, which makes the galaxy a strong outlier in the classical $\mbhm-\sigma$ and $\mbhm-\mbulm$ relation, but not in all scaling relations linking \mbh\,with core properties (central SB, core size) when intrinsic scatters of these relations are taken into account. In these respects, the galaxy resembles NGC1600 \citep{Jens16}.\\
The typical anisotropy profile of cored galaxies, tangential inside the core and then radial at larger radii, is found. Nevertheless, the value of $\beta$ inside the core is typical of galaxy mergers where the two progenitors are, themselves, core-galaxies. This is similar to what it has been observed for Holm15A, which is also an outlier in the scaling relations \citep{Kianusch19}. For NGC~708 we find a $g'$-band \ml$_*$\,value of $3.4 \pm 0.51$, which hints at a uncharacteristic lightweight IMF. However, this seems to be due to a degeneracy in the total mass profile: computing this value at the minimum of the curve yields an estimate in agreement with a Kroupa IMF. \\
We find that the galaxy is observed close to the intermediate axis. Given that the galaxy shows a small isophotal twist as well as minor-axis rotation, it must be triaxial, but the large amount of $z$-tubes found at large radii suggests that the triaxiality is not so pronounced. 
This is what we get when looking at the best-fit shape from the dynamical model: the shape of the galaxy is indeed remarkably close to an oblate spheroid, especially in the outermost regions. Interestingly, also our best-fit DM halo shows a similar geometry. Finally, we also attempt to derive the inclination and the PA of the gas disk by comparing its kinematics with our dynamical modeling-derived rotation curve, finding that the gas disk is not on any of the principal planes, even if the lack of IFU data makes it difficult to say more. \\
The successful modeling of a difficult galaxy such as NGC~708 motivates us to systematically extend this analysis to every BCG for which we have photometric and kinematical data. Given that these objects follow different scaling relations with respect to ordinary ETGs, recovering their intrinsic properties can help us to shred light on their evolution history.

\section*{Acknowledgements}
We thank the anonymous referee for carefully reading the manuscript and providing us with useful comments which helped us improving the paper. \\
Computations were performed on the HPC systems Raven and Cobra at the Max Planck Computing and Data Facility. \\
The LBT is an international collaboration among institutions in the United States, Italy and Germany. LBT Corporation partners are:  LBT Beteiligungsgesellschaft, Germany, representing the Max-Planck Society, the Astrophysical Institute Potsdam, and Heidelberg University; The University of Arizona on behalf of the Arizona university system; Istituto Nazionale di Astrofisica, Italy; The Ohio State University, and The Research Corporation, on behalf of The University of Notre Dame, University of Minnesota and University of Virginia. \\
modsCCDRed was developed for the MODS1 and MODS2 instruments at the Large Binocular Telescope Observatory, which were built with major support provided by grants from the U.S. National Science Foundation's Division of Astronomical Sciences
Advanced Technologies and Instrumentation (AST-9987045), the NSF/NOAO TSIP Program, and matching funds provided by the Ohio State University Office of Research and the Ohio Board of Regents. Additional support for modsCCDRed was provided by NSF Grant AST-1108693.
This paper made use of the modsIDL spectral data reduction reduction pipeline developed in part with funds provided by NSF Grant AST-1108693 and a generous gift from OSU Astronomy alumnus David G. Price through the Price Fellowship in Astronomical Instrumentation. \\
This work makes use of the data products from the HST
image archive.

\section*{Data Availability}
The data underlying this article will be shared on reasonable request to the corresponding author.
	
\bibliographystyle{mnras}
\bibliography{bibl}

\appendix


\section{Dust correction}  \label{App.dust_correction}
\begin{figure*}
\centering

\subfloat[\label{Fig.original_f110}]{\includegraphics[width=.5\linewidth]{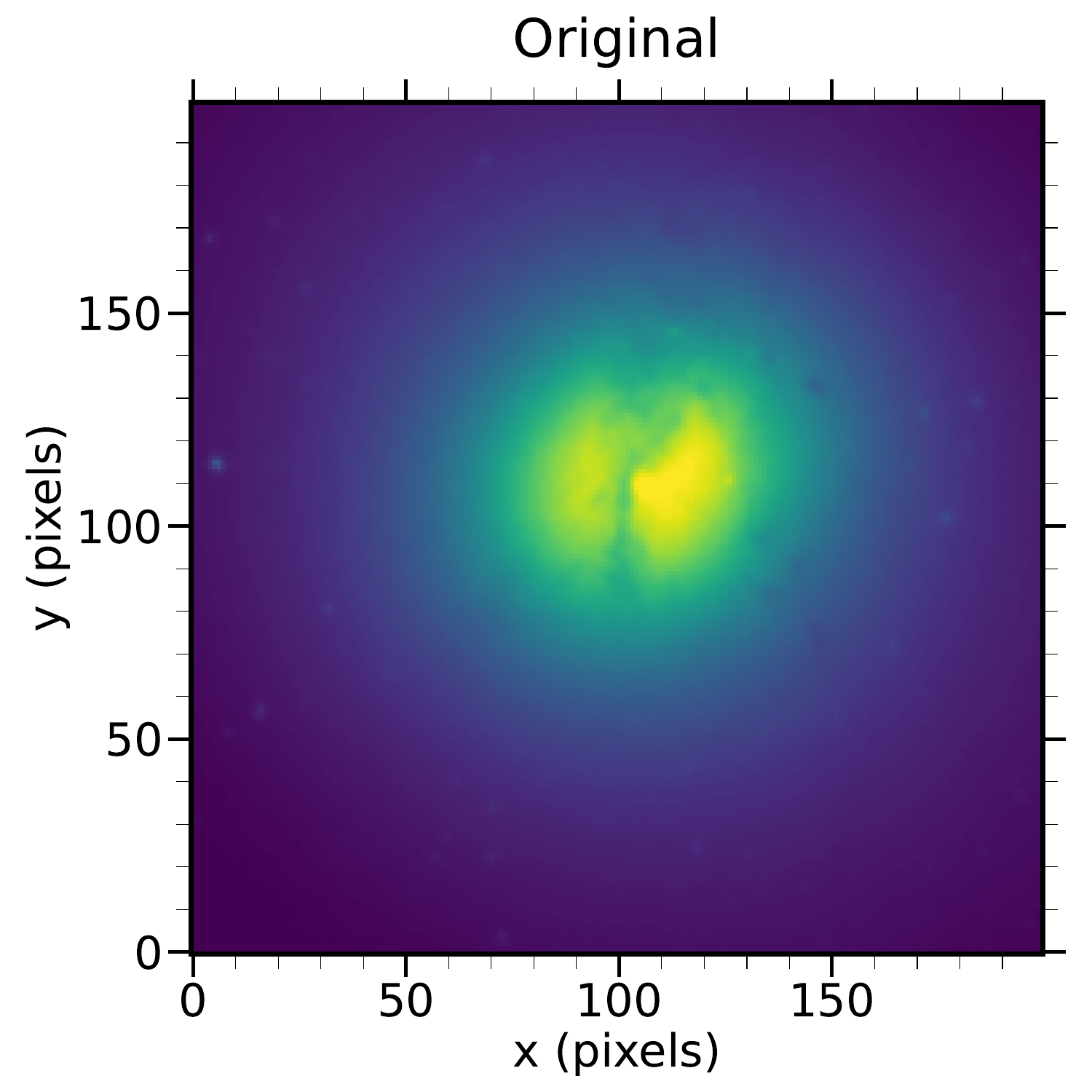}}
\subfloat[\label{Fig.corrected_f110}]{\includegraphics[width=.5\linewidth]{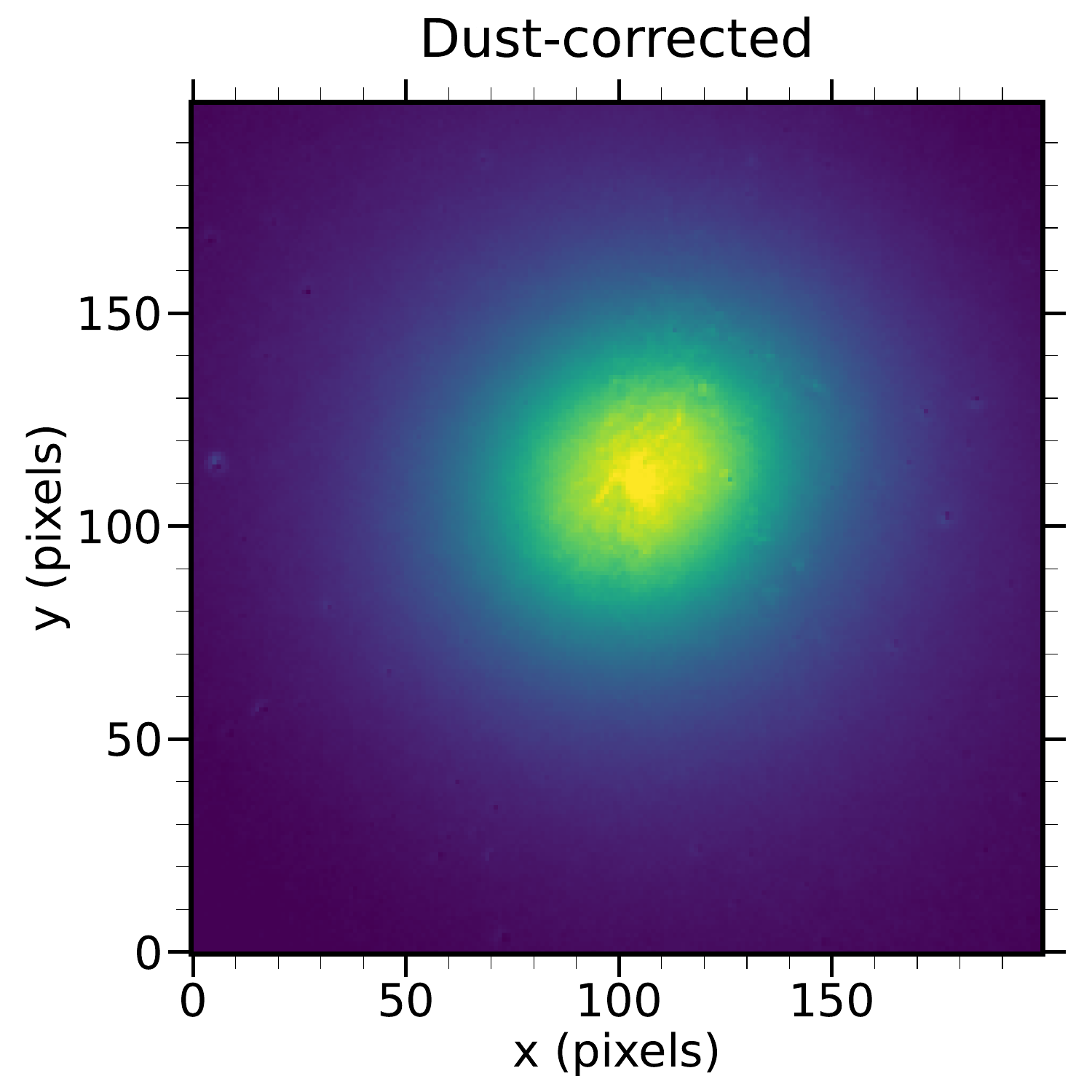}}

    \caption{The central 18" of the HST f110W image of NGC~708 before (left) and after (right) dust-correcting using f814W. A value of 0.9 for the absorption coefficient $\alpha$ was adopted.}
    \label{Fig.dustcorr}
\end{figure*}

\begin{figure*}
\centering
\includegraphics[width=.5\linewidth]{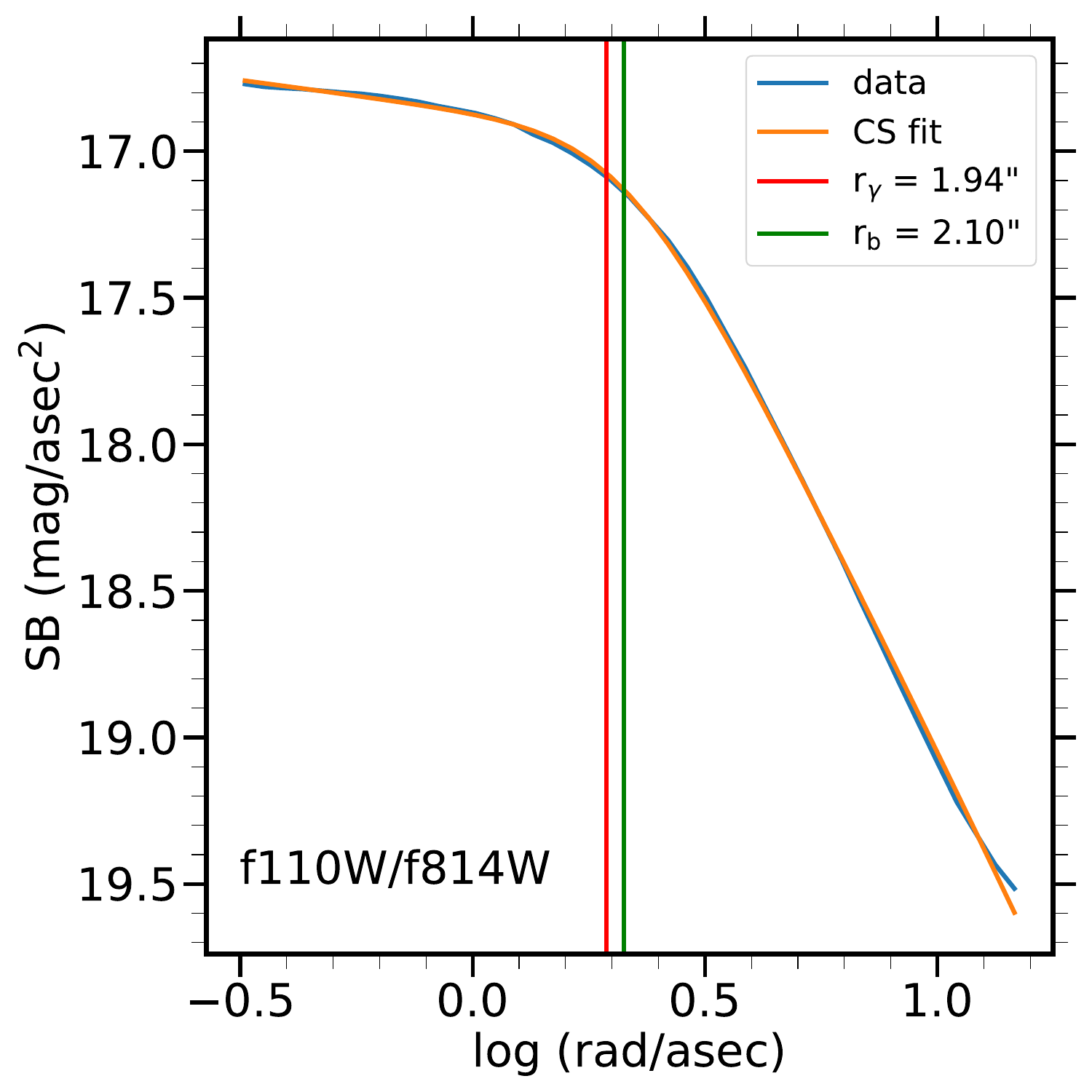}

    \caption{Central SB profile of NGC~708 derived from the dust-corrected f110W image (Fig.~\ref{Fig.corrected_f110}). The blue line is the derived profile using the procedure described in \citet{Matthias23}, while the orange line is the best-fit Core-Sersic (CS) profile. Finally, the vertical lines show the cusp-radius r$_\gamma$ and the break radius r$_\text{b}$.}
    \label{Fig.f814W_SB_corr}
\end{figure*}

The availability of HST images acquired using different filters (in our case f110W, f555W and f814W) allows us to derive dust-corrected images and, hence, multiple estimates of the core radius of NGC~708 for a consistency/stability check with the values derived using the Ks-image acquired at Wendelstein Observatory. The purpose of this appendix is to briefly summarize, following App. A \citet{Nowak08} and Sec. 3.3 of \citet{Bender15}, the adopted procedure. We use as example the correction of the $J$-band f110W image by means of the $I$-band f814W one, but the procedure is the same also for the case where we use f555W to dust-correct f110W. \\
The relation between the $J$-band absorption $A_J \equiv J - J_0$ and the extinction $E(I - J)$ in the $(I - J)$ color is

\begin{equation}
   A_{J} = \alpha E(I - J) \equiv \alpha \left[(I - J) - (I - J)_0 \right] 
    \label{eq.reddening}
\end{equation}

\noindent where the subscript 0 denotes dust-free quantities and $\alpha = 1 / (A_I/A_J - 1)$ is the absorption coefficient. The strongest correction is obtained for $\alpha = 1$, while $\alpha = 0$ implies no correction. Given that the images provide us with fluxes, we can write using eq.~\ref{eq.reddening}

\begin{equation}
    f_{J,0} \propto \frac{f_{J}^{1+\alpha}}{f_I^\alpha}
    \label{eq.reddening_fluxes}
\end{equation}

\noindent where we have assumed that the stellar population gradient is negligible, thus implying $f_{I,0} / f_{J,0} \sim$ constant. \\
After registering the images and interpolating them onto the same pixel grid (f110W has a different pixel scale and a broader field of view), we test different values of $\alpha$, finding that a quite high value of $\alpha = 0.9$ is required to remove the dust (see Fig.~\ref{Fig.dustcorr}) and yield a smooth central SB profile (see Fig.~\ref{Fig.f814W_SB_corr}), which is what we use to estimate the core radius by means of both the cusp-radius r$_\gamma$ and the break radius r$_\text{b}$.

\section{Triaxial models using long-slit kinematics} \label{App.Nbody}
\begin{figure*}

\subfloat{\includegraphics[width=.3\linewidth]{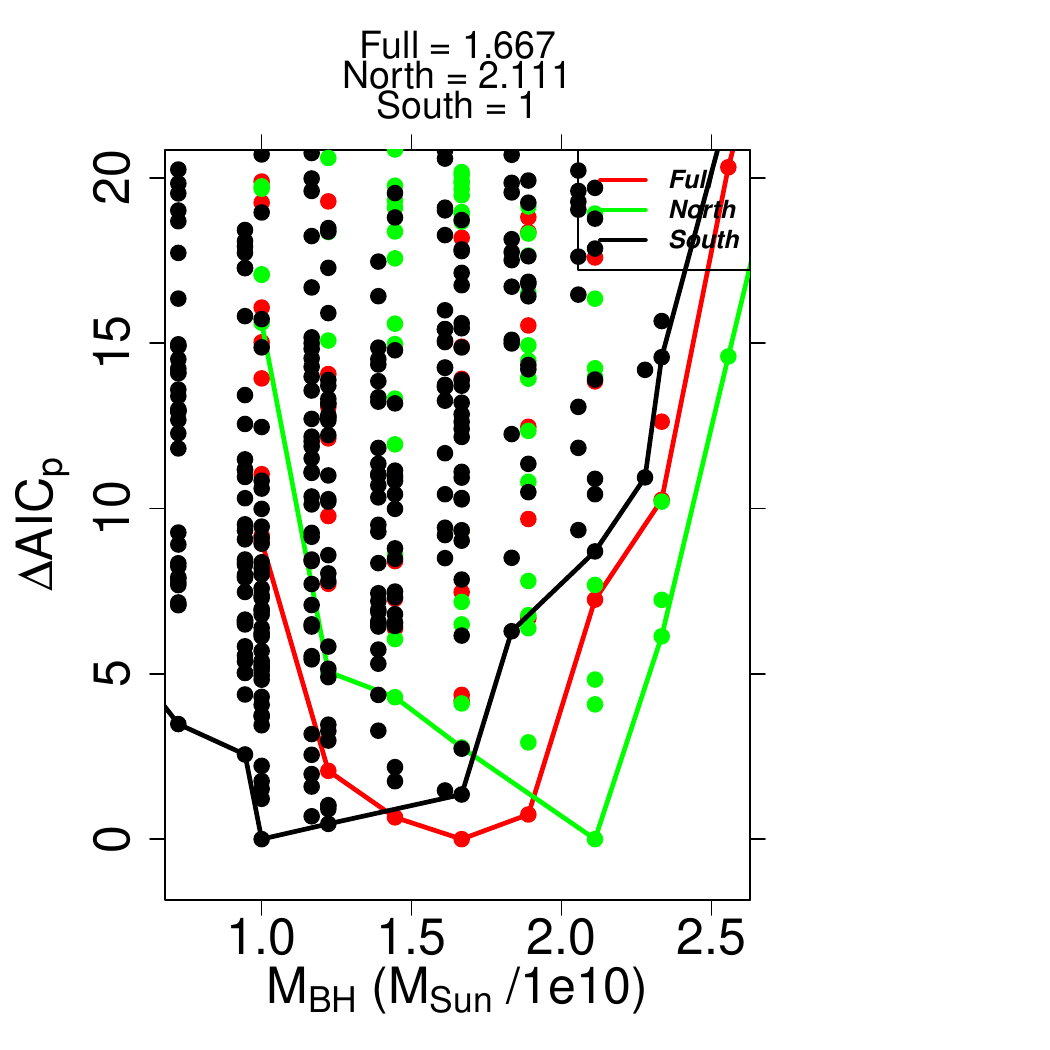}}
\subfloat{\includegraphics[width=.3\linewidth]{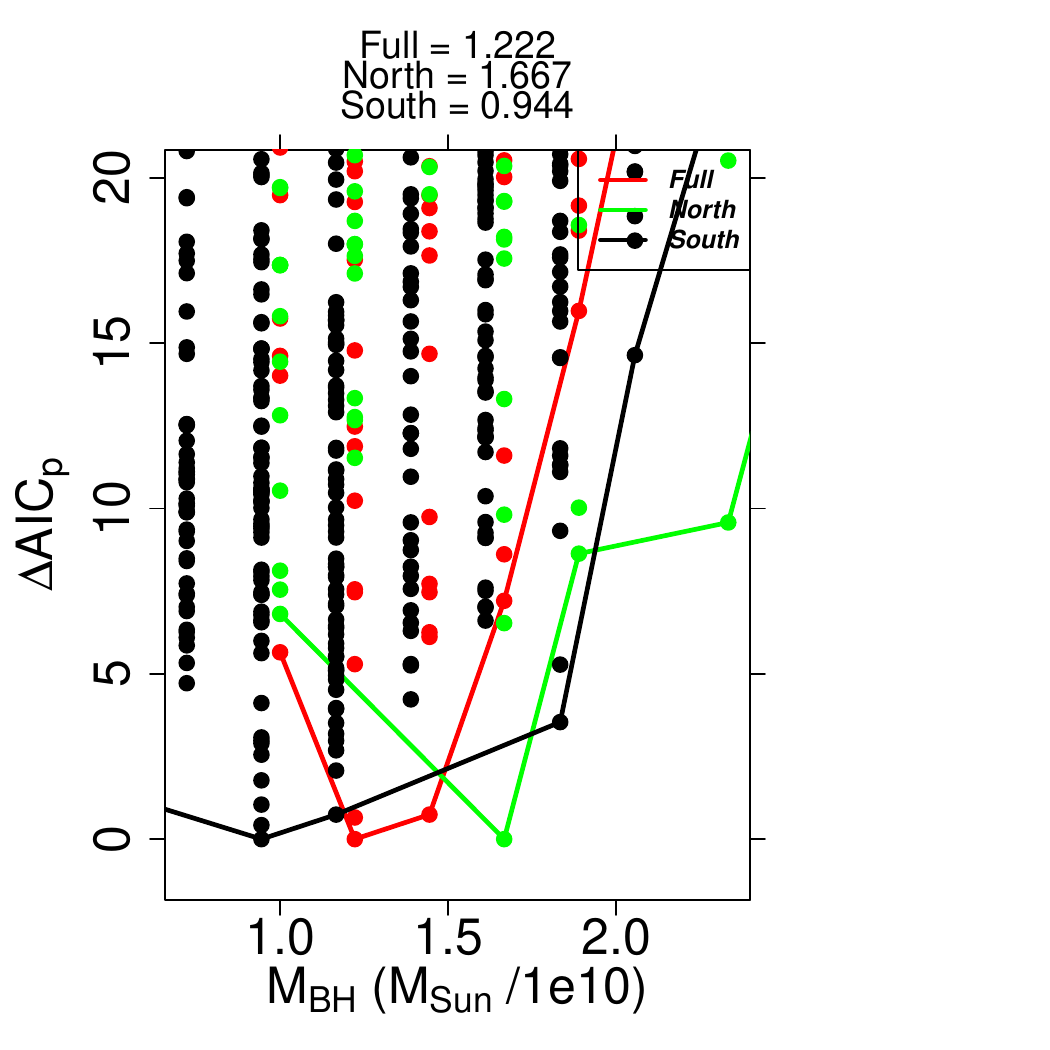}}
\subfloat{\includegraphics[width=.3\linewidth]{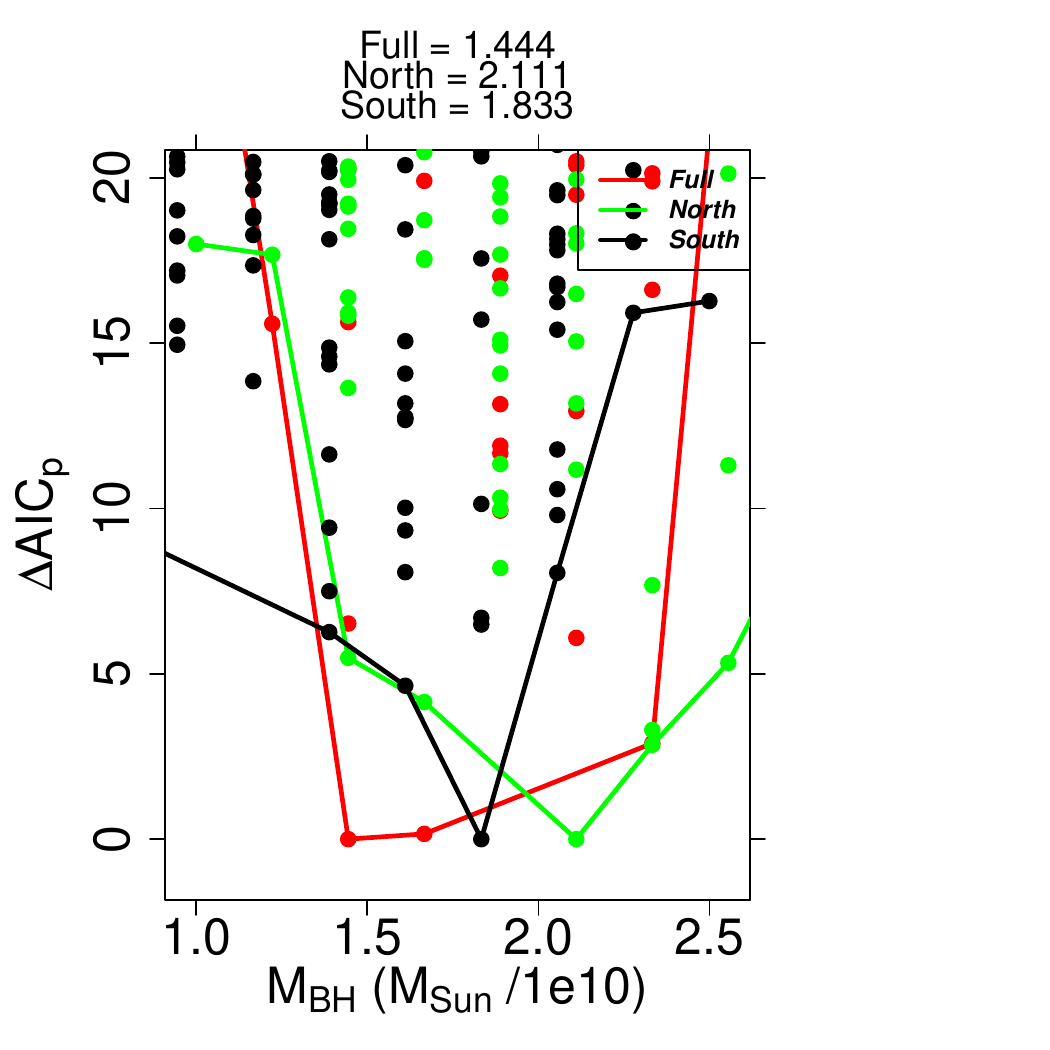}}

\subfloat{\includegraphics[width=.3\linewidth]{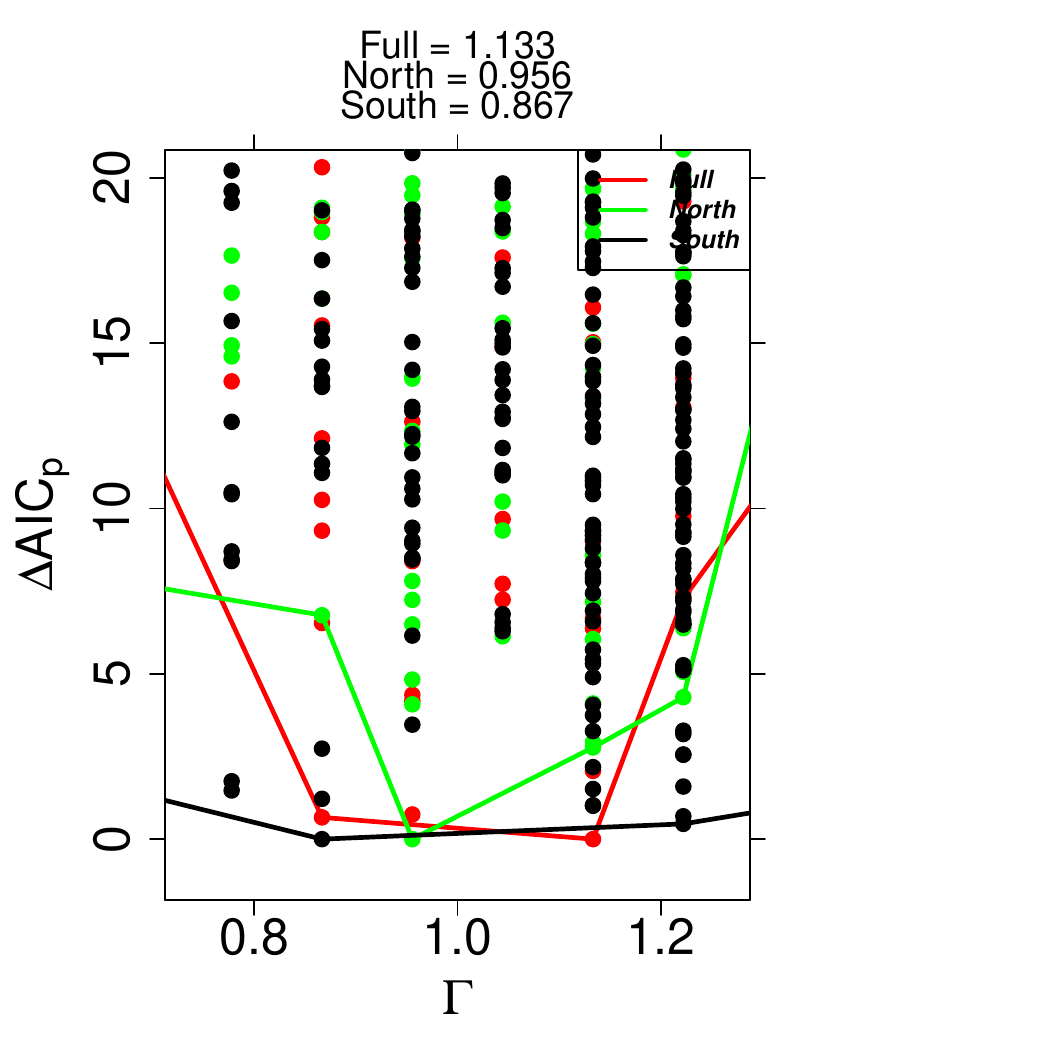}}
\subfloat{\includegraphics[width=.3\linewidth]{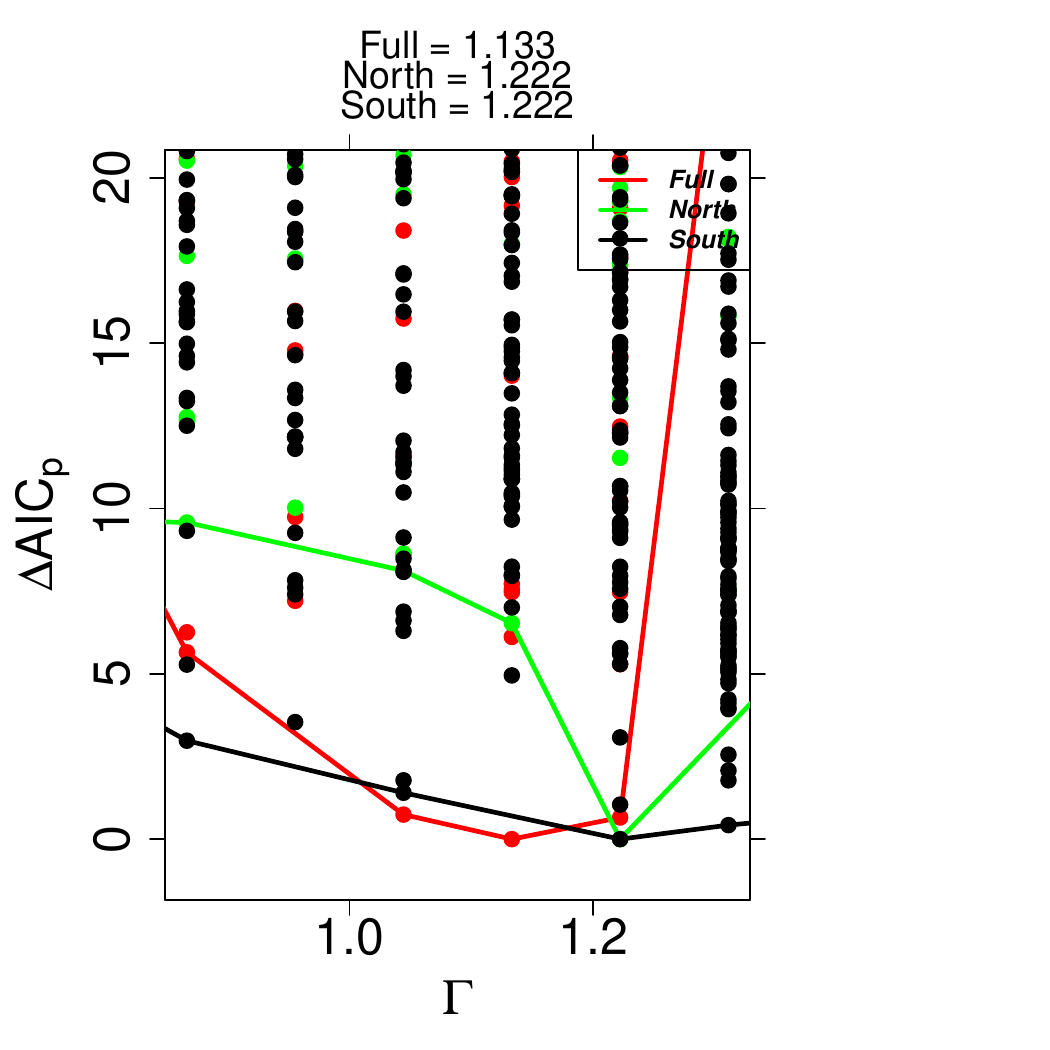}}
\subfloat{\includegraphics[width=.3\linewidth]{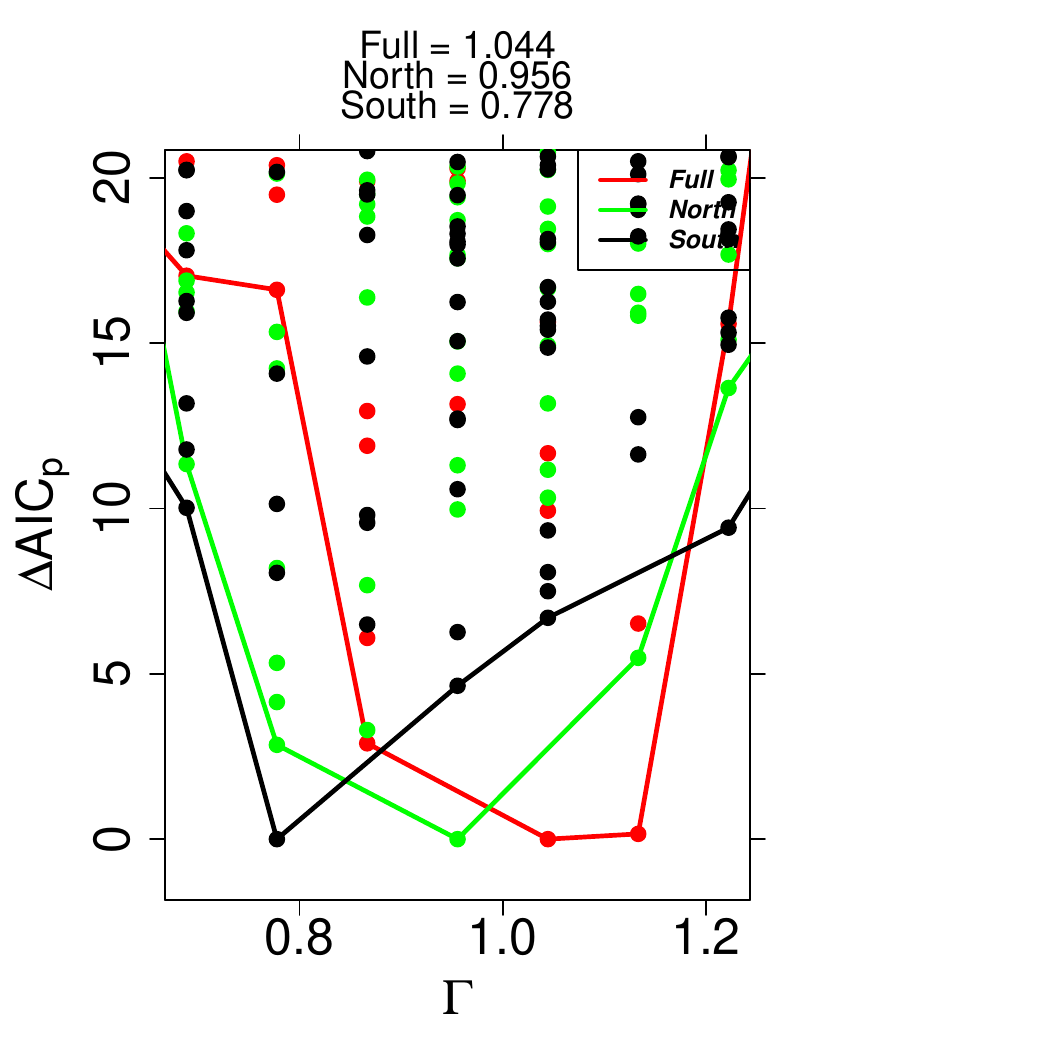}}

    \caption{Similarly as in Fig.~\ref{Fig.Schw_results}, here we show the results of the NOMAD runs for the three mock kinematics, one for each column. The first row shows the \mbh\,recovery whereas the second row shows $\Gamma$. In each plot, the red, green and black lines follow the best-fit model for each tested value for the Full, North and South configurations (see Tab.~\ref{Tab.Nbody}), respectively.}
    \label{Fig.Schw_results_Nbody}
\end{figure*}

\begin{figure}
    \centering
    \includegraphics[width=.8\linewidth]{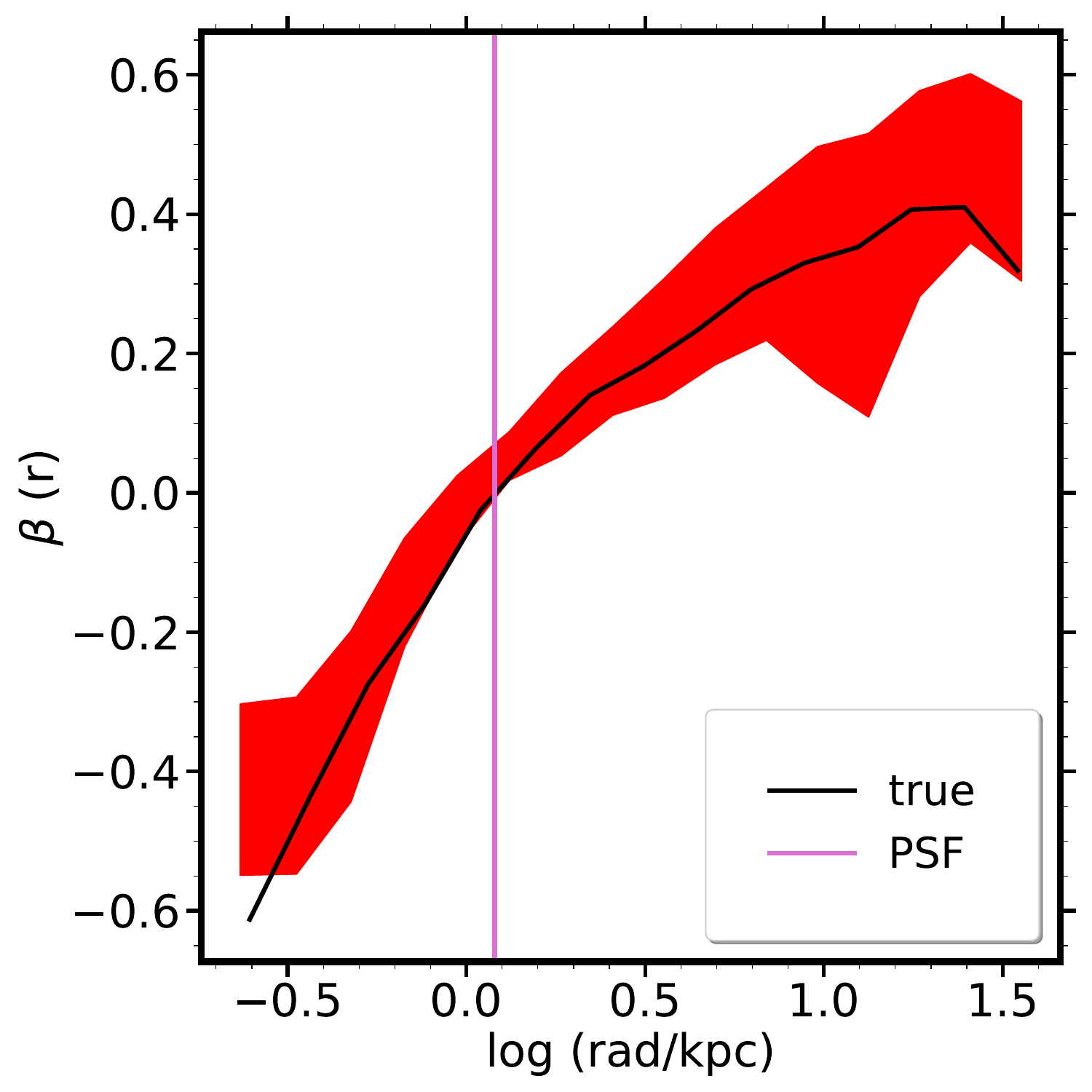}
    \caption{Anisotropy intervals that we get considering all 9 best-fit models. The solid line shows the true $N$-body profile, whereas the vertical line labels the PSF.}
    \label{Fig.beta_Nbdoy}
\end{figure}

This appendix presents the result of the application of SMART to the same $N$-body simulation discussed in \citet{dN22b} and \citet{Bianca23a}, this time using a simulated long-slit kinematics with the same geometry as the one used in this work for NGC~708. This exercise is not only useful to evaluate how accurate our triaxial machinery is with long-slit data, but also to get an estimate of the uncertainties on our mass parameters: given the low number of bins, we need to model the galaxy fitting all bins together rather than split them in two halves and model these separately. Therefore, here we generate three mock kinematics and run independent sets of dynamical models for each one of these. \\
The mock kinematics are generated as follows. The spatial bins for the kinematics are taken at the exact same locations on the plane of the sky as those of NGC~708. After identifying the $N$-body particles found in the $i$-th spatial bin, we sample the $i$-th LOSVD using 15 velocity bins equally spaced in $\left[-1500,1500\right]$ km/s. The particle count inside the $i$-th spatial bin with velocities inside the $j$-th velocity bin gives us $\text{LOSVD}_{\text{data}}^{i,j}$. For a given LOSVD we assume a constant uncertainty for all velocity bins, equal to 3\% of the maximum number of counts among all velocity bins themselves. This gives us error bars comparable to those of NGC~708. We also add random gaussian numbers $\text{N}_{\text{rand}}^{\text{i,j}}$ to each $ij$-th bin to add noise to the data. Finally, we convolve the kinematics with a Moffat profile (see eq. 3 of \citep{Rob93}) with FWMH = 1.4" and $\beta = 2.5$, similar to our MODS observations. Repeating this procedure thrice gives us the three kinematics sets.  \\
Besides modeling all bins together as we did for NGC~708, we also modeled the two halves separately, given that for the simulated kinematics the S/N is much higher, for a total of 9 model runs. Modeling the two halves separately would be the usual strategy provided that the number of bins and the signal-to-noise are large enough.
For each one of the 9 model runs we compute a total of $\sim$1000 models. We do not fit the orientation, but assume it to be $\left(\theta, \phi, \psi\right) = (80,90,135)^\circ$, i.e. the best-fit orientation yielded by SMART for NGC~708. \\
The results of the NOMAD runs are shown in Fig.~\ref{Fig.Schw_results_Nbody} and summarized in Tab.~\ref{Tab.Nbody}. As it can be seen, averaging the results of the three mocks gives estimates which are well within 10\% of the true value. Moreover, having 9 estimates per variable coming from the three mocks, we can also get an estimate on the statistical uncertainties on each variable by taking the standard deviation of the 9 values we got. These uncertainties are what we assume for our NGC~708 mass parameter estimates. Finally, in Fig.~\ref{Fig.beta_Nbdoy} we consider all 9 best-fit models and plot the resulting anisotropy intervals comparing this to the true anisotropy profile of the $N$-body simulation. The fact that the true profile lies well in between the interval is a further proof of the goodness of our triaxial machinery even using long-slit data only.

\begin{table}
    \centering
    \begin{tabular}{c c c c c}
       & Configuration & \mbh / $10^{10}$M$_\odot$ & $\Gamma$ \\ \hline \hline
       true values & - & 1.7 & 1.0 \\ \hline
      mock 1 & Full & 1.67 & 1.13 \\ 
            & North & 2.11 & 0.95 \\ 
            & South & 1.00 & 0.95 \\ \hline
      mock 2 & Full & 1.22 & 1.13 \\ 
             & North & 1.67 & 1.22 \\ 
            & South & 0.94 & 1.22 \\ \hline
        mock 3 & Full & 1.44 & 1.04 \\ 
             & North & 2.11 & 0.96 \\ 
            & South & 1.89 & 0.78 \\ \hline
         \hline
    Result & - &  $1.56 \pm 0.44$ & $1.03 \pm 0.15$ \\ \hline 
    \hline
    \end{tabular}

    \caption{Best-fit \mbh\,and \ml\,values for the $N$-body simulation when modelled using simulated long-slit kinematics. The true values are \\ \mbh\,= 1.7 $\times 10^{10}$ M$_\odot$ and \ml = 1.0. \textit{Col. 1:} Setup name. \textit{Col. 2:} Whether we model all bins together (Full) or we split the bins in two subsets, one for each galaxy half (North and South). \textit{Col. 3-4:} Best-fit \mbh, \ml\,values.}
    \label{Tab.Nbody}
\end{table}


\bsp	
\label{lastpage}
\end{document}